\documentclass[superscriptaddress,showpacs,twocolumn,notitlepage,nofootinbib,aps,pre]{revtex4-2}
\usepackage[normalem]{ulem}
\usepackage{graphicx}
\usepackage{color}
\usepackage{xcolor}
\usepackage{verbatim}
\usepackage{subfigure}
\usepackage{amssymb}
\usepackage{amsmath}
\usepackage{comment}
\usepackage{chngcntr}
\usepackage{appendix}
\usepackage{lipsum}
\usepackage{dsfont}
\usepackage{xfrac}
\usepackage[makeroom]{cancel}
\usepackage{makecell}
\usepackage{lipsum}
\usepackage{mathrsfs}  

\newcommand{\bxi}{\boldsymbol{\xi}}

\newcommand{\bGamma}{\boldsymbol{\Gamma}}
\newcommand{\blambda}{\boldsymbol{\lambda}}
\newcommand{\beeta}{\boldsymbol{\eta}}

\newcommand{\bzeta}{\boldsymbol{\zeta}}
\newcommand{\dd}{\text{d}}

\newcommand{\ee}{\text{e}}

\newcommand{\p}{\partial}

\newcommand{\bR}{{\bm{R}}}

\newcommand{\bT}{{\bf T}}

\newcommand{\bg}{\text{\bf g}}

\newcommand{\bu}{{\bm{u}}}
\newcommand{\br}{{\bm{r}}}

\newcommand{\bW}{{\bm{W}}}

\newcommand{\bv}{\text{\bf v}}

\newcommand{\bnabla}{\boldsymbol{\nabla}}
\newcommand{\bmu}{\boldsymbol{\mu}}

\newcommand{\id}{\mathbf{1}}

\newcommand{\bF}{{\bm{F}}}
\newcommand{\bG}{{\bm{G}}}
\newcommand{\bB}{\text{\bf B}}

\newcommand{\bA}{\text{\bf A}}
\newcommand{\bD}{\text{\bf D}}
\newcommand{\bJ}{{\bm{J}}}
\newcommand{\bP}{{\bm{P}}}
\newcommand{\bO}{{\bm{O}}}

\newcommand{\bM}{\text{\bf M}}

\newcommand{\bnu}{\boldsymbol{\nu}}

\newcommand{\mP}{\mathcal{P}}
\newcommand{\mF}{\mathcal{F}}
\newcommand{\mA}{\mathcal{A}}
\newcommand{\mO}{\mathcal{O}}
\newcommand{\mB}{\mathcal{B}}
\newcommand{\mT}{\mathcal{T}}
\newcommand{\mU}{\mathcal{U}}
\newcommand{\mQ}{\mathcal{Q}}
\newcommand{\mLB}{\mathcal{L}_\text{b}}
\newcommand{\mLo}{\mathcal{L}_\text{o}}

\newcommand{\Teffrot}{T_{\Theta}^{\text{eff}}}
\newcommand{\Tefftrans}{T_{\bR}^\text{eff}}
\newcommand{\Gammaavg}{\langle\Gamma\rangle_{\text{b}}}
\newcommand{\bFavg}{\langle\bF\rangle_{\text{b}}}
\newcommand{\rhob}{\rho_\text{b}}
\newcommand{\rhoo}{\rho_\text{o}}
\newcommand{\rhooss}{\rho_{\text{o}}^\text{ss}}

\usepackage{booktabs}
\usepackage{pifont}
\usepackage{xcolor}
\newcommand{\cmark}{\textcolor{green!80!black}{\ding{51}}}
\newcommand{\xmark}{\textcolor{red}{\ding{55}}}

\newcommand{\subfigref}[2]{\hyperref[#1]{\ref*{#1}#2}} 
\DeclareMathOperator{\Tr}{Tr}
\usepackage{amsthm}
\usepackage{bm}
\usepackage[hang, flushmargin]{footmisc}
\usepackage[pdftex]{hyperref}
\begin{document}

\title{Passive objects in a chiral active bath: from micro to macro}

\author{Cory Hargus}
\affiliation{Laboratoire Mati\`ere et Syst\`emes Complexes (MSC), Université Paris Cité  \& CNRS (UMR 7057), 75013 Paris, France}
\author{Federico Ghimenti}
\affiliation{Laboratoire Mati\`ere et Syst\`emes Complexes (MSC), Université Paris Cité  \& CNRS (UMR 7057), 75013 Paris, France}
\affiliation{Department of Applied Physics, Stanford University, Stanford, CA 94305, USA}
\author{Julien Tailleur}
\affiliation{Laboratoire Mati\`ere et Syst\`emes Complexes (MSC), Université Paris Cité  \& CNRS (UMR 7057), 75013 Paris, France}
\affiliation{Department of Physics, Massachusetts Institute of Technology, Cambridge, MA 02139, USA}
\author{Frédéric van Wijland}
\affiliation{Laboratoire Mati\`ere et Syst\`emes Complexes (MSC), Université Paris Cité  \& CNRS (UMR 7057), 75013 Paris, France}
\affiliation{Yukawa Institute for Theoretical Physics, Kyoto University,
Kitashirakawaoiwake-cho, Sakyo-ku, Kyoto 606-8502, Japan}
\begin{abstract}
We present a detailed derivation of the  Langevin dynamics obeyed by a massive rigid body immersed in a chiral active bath. We show how the antisymmetric nature of the noise leads to an unusual relationship between the Langevin equation describing stochastic trajectories and the Fokker-Planck equation governing the time-evolution of the probability density. The chirality of the bath endows the object dynamics with odd diffusivity, odd mobility, and rotational ratchet effects that depend on the object symmetries. For rotationally-symmetric objects, we show that a hidden time-reversal symmetry leads to separate effective equilibrium descriptions for the translational and rotational degrees of freedom. Finally, starting from the bath dynamics, we construct a multipole expansion to quadrupolar order that allows predicting the far-field current and density modulation induced by the object on the bath.
\end{abstract}

\maketitle

\tableofcontents{}  

\section{Introduction}\label{sec:introduction}
The random motion of particles immersed in fluids has been a topic of fundamental interest in statistical mechanics since the work of Einstein, Smoluchowski and Langevin~\cite{Einstein1905, Smoluchowski1906,Langevin1908}.
For equilibrium fluids, projection methods have been developed that, by eliminating the detailed dynamics of the fluid bath, result in an effective stochastic description for the object motion~\cite{nakajimaQuantumTheoryTransport1958,zwanzigEnsembleMethodTheory1960,moriTransportCollectiveMotion1965,van1986brownian}. Equilibrium is then reflected in the relations between fluctuations of a particle's motion and its response to external perturbation~\cite{zwanzigNonequilibriumStatisticalMechanics2001,VanKampen1981,kuboStatisticalPhysicsII1991}.

For nonequilibrium baths, the motion of passive objects is still driven by collisions with the bath particles, but the irreversible nature of the bath necessitates distinct theoretical approaches~\cite{agarwal1972,Speck2006,Seifert2010,Baiesi2009,Baiesi2013,bertiniMacroscopicFluctuationTheory2015}.
Recently this topic has attracted considerable attention, due the combined use of microrheology to characterize biological media~\cite{MasonWeitz1995,CrockerWeitz2000,wilhelmOutofEquilibriumMicrorheologyLiving2008,robertVivoDeterminationFluctuating2010,puertasMicrorheologyColloidalSystems2014,reichhardtActiveMicrorheologyActive2015} and the need to account for the rich phenomenology observed in active-matter experiments~\cite{wu,di2010bacterial,sokolov2010swimming,Sokolov2012,bechingerActiveParticlesComplex2016,anandTransportEnergeticsBacterial2024a}. Further, chiral active fluids---composed of biological~\cite{Diluzio2005,Riedel2005,Drescher2009,beppu2021edge} or synthetic~\cite{Kummel2013,Nourhani2016,Soni2019,Witten2020} active particles that spin or turn with a favored direction---are known to exhibit odd viscosity due to the breaking of time-reversal and parity symmetries~\cite{avron1995viscosity,Avr98,banerjee2017odd,Epstein2020,Hargus2020,han2021fluctuating,fruchartOddViscosityOdd2023}, and to impart odd diffusivity~\cite{Hargus2021,Muzzeddu2022,Kalz2022,vegareyesDiffusiveRegimesTwodimensional2022,Yasuda2022,ghimentiSamplingEfficiencyTransverse2023, ghimenti2024irreversible, ghimenti2024transverse, langerDanceOdddiffusiveParticles2024,muzzedduSelfdiffusionAnomaliesOdd2024,kalz2024oscillatory,guoDiffusionActiveParticles2025} and odd mobility~\cite{Kogan2016,Reichhardt2019,hosakaNonreciprocalResponseTwodimensional2021,Poggioli2023odd,battonMicroscopicOriginTunable2024,yangTopologicallyProtectedTransport2021,lierLiftForceOdd2023,kalzReversalTracerAdvection2025} to immersed objects.
Yet, although these odd transport and response coefficients arise from altogether widespread microscopic origins, a formalism describing their relationship is missing.
In particular, because of their  non-relaxational nature~\cite{hargusFluxHypothesisOdd2025}, it has remained an open question whether such odd coefficients should obey generalized fluctuation-dissipation and Einstein relations, including in circumstances where their even ({i.e.~symmetric}) counterparts do.

In our companion Letter~\cite{companionPRL}, we have shown how an effective Langevin description of the motion of massive objects accounts for the stochastic dynamics emerging from interactions with a chiral active bath.
This revealed the existence of effective equilibrium regimes for rotationally-symmetric objects and how breaking geometric symmetries of the object leads to richer and increasingly irreversible dynamics. These results are summarized in Table~\ref{tab:object-symmetries}.
In this article, we first detail in Section~\ref{section:effective-langevin} how to apply a projection method to derive the Langevin dynamics of the object, together with Kubo and Agarwal formulas for the transport and response coefficients, respectively. In Section~\ref{section:tracer-symmetries}, we then study how the structure of this Langevin equation is determined by the object symmetries. We then characterize the consequences of this structure in Section~\ref{section:diffusion-mobility-einstein}, showing that rotationally-symmetric objects admit an effective equilibrium description with distinct translational and rotational temperatures. In Section~\ref{section:confined-density-and-flux}, we study these equilibrium dynamics in the presence of external traps. In Section~\ref{section:entropy-production-and-hidden-TRS}, we rationalize the existence of the effective equilibrium dynamics by showing that the entropy production vanishes under a hidden time-reversal symmetry, drawing parallels with the thermal motion of a charged particle in a magnetic field. Finally, in Section~\ref{section:multipole}, we turn our attention to the bath itself, developing a multipole approximation that connects the local active ratchet properties of the object to its long-range influence on the structure and flows of the bath.

\section{Effective Langevin equation for the object}\label{section:effective-langevin}
\subsection{Setting the stage}

Given a rigid object of arbitrary shape immersed in a two-dimensional bath of chiral active particles, we build a stochastic description of the object dynamics by integrating out the bath degrees of freedom in the adiabatic limit of large object mass $M$ and moment of inertial $I$. To do so, we consider the Newtonian dynamics of the object given by
\begin{equation}\label{eq:newton}
	\begin{split}
		\dot \bR &= \frac{1}{M}\bP\,, \\
        \dot \bP &= \bF\,, \\
     \end{split}
     \quad
     \begin{split}
        \dot \Theta &= \frac{1}{I}L\,, \\
		\dot L &= \Gamma\,.
	\end{split}
\end{equation}
Here, $\bR$ is the position of the object's center of mass, $\bP$ its momentum, $\Theta$ its orientation with respect to the $x$-axis, and $L$ its angular momentum. $\bF$ and $\Gamma$ are the total force and torque exerted on the object due to pairwise interactions between the object and the bath particles. We model the latter as chiral active Brownian particles, which evolve as
\begin{equation}\label{eq:cABP-EOM}
    \begin{split}
        \gamma \dot\br_i &= \bF_i + f_0 \bu(\theta_i) \,,\\
        \dot{\theta}_i &= \omega_0 + \sqrt{2 D_r} \xi_i\,,
    \end{split}
\end{equation}
where $\gamma$ is the friction from the substrate, $f_0 \bu$ is the self-propulsion force oriented along $\bu(\theta_i) = [\cos(\theta_i), \sin(\theta_i)]^T$, $\omega_0$ is a constant angular drift, $D_r$ is the rotational diffusivity, and $\xi_i$ is a centered, unit-variance Gaussian white noise.
Finally, $\bF_i$ is the force on bath particle $i$ due to interactions with the object, satisfying $\sum_i \bF_i = -\bF$. In the absence of collisions with the object, the bath particles self-propel with speed $v_0 = f_0/\gamma$, and are characterized by two active lengthscales. The persistence length $\ell_p = \frac{f_0}{D_r \gamma}$ describes how far the particle travels before losing memory of its direction, while the gyroradius $\ell_g = \frac{f_0}{|\omega_0| \gamma}$ describes the average curvature of trajectories due to chirality.

The evolution of an observable 
\begin{equation}\label{eq:observables}
    O(t) = O\big(\bR(t), \bP(t), \Theta(t), L(t), \br(t)^N,\theta(t)^N\big)\,,
\end{equation} 
which depends on the object and bath degrees of freedom can be computed using the It\=o formula. Following~\cite{liu2021dynamics}, it can be written in terms of the linear equation
\begin{equation}\label{eq:dynamics_observable}
    \p_t O(t) = \left[\mLo^\dagger + \mLB(t)^\dagger\right]O(t)\,,
\end{equation}
where $\mLo^\dagger$ and $\mLB(t)^\dagger$ are evolution operators acting respectively on the object and the bath degrees of freedom: 
\begin{align}\label{eq:operators}
    \begin{split}
            \mLo^\dagger &= \frac{\bP}{M}\cdot\bnabla_\bR + \bF\cdot\bnabla_{\bP} + \frac{L}{I}\p_{\Theta} + \Gamma \p_L\,, \\
             \mLB(t)^\dagger &= \sum_i \frac{1}{\gamma} [\bF_{i} + f_0\bu(\theta_i)]\cdot \bnabla_{\br_i} \\
             &+\sum_i  [D_r \p_{\theta_i} + \omega_0 +\xi_i(t)]\p_{\theta_i}\,.
     \end{split}   \raisetag{2.5\baselineskip}
\end{align}
The operator $\mLB(t)^\dagger$ depends on time through the realizations of the noise $\xi_i(t)$. The Laplacian terms $\p_{\theta_i}$ appearing in the definition of $\mLB(t)^\dagger$ are a consequence of It\={o}' s lemma. This formalism is analogous to the one involving Liouville operators for Hamiltonian dynamics in the original Mori-Zwanzig formalism~\cite{zwanzigEnsembleMethodTheory1960, moriTransportCollectiveMotion1965}, where the noise is treated as an external time-dependent force.

Writing the evolution of $O(t)$ by means of linear operators allows us to give a formal solution to Eq.~\eqref{eq:dynamics_observable}, namely
\begin{equation}\label{eq:O_t_formal}
    O(t) = \mU[\mLB^\dagger + \mLo^\dagger](t,0)O(0)\,,
\end{equation}
where the initial condition $O(0)$ is propagated in time by means of the evolution operator $\mU[\mLB^\dagger + \mLo^\dagger](t,0)$. The latter is defined as
\begin{equation}
    \mU[\mLB^\dagger + \mLo^\dagger](t,0) \equiv \exp\left[\mT\int_0^t \dd\tau\,[\mLB^\dagger(\tau) + \mLo^\dagger]\right]\,, 
\end{equation}
where $\mT$ is a time-ordering operator, placing earliest times to the right and latest ones to the left~\cite{j2007statistical}. This is consistent with, Eq.~\eqref{eq:O_t_formal} being a solution to Eq.~\eqref{eq:dynamics_observable}.  Within this formalism, the equations of motion of the translational and angular momentum of the object become
\begin{equation}\label{eq:Pdot_Ldot_formal}
    \begin{split}
        \dot\bP(t) &=\mU[\mLB^\dagger + \mLo^\dagger](t,0)\bF(0) \\
        \dot L(t) &= \mU[\mLB^\dagger + \mLo^\dagger](t,0)\Gamma(0)\,.
    \end{split}
\end{equation}
In the spirit of Mazur and Oppenheim~\cite{mazur1970molecular, kim1971molecular, kim1972molecular} and van Kampen~\cite{van1986brownian}, we now integrate out the bath degrees of freedom in Eq.~\eqref{eq:Pdot_Ldot_formal}, transforming it into a closed non-Markovian equation of motion for the object. We work in the regime where the motion of bath is fast compared to motion of the object, such that the timescales associated with $\mLo^\dagger$ and $\mLB(t)^\dagger$ are well-separated. In this regime, we will show that the non-Markovian equation takes the form of a stochastic differential equation, where the action of the bath on the object is decomposed into average forces and torques, an effective friction, and fluctuating forces and torques with Gaussian statistics. As the separation of timescales becomes very large, the effect of the bath on the object is completely determined by the instantaneous configuration of the latter. The effective Langevin equation then becomes Markovian. In this regime, the friction coefficient and the statistics of the random force can be computed from the long-time dynamics of the bath particles with the object held fixed. 

To formalize this adiabatic regime, we introduce the dimensionless parameter
\begin{equation}\label{eq:epsilon}
    \epsilon\equiv \sqrt{\frac{\gamma}{M D_r}}\,,
\end{equation}
which we will make small by taking $M$ large to reach the adiabatic limit\footnote{We can define an effective mass of an achiral ABP as follows: an ABP with self-propulsion speed $v_0 = f_0/\gamma$ transfers a momentum $\Delta p=\gamma v_0/D_r$ during a collision with an obstacle~\cite{Fily2018}, which suggests and effective mass of the bath particles $m_\text{b}\equiv \Delta p/v_0=\gamma/D_r$, and thus $\epsilon \equiv \sqrt{\frac{m_\text{b}}{M}}$.}.
We assume that $M \propto I$, so that adiabaticity in the translational motion of the object coincides with adiabaticity in its rotational motion.
We work for convenience with units such that $\gamma = D_r = 1$. We then introduce a rescaled time $t^*\equiv \epsilon^2 t$, a rescaled mass $M^{*}\equiv \epsilon^2 M $, a rescaled momentum $\bP^*(t^*)  \equiv \epsilon \bP(\epsilon^{-2} t^*)$, a rescaled angular momentum $L^*(t^*) \equiv \epsilon L(\epsilon^{-2}t^*)$, and a rescaled moment of inertia $I^* = \epsilon^2 I$. With these rescaled variables Eq.~\eqref{eq:Pdot_Ldot_formal} becomes
\begin{equation}\label{eq:Pdot_Ldot_formal_rescaled}
    \begin{split}
        \frac{\dd}{\dd t^*}\bP^*(t^*) &\equiv \dot\bP^* =\epsilon^{-1} \mU[\mLB^\dagger + \epsilon \mLo^{*\dagger}](\epsilon^{-2}t^*,0)\bF(0) \\
        \frac{\dd}{\dd t^*} L^*(t^*) &\equiv \dot L^* =\epsilon^{-1} \mU[\mLB^\dagger + \epsilon \mLo^{*\dagger}](\epsilon^{-2}t^*,0)\Gamma(0)\,.
    \end{split}
\end{equation}
with $\mLo^{*\dagger} = \frac{\bP^*}{M^*}\cdot\bnabla_\bR + \bF\cdot\bnabla_{\bP^*} + \frac{L^*}{I^*}\p_{\Theta} + \Gamma \p_{L^*}$. Equation~\eqref{eq:Pdot_Ldot_formal_rescaled} is the starting point of our coarse-graining procedure. At the end of the analysis we will revert back to the unscaled variables. Sections~\ref{sec:projop} through~\ref{sec:friction} contain a pedagogical derivation of our results and the reader is invited, in a first reading, to jump directly to Sec.~\ref{sec:final}, where the resulting dynamics for the object is presented.

\subsection{Projection operator formalism}\label{sec:projop}

To obtain an effective Langevin equation from Eq.~\eqref{eq:Pdot_Ldot_formal_rescaled} we need to integrate out the degrees of freedom of the active chiral bath. To proceed, we note that the observable $O(t)$ defined  in Eq.~\eqref{eq:observables} can be seen using Eq.~\eqref{eq:O_t_formal} as a function of $\bR(0), \bP(0), \Theta(0), L(0), \br(0)^N,\theta(0)^N$ at an initial time $t=0$ and of the realization of the noises $\xi_i(t')$ for $t'\in [0,t]$. We then define an  operator $\mP$  which acts on  $O(t)$ as
\begin{equation}\label{eq:P_def}
    \begin{split}
        \mP O(t) &\equiv \Biggl\langle \int \dd \br^N \dd\theta^N \rhob(\br^N,\theta^N|\bR, \Theta) O(t)\Biggr\rangle  \\
        &\equiv \langle O(t) \rangle_\text{b}\,,
    \end{split}
\end{equation}
where $\rhob(\br^N,\theta^N|\bR, \Theta)$ weights the bath initial conditions at $t=0$ and the average $\langle\ldots\rangle$ is carried over the realizations of the noises $\xi_i(t'\in[0,t])$. As a result, $\mP O(t)$ is a function of $\bR(0), \bP(0), \Theta(0), L(0)$. 
We choose $\rhob$ such that it corresponds to the steady-state distribution of the bath for a fixed position of the object, i.e.
\begin{equation}
    \langle \mLB(t) \rhob\rangle=0\,,
\end{equation}
where $\mLB(t)$ is the operator adjoint to $\mLB(t)^\dagger$. Further, the operator $\mP$ is a projector because $\mP\mP =\mP$. Noting that
\begin{equation}\label{eq:PL_B}
    \mathcal{P}\mLB(t)^\dagger = \mLB(t)^\dagger \mathcal{P} = 0\,,
\end{equation}
we see that $\mP$ can be physically interpreted as a projector along a direction orthogonal to that of the evolution of the bath degrees of freedom induced by $\mLB$. We denote by $\mQ$ the projection operator orthogonal to $\mP$, defined by $\mQ \equiv 1 - \mP$.

With this choice of $\mP$, the projected observable behaves as though the bath relaxes instantaneously to its steady state distribution with the object degrees of freedom effectively frozen. The space over which $\mP$ projects is thus tailored to capture the dynamical regime with widely separated timescales.

Our goal is to transform Eq.~\eqref{eq:Pdot_Ldot_formal_rescaled} into an effective Langevin equation for the object.   Using the operator identity~\cite{j2007statistical}
\begin{equation}\label{eq:operator_identity}
    \begin{split}
        \mU[\mA](t,0)  &=  \mU[\mA + \mB](t,0)\\
        &-\int_0^t\dd\tau\, \mU[\mA]( t,\tau)\mB\mU[\mA+\mB](\tau,0)\,,
    \end{split}
\end{equation}
with $\mA = \mLB^\dagger + \epsilon\mLo^{*\dagger}$, $\mB = -\mP(\mLB(t)^\dagger + \epsilon\mLo^{*\dagger}) = -\epsilon\mP\mLo^{*\dagger}$, we then obtain
\begin{align}\label{eq:Pdot_effective_Langevin}
        &\dot \bP^*(t^*) = \epsilon^{-1}\bF^+(\epsilon^{-2} t^*) \\
        &+ \epsilon^{-2}\int_0^{t^*} \dd\tau\, \mU[\mLB^\dagger + \epsilon\mLo^{*\dagger}](\epsilon^{-2}t^*,\epsilon^{-2}\tau)\mP\mLo^{*\dagger}\bF^+(\epsilon^{-2}\tau)\notag\,,
\end{align}
and
\begin{align}\label{eq:Ldot_effective_Langevin}
        &\dot L^* (t^*) = \epsilon^{-1}\Gamma^+(\epsilon^{-2} t^*) \\
        &+ \epsilon^{-2}\int_0^{t^*} \dd\tau\, \mU[\mLB^\dagger + \epsilon\mLo^{*\dagger}](\epsilon^{-2}t^*,\epsilon^{-2}\tau)\mP\mLo^{*\dagger}\Gamma^+(\epsilon^{-2}\tau)\notag\,.
\end{align}
We have thus transformed the dynamics of $\dot\bP^*$ and $\dot L^*$ into closed non-Markovian ones, involving a projected force $\bF^+(\tau) \equiv  \mU[\mQ(\mLB +\epsilon\mLo)](\tau,0)\bF(0)$ and a projected torque $\Gamma^+(\tau)\equiv  \mU[\mQ(\mLB +\epsilon\mLo)](\tau,0)\Gamma(0)$. The projected force and torque can be thought of as stochastic variables, whose randomness stems both from the bath initial condition and from the realizations of the noises acting on the bath in $[0,t]$. 

The statistics of $\bF^+$ and $\Gamma^+$ are very complicated in the general case. However, in the adiabatic limit, we will show that the instantaneous statistics of $\bF^+(\tau)$ and $\Gamma^+(\tau)$ become identical to the statistics of $\bF_0(\tau) \equiv \mU[\mLB^\dagger](\tau,0)\bF(0)$ and $\Gamma_0(\tau) \equiv \mU[\mLB^\dagger](\tau,0)\Gamma(0)$, which are respectively the force and the torque exerted by the bath on the object when the latter is held fixed and the bath degrees of freedom are allowed to evolve for a time interval $\tau$. The breaking of time-reversal and parity symmetry by the bath allows a nonzero average value of $\bF_0$ and $\Gamma_0$, yielding translational and rotational ratchet effects. In the adiabatic limit, we will show that the fluctuations of the stochastic force and torque around these mean values are Gaussian and that the second term in Eqs.~\eqref{eq:Pdot_effective_Langevin} and \eqref{eq:Ldot_effective_Langevin} are  friction terms.  We now substantiate this picture with the calculation of the statistics of $\bF^+$ and $\Gamma^+$ in the adiabatic limit. 

\subsection{The projected force}
The deviation of the projected force $\bF^{+}$ from the force acting on a fixed object $\bF_0$ can be systematically expanded in powers of the adiabaticity parameter $\epsilon$. The details of this derivation are given in Appendix \ref{app:random_force}. To lowest order in $\epsilon$, we find
\begin{equation}
    \begin{split}
        \langle \bF^+(t)\rangle_\text{b} &= \langle \bF_0(t)\rangle_\text{b} + O(\epsilon)\\
        &= \langle\bF_0\rangle_\text{b}[\Theta(0)] + O(\epsilon)\,,
    \end{split}
\end{equation}
where we have made explicit that $\langle\bF_0\rangle_\text{b}$ solely depends on the orientation of the object at time $0$, due to translational invariance. 
We also note that $\langle\bF_0\rangle_\text{b}[\Theta(0)]$ does not depend on the noises $\xi(t'\in[0,t])$ so that $\langle\bF_0\rangle_\text{b}[\Theta(0)]$ reduces to a steady-state average computation of $\bF_0$ with the object held fixed.
The average value of $\bF^+(t)$ thus matches the average, steady-state force exerted on the object by the bath when the object is held fixed. We obtain similarly, for the torque,
\begin{equation}
    \langle\Gamma^+\rangle_\text{b} = \langle \Gamma_0\rangle_\text{b} + O(\epsilon)\,\,,
\end{equation}
where we note that $\langle \Gamma_0\rangle_\text{b}$ is independent of both $\Theta(0)$ and $\bR(0)$. 

To characterize the fluctuations of $\bF^+(t)$ and $\Gamma^+$, it is useful to introduce
\begin{align}\label{eq:deltas}
    \delta\bF_0 &\equiv\bF_0-\left\langle\bF_0\right\rangle_\text{b},\qquad \delta\Gamma_0 \equiv \Gamma_0-\left\langle\Gamma\right\rangle_\text{b}\\
\label{eq:deltas_+}
    \delta\bF^+ &\equiv\bF^+-\left\langle\bF_0\right\rangle_\text{b},\qquad \delta\Gamma^+ \equiv \Gamma^+-\left\langle\Gamma_0\right\rangle_\text{b}\,.
\end{align}
The two-point correlator of the random force fluctuations $\delta \bF^+$ is shown in Appendix \ref{app:random_force} to obey
\begin{equation}
    \langle \delta \bF^+(0) \otimes \delta\bF^+(\tau)\rangle_\text{b} = \langle\delta \bF_0(0) \otimes \delta\bF_0(\tau)\rangle_\text{b} + O(\epsilon)\,.
\end{equation}\\
Similar results hold for higher-order moments of $\delta\bF$, for $\delta\Gamma^+$, and for their cross-correlations.

The results presented here rely on the assumption that the dynamics of the bath around a fixed object is ergodic, and that the decay of correlation functions of physical observables happens sufficiently fast. This assumption breaks down in the thermodynamic limit of an infinitely large bath, due to the slow relaxation of its hydrodynamic modes~\cite{VanBeijeren1982,Granek2022}. We thus expect our theory to apply when the large-mass limit is taken before the large-system-size limit.

The effective dynamics of Eqs.~\eqref{eq:Pdot_effective_Langevin} and~\eqref{eq:Ldot_effective_Langevin} can then be rewritten as
\begin{widetext}
\begin{align}\label{eq:Pdot_effective_Langevin_F+_as_F0}
    \begin{split}
        \dot \bP^*(t^*) &= \epsilon^{-1}\langle \bF_0 \rangle_\text{b}(\Theta(0))+  \epsilon^{-1}\delta\bF_0(\epsilon^{-2} t^*) + \epsilon^{-2}\int_0^{t^*} \dd\tau\, \mU[\mLB^\dagger + \epsilon\mLo^{*\dagger}](\epsilon^{-2}t^*,\epsilon^{-2}\tau)\mP\mLo^{*\dagger}\bF^+(\epsilon^{-2}\tau)\,\\
        \dot L^*(t^*) &= \epsilon^{-1}\langle \Gamma_0\rangle_\text{b} + \epsilon^{-1} \delta\Gamma_0(\epsilon^{-2}t^*) + \epsilon^{-2}\int_0^{t^*} \dd\tau\, \mU[\mLB^\dagger + \epsilon\mLo^{*\dagger}](\epsilon^{-2}t^*,\epsilon^{-2}\tau)\mP\mLo^{*\dagger}\Gamma^+(\epsilon^{-2}\tau)\,.
    \end{split}
\end{align}
\end{widetext}
The statistics of the random force $\delta\bF_0$ and torque $\delta\Gamma_0$ in the adiabatic limit is what we determine next.
\subsection{Statistics of the random force in the adiabatic limit}\label{sec:noise}
By definition, $\langle\delta\bF_0(\epsilon^{-2}t^*)\rangle_\text{b} = \mathbf{0}$. The two-point correlation functions evaluated at different times $t_1^*$ and  $t_2^*$ are, in the adiabatic limit,
\begin{equation}\label{eq:corr_deltaF}
    \begin{split}
        &\lim_{\epsilon \to 0} \epsilon^{-2}\langle \delta \bF_0(\epsilon^{-2}t^*_1)\otimes \delta\bF_0(\epsilon^{-2}t^*_2)\rangle_\text{b} \\
        &= \lim_{\epsilon \to 0} \epsilon^{-2}\langle \delta \bF_0(0)\otimes \delta\bF_0(\epsilon^{-2}(t^*_2-t^*_1))\rangle_\text{b}\\
        &=\lim_{\epsilon\to 0} \epsilon^{-2} \langle\delta\bF_0 \rangle_\text{b}\otimes \langle \delta \bF_0\rangle_{\text{b}} = \mathbf{0}\,.
    \end{split}
\end{equation}
Here we made use of the stationarity of the dynamics of the bath and of the assumption that for widely separated times the correlations in the bath vanish. Note that the last equality is valid as long as the relaxation time of the bath  diverges slower than $\epsilon^{-2}$ as $\epsilon\to 0$. This is trivially true in our case, since the adiabatic limit is implemented through a large object mass so that the bath relaxation time remains finite in this limit. 

Equation~\eqref{eq:corr_deltaF} shows that the two-point correlator of $\epsilon^{-1}\delta\bF_0$ vanishes when evaluated at different times in the adiabatic limit. In addition, integrating the autocorrelation functions over the intervals $(-\infty,0]$ and $[0,+\infty)$ gives
\begin{widetext}
\begin{equation}\label{eq:noisecorrmat}
    \begin{split}
        \epsilon^{-2}\int_0^{+\infty}\dd \tau \langle \delta\bF_0(\epsilon^{-2}\tau)\otimes \delta\bF_0(0)\rangle_\text{b} &= \int_0^{+\infty}\dd \tau \langle \delta\bF_0(\tau)\otimes \delta\bF_0(0)\rangle_\text{b} \equiv \blambda_{\bP\bP}\,,\\
        \epsilon^{-2} \int_{-\infty}^{0}\dd \tau \langle \delta\bF_0(\epsilon^{-2}\tau)\otimes \delta\bF_0(0)\rangle_\text{b} &= \int_0^{+\infty}\dd \tau \langle \delta\bF_0(0)\otimes \delta\bF_0(\tau)\rangle_\text{b} \equiv \blambda^\mathrm{T}_{\bP\bP}\,.
    \end{split}
\end{equation}
\end{widetext}
This defines the noise correlation matrix $\blambda_{\bP\bP}$, which need not be symmetric in general so that one may have $\blambda_{\bP\bP}^\mathrm{T} \neq \blambda_{\bP\bP}$. Equations~\eqref{eq:corr_deltaF} and~\eqref{eq:noisecorrmat} can be written in a compact form as
\begin{equation}\label{eq:xi_PP_corr}      \lim_{\epsilon\to 0}\epsilon^{-2}\langle\delta\bF_0(\epsilon^{-2} \tau) \otimes \delta\bF_0(0)\rangle_\text{b} =  \blambda_{\bP\bP}\delta_+(\tau) + \blambda^\mathrm{T}_{\bP\bP}\delta_-(\tau)\,.
\end{equation}
Here the distributions $\delta_\pm(t)$ are the two halves of the Dirac distribution. With a test function $f(t)$ which is continuous at $t=0$, these can be defined by
\begin{align}
    2\delta(\tau)&=\delta_-(\tau) + \delta_+(\tau)\,,\label{eq:symmetricdelta}\\
    \int_0^\infty \dd \tau\, \delta_+(\tau)f(\tau) &=\int_{-\infty}^0 \dd \tau\, \delta_-(\tau)f(\tau)=f(0)\,,\\
    \int_{-\infty}^0 \dd \tau\, \delta_+(\tau)f(\tau) &=\int_{0}^{\infty} \dd \tau\, \delta_-(\tau)f(\tau)=0\,.
\end{align}

Analogously, we can compute the torque correlations and the cross correlations
\begin{equation}
    \begin{split}
        \lim_{\epsilon\to0} \epsilon^{-2} \langle\delta\Gamma(\epsilon^{-2}\tau)\delta\Gamma_0(0)\rangle_\text{b}&= \lambda_{LL}\delta(\tau)\,,\\    
        \lim_{\epsilon\to 0} \epsilon^{-2}\langle\delta \bF(\epsilon^{-2}\tau) \delta\Gamma(0)\rangle_\text{b} &= \blambda_{\bP L}\delta_+(\tau) + \blambda_{L \bP}^\mathrm{T}\delta_-(\tau)\,,
    \end{split}
\end{equation}
with the noise correlations $\blambda_{\bP L}$, $\blambda^\mathrm{T}_{L\bP}$, and $\lambda_{LL}$ defined as
\begin{equation}
    \begin{split}
        \blambda_{\bP L} &\equiv \int_0^{+\infty}\dd \tau \langle \delta\bF_0(\tau)\delta\Gamma_0(0)\rangle_\text{b}\,,\\
        \blambda_{L\bP}^\mathrm{T} &\equiv \int_0^{+\infty} \dd \tau \langle \delta\Gamma_0(\tau)\delta\bF_0(0)\rangle_\text{b}\,,\\
        \lambda_{LL} &\equiv \int_0^{+\infty}\dd \tau \langle\delta\Gamma_0(\tau)\delta\Gamma_0(0)\rangle_\text{b}\,.
    \end{split}
\end{equation}

To express the fluctuating force and torque in a compact way we introduce a stochastic noise  $\epsilon^{-1}\bxi(\epsilon^{-2}\tau)$ in the extended space of the object's linear and angular momentum
\begin{equation}\label{eq:noise_rescaled}
    \begin{split}
        \epsilon^{-1}\bxi(\epsilon^{-2}\tau) &\equiv 
        \begin{bmatrix}
        \epsilon^{-1}\delta\bF_0(\epsilon^{-2}\tau)\\ \epsilon^{-1}\delta\Gamma(\epsilon^{-2}\tau)
        \end{bmatrix} \equiv \begin{bmatrix} \epsilon^{-1}\bxi_\bP(\epsilon^{-2}\tau) \\ \epsilon^{-1}\xi_L(\epsilon^{-2}\tau)\end{bmatrix}\\
        &\equiv \begin{bmatrix} \bxi^*_\bP(\tau) \\ \xi^*_L(\tau)\end{bmatrix} \equiv \bxi^*(\tau)\,,
    \end{split}
\end{equation}
with correlations
\begin{equation}\label{eq:corr_xi_gen}
        \lim_{\epsilon\to0}\langle \epsilon^{-2}\bxi(0)\otimes\bxi(\epsilon^{-2}\tau)\rangle_\text{b} = \blambda\delta_+(\tau) + \blambda^\mathrm{T}\delta_-(\tau)\,,\\  
\end{equation}
which define the diffusivity matrix $\blambda$ as
\begin{equation}
    \blambda \equiv \begin{bmatrix} \blambda_{\bP\bP} & \blambda_{\bP L} \\ \blambda_{L\bP} & \lambda_{LL}
    \end{bmatrix}\,.
\end{equation}
Finally, in the adiabatic limit the statistics of the noise $\bxi$ is Gaussian as shown in Appendix \ref{app:force_is_gaussian}. Note that Eq.~\eqref{eq:symmetricdelta} shows that when $\blambda$ is symmetric, $\bxi$ is a standard Gaussian white noise. With the statistics of $\bxi$ at hand, we now show that, in the adiabatic limit, the second term on the right-hand side of Eq.~\eqref{eq:Pdot_effective_Langevin_F+_as_F0} contributes a friction term to the object dynamics.
\subsection{The friction matrix}\label{sec:friction}

We now consider the time integral appearing in Eq.~\eqref{eq:Pdot_effective_Langevin_F+_as_F0} for the translational momentum $\bP^*$, namely
\begin{equation}\label{eq:integral_P}
    \epsilon^{-2}\int_0^{t^*} \dd\tau\, \mU[\mLB^\dagger + \epsilon\mLo^{*\dagger}](\epsilon^{-2}t^*,\epsilon^{-2}\tau)\mP\mLo^{*\dagger}\bF^+(\epsilon^{-2}\tau)\,.
\end{equation}
In Appendix \ref{app:PLF+}, we show that, in the adiabatic limit, this integral becomes
\begin{align}
\epsilon^{-1}&\big[\langle\bF_0\rangle_\text{b}(\Theta^*(t^*)) -\langle\bF_0\rangle_\text{b}(\Theta^*(0)) \big]\notag\\
&-\bzeta_{\bP\bP}(\Theta^*(t^*))\cdot \frac{\bP^*(t^*)}{M^*}-\bzeta_{\bP L}(\Theta^*(t^*))\frac{L^*(t^*)}{I^*}\,,
\end{align}
where $\Theta^*(t^*)\equiv \Theta(\epsilon^{-2}t^*)$ is the orientation of the object measured in rescaled time, and  the linear momentum and torque friction coefficients $\bzeta_{\bP\bP}$ and $\bzeta_{\bP L}$ are given by Agarwal formulae
\begin{align}
    \bzeta_{\bP\bP}(\Theta^*(t^*)) &= \int_0^{+\infty} \dd\tau\, \langle \delta\bF_0( \tau)\otimes\bnabla_{\bR} \ln\rhob(0)\rangle_\text{b}\,,\\
    \bzeta_{\bP L}(\Theta^*(t^*)) &=\int_0^\infty \dd \tau \langle \delta\bF_0( \tau)\partial_{\Theta^*}\ln\rhob(0)\rangle_\text{b}\;.
\end{align}
Using this and Eq.~\eqref{eq:noise_rescaled}, we rewrite Eq.~\eqref{eq:Pdot_effective_Langevin_F+_as_F0} as a Langevin equation for the linear momentum:
\begin{equation}\label{eq:dynmomrescaled}
    \begin{split}
        \dot \bP^*(t^*)&=\langle \bF_0\rangle_\text{b}(\Theta^*(t^*)) -  \bzeta_{\bP\bP} \frac {\bP^*(t^*)}{M^*} - \bzeta_{\bP L} \frac{L^*(t^*)}{I^*} \\
        &+ \bxi^*_{\bP}(t^*)\;.
    \end{split}
\end{equation}

 A similar analysis can be carried out to study the memory term in the angular-momentum dynamics, to find
\begin{equation}\label{eq:dynandmomrescaled}
    \dot L^*(t^*) = \langle \Gamma_0 \rangle_\text{b} - \bzeta^\mathrm{T}_{L\bP}\cdot\frac{\bP^*(t^*)}{M^*} - \zeta_{LL}\frac{L^*(t^*)}{I^*} + \xi^*_L(t^*)\,,
\end{equation}
where the friction coefficients $\bzeta_{L\bP}$ and  $\zeta_{LL}$ read, respectively
\begin{equation}
    \begin{split}
        \zeta_{LL} &\equiv \int_0^{+\infty}\dd \tau \langle \delta\Gamma_0(\tau)\p_{\Theta^*}\ln\rhob(0)\rangle_\text{b} \\
        \bzeta_{L\bP}^\mathrm{T} &\equiv \int_0^{+\infty}\dd \tau\langle \delta\Gamma_0(\tau)\bnabla_{\bR} \ln\rhob(0)\rangle_\text{b}\,.
    \end{split}
\end{equation}

\if{With the friction coefficients and the statistics of the random force and torques at hand, we can present a  final, compact form of the effective Langevin equation for the object. In the rescaled coordinates and adopting the generalized noise defined in Eq.~\eqref{eq:noise_rescaled}, it reads
\begin{equation}\label{eq:tracerdynamics_rescaled}
  \begin{split}
    \begin{bmatrix}\dot\bP^*(t^*) \\  \dot L^*(t^*) \end{bmatrix} &= \begin{bmatrix} \bFavg(\Theta^*(t^*)) \\ \Gammaavg\end{bmatrix} - \begin{bmatrix} \bzeta_{\bP\bP} & \bzeta_{\bP L} \\ \bzeta_{L\bP} & \zeta_{LL} \end{bmatrix} \begin{bmatrix} \frac{1}{M^*}\bP^* \\ \frac{1}{I^*} L^* \end{bmatrix} \\&+\begin{bmatrix}\epsilon^{-1}\bxi^*_P(\epsilon^{-2}t^*) \\ \epsilon^{-1}\xi^*_L(\epsilon^{-2} t^*)\end{bmatrix}\,.
  \end{split}
\end{equation}}\fi 
In the next section, we revert to the original coordinates of the system and we summarize the results of the projection operator approach.

\subsection{Final result}\label{sec:final}
We reintroduce the original momenta $\bP(t) =\epsilon^{-1}\bP^*(\epsilon^2 t)$, $L(t)=\epsilon^{-1}L^*(\epsilon^2 t)$, mass and moment of inertia $M=\epsilon^{-2}M^*$, $I=\epsilon^{-2}I$ and time $t\equiv \epsilon^{-2}t^*$. We drop the index $0$ from the stochastic force $\bF_0$ and torque $\Gamma_0$. We can thus rewrite  Eqs.~\eqref{eq:dynmomrescaled} and~\eqref{eq:dynandmomrescaled} as
\begin{equation}\label{eq:tracerdynamics}
  \begin{split}
    \begin{bmatrix}\dot\bP(t) \\  \dot L(t) \end{bmatrix} &= \begin{bmatrix} \bFavg \\ \Gammaavg\end{bmatrix} - \underbrace{\begin{bmatrix} \bzeta_{\bP\bP} & \bzeta_{\bP L} \\ \bzeta_{L\bP} & \zeta_{LL} \end{bmatrix} }_{\bzeta}\begin{bmatrix} \frac{1}{M}\bP \\ \frac{1}{I} L \end{bmatrix} +\underbrace{\begin{bmatrix}\bxi_\bP(t) \\ \xi_L(t)\end{bmatrix}}_{\bxi}\,,
  \end{split}
\end{equation}
where $\bFavg$ and $\Gammaavg$ are the steady-state average force and torque exerted on an object held fixed.

The noise $\bxi(t)$ is a Gaussian noise with mean $\langle \bxi(t) \rangle=\mathbf{0}$ and correlations
\begin{equation}\label{eq:noise}
    \langle \bxi(t) \otimes \bxi(t') \rangle = \blambda \delta_+(t-t') + \blambda^\mathrm{T}\delta_-(t-t')\,,
\end{equation}
where the functions $\delta_\pm(t)$ have been defined below Eq.~\eqref{eq:xi_PP_corr}. As shown in Sec. \ref{sec:noise}, the noise correlations are characterized by a $3 \times 3$ momentum diffusivity matrix $\blambda$ given by the Green-Kubo formula~\cite{kubo1957a,Kubo1957b}
\begin{equation}\label{eq:green-kubo}
    \blambda = \int_0^{\infty}\dd \tau \begin{bmatrix} \langle \delta\bF(\tau) \otimes \delta\bF(0)\rangle_\text{b} & \langle\delta\bF(\tau) \delta\Gamma(0)\rangle_\text{b} \\ \langle \delta \Gamma(\tau) \delta\bF^\mathrm{T}(0)\rangle_\text{b} & \langle\delta\Gamma(\tau) \delta\Gamma(0)\rangle_\text{b}\end{bmatrix}\,.
\end{equation}
Likewise, as shown in Sec. \ref{sec:friction} the $3\times 3$ friction matrix $\bzeta(\Theta)$ is given by an Agarwal formula~\cite{agarwal1972}
\begin{equation}\label{eq:kubo-agarwal}
    \bzeta = \int_0^{\infty}\hspace{-3mm}\dd \tau\begin{bmatrix} \langle \delta\bF(\tau) \otimes \bnabla_{\bR} \ln \rho_\text{b} (0)\rangle_\text{b} & \langle \delta\bF(\tau)\partial_\Theta\ln \rho_\text{b} (0)\rangle_\text{b}\\ \langle \delta\Gamma(\tau)\bnabla_\bR^\mathrm{T}\ln \rho_\text{b} (0)\rangle_\text{b} & \langle \delta\Gamma(\tau)\partial_\Theta\ln \rho_\text{b} (0)\rangle_\text{b} \end{bmatrix}\hspace{-1mm}\,.
\end{equation}
Note that the average force $\bFavg(t)$, and the friction and momentum diffusivity matrices depend on the object orientation at time $t$, $\Theta(t)$.

The effective Langevin equation that describes the adiabatic dynamics of an object in a chiral active bath given by Eq.~\eqref{eq:tracerdynamics}, together with the expressions of the friction and diffusion matrices in terms of correlation functions, is one of the main results of this paper. 

Let us now discuss the properties of the probability current of the object's degrees of freedom, and the evolution of their associated probability density predicted by our theory. 

\subsection{Noise, currents and probabilities}\label{sec:current}
The noise described by Eq.~\eqref{eq:noise} differs from the usual Gaussian white noise appearing in the classical derivation of Brownian motion~\cite{mazur1970molecular}. This is due to the chiral nature of the bath, and requires some careful considerations when the adiabatic limit is taken, as also happens when considering the overdamped limit of equilibrium Langevin dynamics in an external magnetic field \cite{chun2018emergence}. The correlator of Eq.~\eqref{eq:noise} is to be understood as the Markovian limit, implemented by sending the parameter $\epsilon$ in Eq.~\eqref{eq:epsilon} to 0, of a Gaussian process $\bxi_\epsilon(t)$:
\begin{equation}\label{eq:markovian_from_nonmarkovian}
    \langle \bxi(t) \otimes \bxi(t') \rangle = \lim_{\epsilon \to 0} \langle \bxi_\epsilon(t)\otimes \bxi_\epsilon(t')\rangle\,.
\end{equation}
The memory kernel of $\bxi_\epsilon(t)$ decays over a time scaled by $\epsilon^2$, consistently with the scaling adopted in the projection operator approach. In many cases, computations have to be carried out using a finite $\epsilon$, before taking the $\epsilon\to 0$ limit. This is for instance the case when computing the entropy production rate as done  in Section~\ref{section:entropy-production-and-hidden-TRS}. We expect our results to be independent of the specific form of the correlations of $\bxi_\epsilon$, as long as Eq.~\eqref{eq:markovian_from_nonmarkovian} is respected.

Next, from Eq.~\eqref{eq:tracerdynamics}, we can define and derive a probability current. We introduce a generalized coordinates system encompassing both translational and rotational degrees of freedom, defining the momentum $\bW = [\bP, L]^\mathrm{T}$, force $\langle\bG\rangle = [\bFavg, \Gammaavg]^\mathrm{T}$, and mass matrix $\mathcal{M} = \mathrm{diag}([M, M, I])$.
Equation~\eqref{eq:tracerdynamics} is then expressed as
\begin{equation}\label{eq:tracerdynamics-vectorial}
    \begin{split}
    \dot\bW &= \langle\bG\rangle_\text{b} - \mathcal{M}^{-1} \bzeta \cdot \bW +\bxi_\epsilon(t) \\
    &= \langle\bG\rangle_\text{b} - \tilde{\bzeta} \cdot \bW +\bxi_\epsilon(t)\,,
    \end{split}
\end{equation}
where in the second line we absorb the mass matrix into the friction matrix, defining $\tilde{\bzeta} \equiv \mathcal{M}^{-1}\bzeta$. Let $\bnu(t) \equiv [\bR(t), \Theta(t), \bW(t)]^\mathrm{T}$ denote the set of degrees of freedom of the object. We define the current of the system $\bJ(\bnu,t)$ as
\begin{equation}\label{eq:bJ_definition}
    \bJ(\bnu,t) \equiv  \langle \dot\bnu(t) \delta(\bnu -\bnu(t))\rangle\,, 
\end{equation}
The current $\bJ(\bnu,t)$ contains a term of the form $\langle\bxi_\epsilon(t)\delta(\bnu - \bnu(t))\rangle$, which can be computed using the Novikov relation \cite{novikov1965functionals}
\begin{equation}\label{eq:novikov}
    \begin{split}
        \langle\bxi_\epsilon(t)\delta(\bnu - \bnu(t))\rangle &= -\int_0^t \dd \tau\, \langle\bxi_\epsilon(t) \otimes \bxi_\epsilon(\tau)\rangle\\
        &\cdot \bnabla_{\bW}\left\langle\frac{\delta\bW(t)}{\delta\bxi_\epsilon(\tau)}\delta(\bnu - \bnu(t))\right\rangle\,.
    \end{split}
\end{equation}
We have introduced the notation $\left[\frac{\delta \bW(t)}{\delta\bxi_{\epsilon}(\tau)}\right]_{ij} \equiv \frac{\delta W_{i}(t)}{\delta \xi_{\epsilon,j}(\tau)}$. Using the fact that in the adiabatic limit the statistics of $\bxi_\epsilon$ converges to the one of $\bxi(t)$, we show in Appendix \ref{app:novikov} that, as $\epsilon\to0$, Eq.~\eqref{eq:novikov} reduces to
\begin{equation}\label{eq:Novikov_to_result}
    \lim_{\epsilon\to0} \langle\bxi_\epsilon(t)\delta(\bnu - \bnu(t))\rangle = -\blambda \bnabla_\bW \rhoo(\bnu,t)\,,
\end{equation}
where $\rhoo(\bnu,t)\equiv \langle \delta(\bnu - \bnu(t)\rangle$ is the probability density of the object degrees of freedom.
The current associated with Eq.~\eqref{eq:tracerdynamics} in the adiabatic limit is thus
\begin{equation}\label{eq:J_expression}
    \bJ(\bnu,t) \equiv \begin{bmatrix} \bP/M\\L/I\\\langle \bG \rangle_\text{b} - \tilde{\bzeta}\bW -\blambda\bnabla_\bW \end{bmatrix}\rhoo(\bnu,t)\,.
\end{equation}
In situations with Gaussian white noise with symmetric correlation matrices, the current $\bJ$ is readily related to the probability density $\rhoo$ through a Fokker-Planck equation, $\p_t\rhoo=-\bnabla_{\bnu} \cdot \bJ$. In this case, however, the singular nature of the adiabatic limit demands  a more careful treatment. To obtain a Fokker-Planck equation for $\rhoo$ we consider a time-coarse graining of the effective Langevin equation~\eqref{eq:tracerdynamics-vectorial}
\begin{equation}\label{eq:tracer_coarse_grained_time}
    \overline{\dot\bW} = \overline{\langle\bG\rangle_\text{b}} - \overline{\tilde{\bzeta}\bW} + \overline{\bxi(t)}\,, 
\end{equation}
where, for any time dependent vector $\bv(t)$, we have introduced a time-coarse grained vector $\overline{\bv(t)} \equiv \lim_{\delta t\to 0} \frac{1}{2\delta t}\int_{t-\delta t}^{t+\delta t}\dd \tau \bv(\tau)$. Using Eq.~\eqref{eq:noise}, we see that the correlations of the time-coarse grained noise are given by
\begin{equation}
    \langle\overline{\bxi(t)} \otimes \overline{\bxi(t')}\rangle = \delta(t-t')\left[\blambda + \blambda^\mathrm{T}\right]\equiv 2\delta(t-t')\blambda_S\,,
\end{equation}
The time-coarse grained noise has a symmetric correlation matrix $\blambda_S \equiv \frac{1}{2}\left[\blambda + \blambda^{\mathrm{T}}\right]$. We can thus write a Fokker-Planck equation \cite{van1992stochastic} for the probability density associated with the time-coarse grained-variables
\begin{equation}\label{eq:FPrhoT}
    \p_t\rhoo(\overline{\bnu},t) = -\bnabla_{\overline{\bnu}} \cdot\overline{\bJ}(\overline{\bnu},t)\,,
\end{equation}
where the coarse-grained probability current is defined as
\begin{equation}
    \overline{\bJ}(\overline{\bnu},t) \equiv \begin{bmatrix} \overline{\bP}/M \\ \overline{L}/I \\ \overline{\langle\bG\rangle_\text{b}} - \tilde{\bzeta}\overline{\bW} + \blambda_S\bnabla_\bW\end{bmatrix}\rhoo(\overline{\bnu},t)\,.
\end{equation}
Only the symmetric part of the correlation matrix thus contributes to the Fokker-Planck equation. This loss of information is a byproduct of the time-coarse-graining procedure. The Fokker-Planck equation~\eqref{eq:FPrhoT} correctly describes the evolution of the probability density of the object degrees of freedom. However, since the dynamics of the object is, in essence, non Markovian, the correct expression for the probability current  is instead given by Eq.~\eqref{eq:J_expression}. We note again that these subtleties are very similar to those reported in~\cite{chun2018emergence} for the study of an underdamped charged particle in a magnetic field.

With the Fokker-Planck equation and an expression for the current of the system at hand, we now have a complete framework to study the motion of the object. 
\begin{table*}[t]
  \centering
    \setlength{\tabcolsep}{.8em}
    \setlength{\heavyrulewidth}{1.5pt}
    \renewcommand{\arraystretch}{2.5}
    \normalsize
    \begin{tabular}{l c c c c c c c c}\toprule
    \makecell[l]{\textbf{Symmetry group}} & \multicolumn{2}{c}{$SO(2)$} & \multicolumn{2}{c}{$C_n \cup \Pi$} &
    \multicolumn{2}{c}{$C_n$} &
    \multicolumn{2}{c}{$\Pi$} \\

    \makecell[l]{\textbf{Examples}} &
    \multicolumn{2}{c}{\raisebox{-.5\totalheight}{\includegraphics[width=0.08\textwidth]{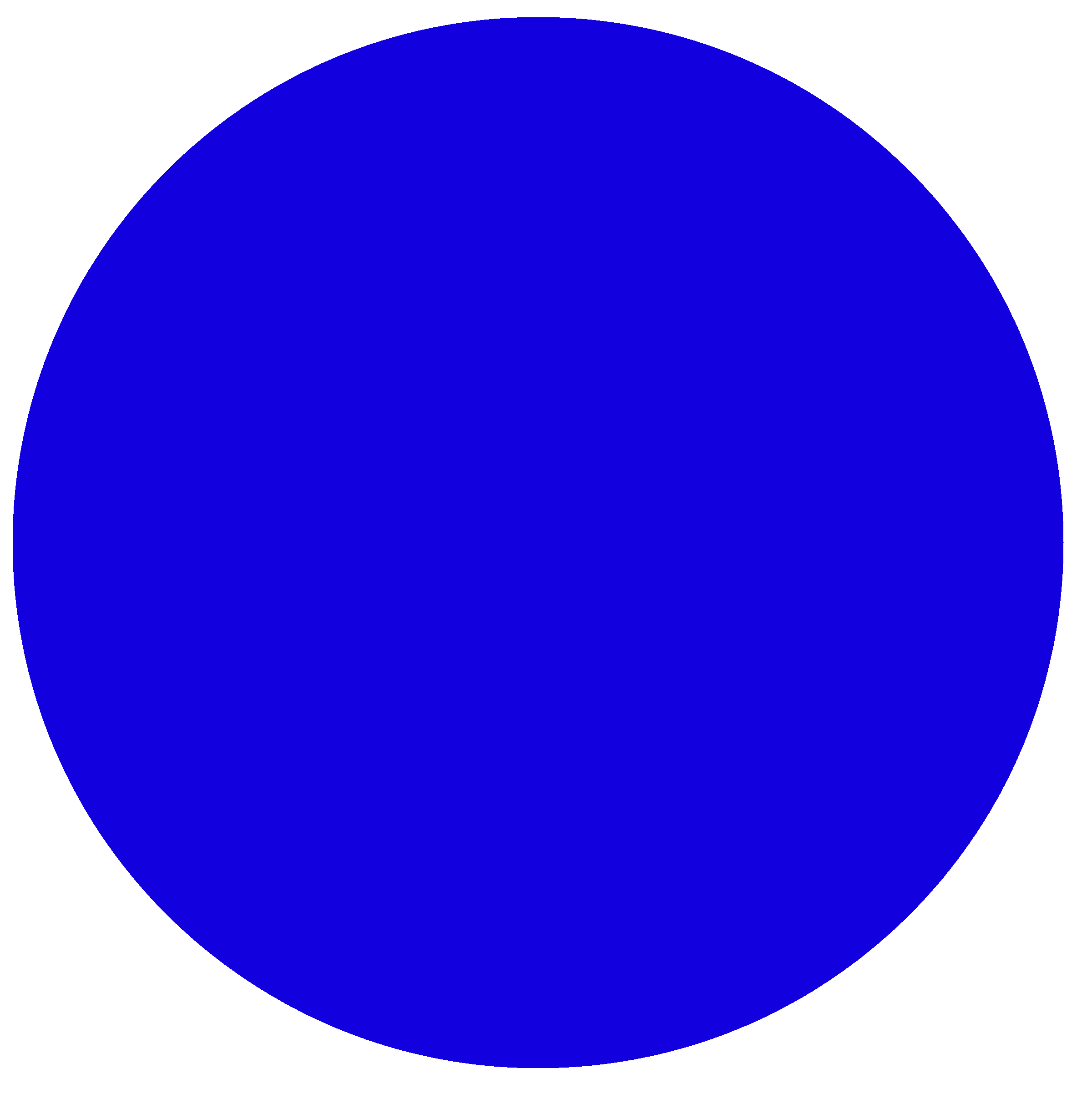}}} &

     \multicolumn{2}{c}{
     \makecell{\raisebox{-.5\totalheight}{\includegraphics[width=0.1\textwidth]{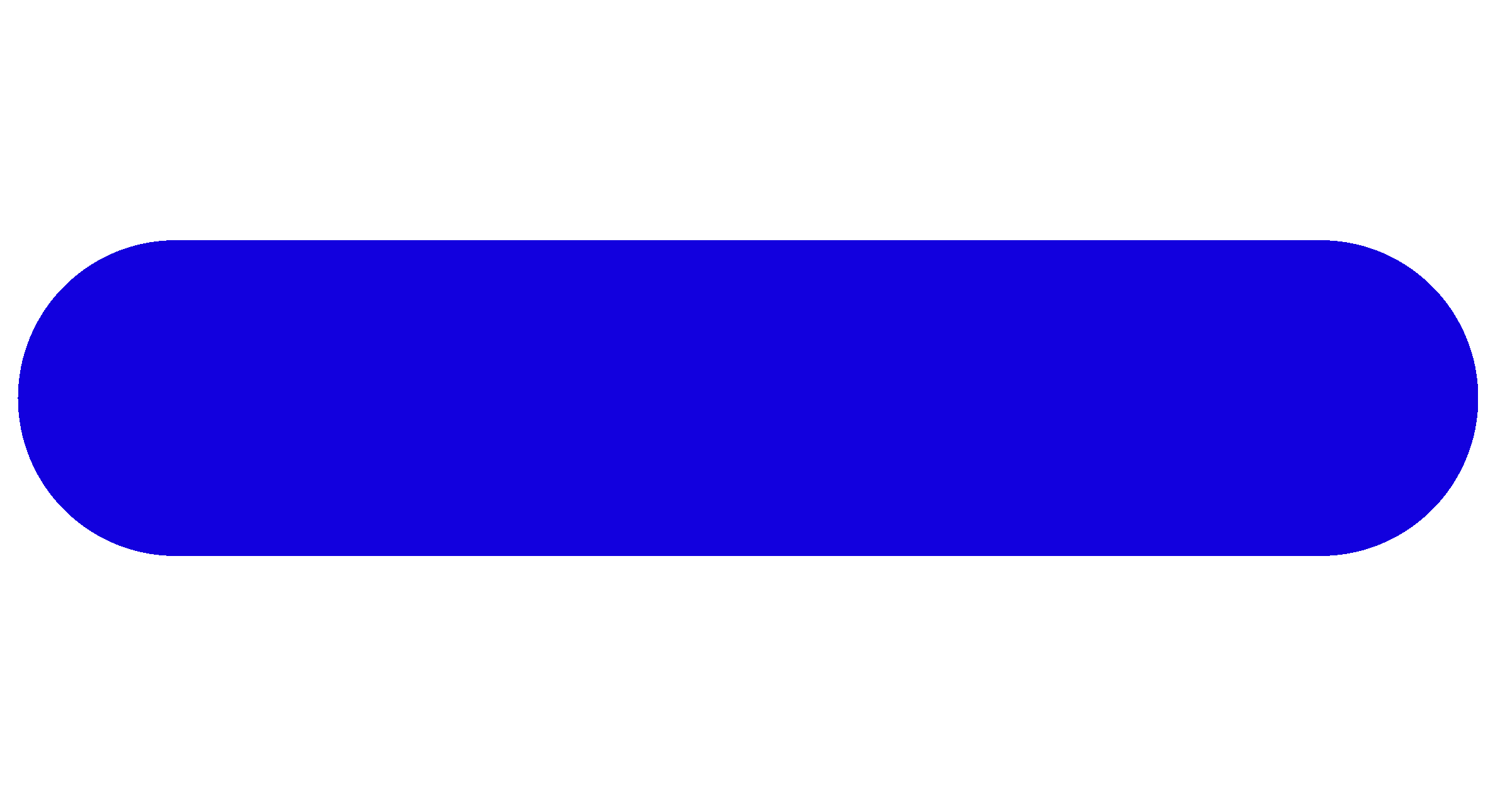}}\\
     \raisebox{-.5\totalheight}{\includegraphics[width=0.1\textwidth]{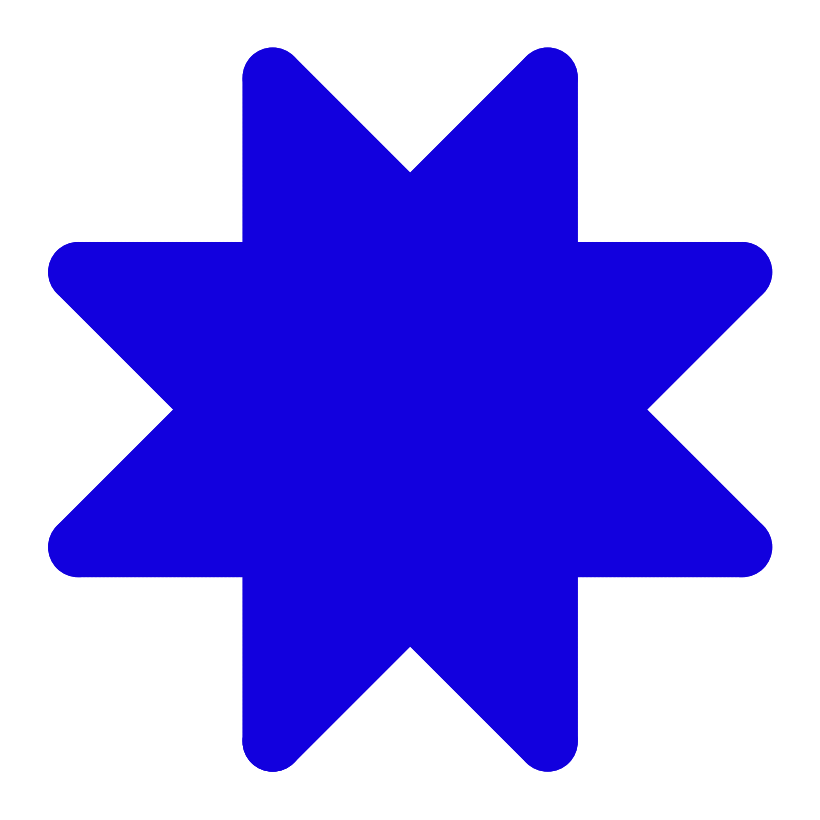}}}} &

     \multicolumn{2}{c}{
     \makecell{\raisebox{-.5\totalheight}{\includegraphics[width=0.1\textwidth]{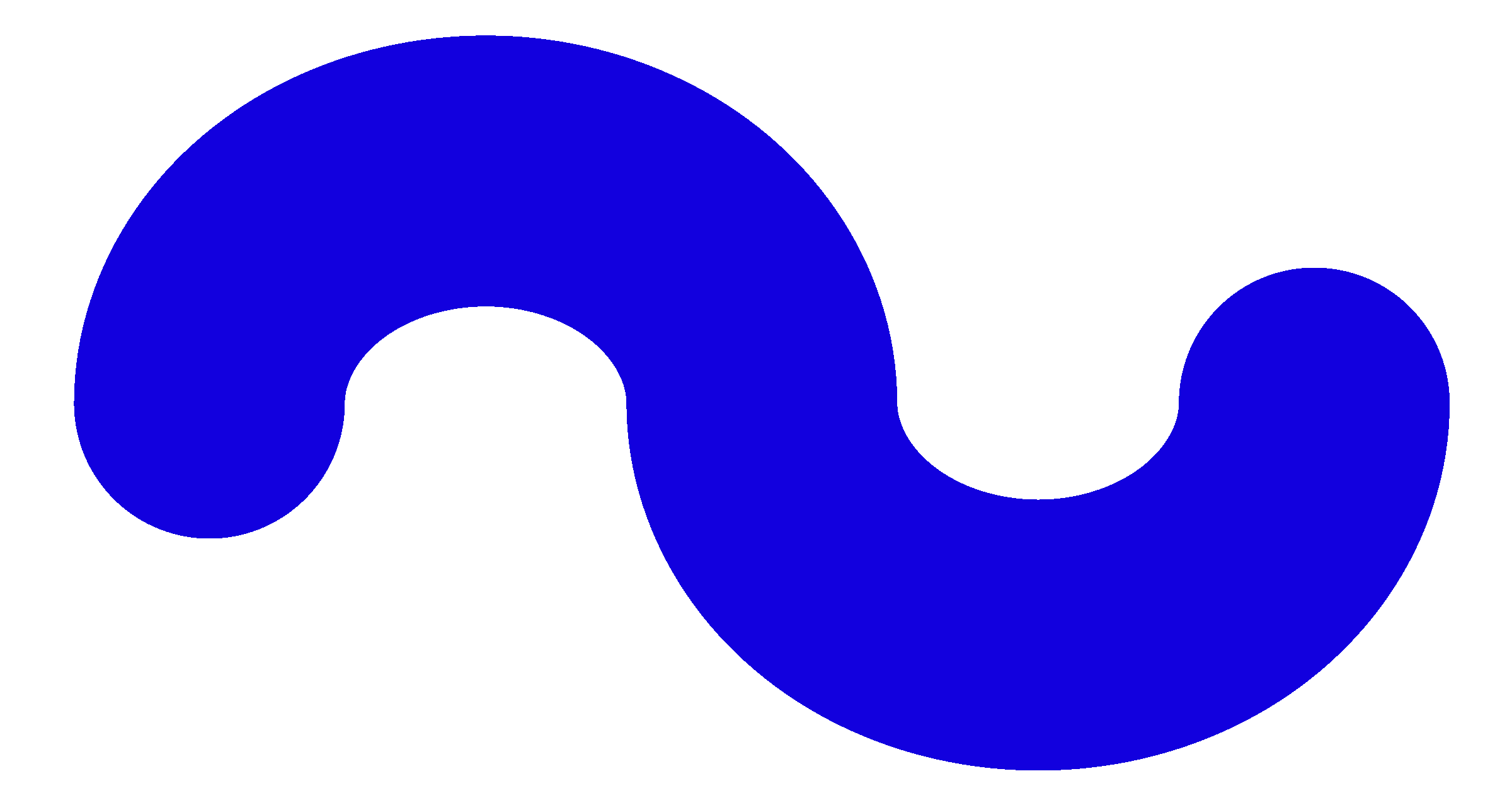}}\vspace{2mm}\\
     \raisebox{-.5\totalheight}{\includegraphics[width=0.1\textwidth]{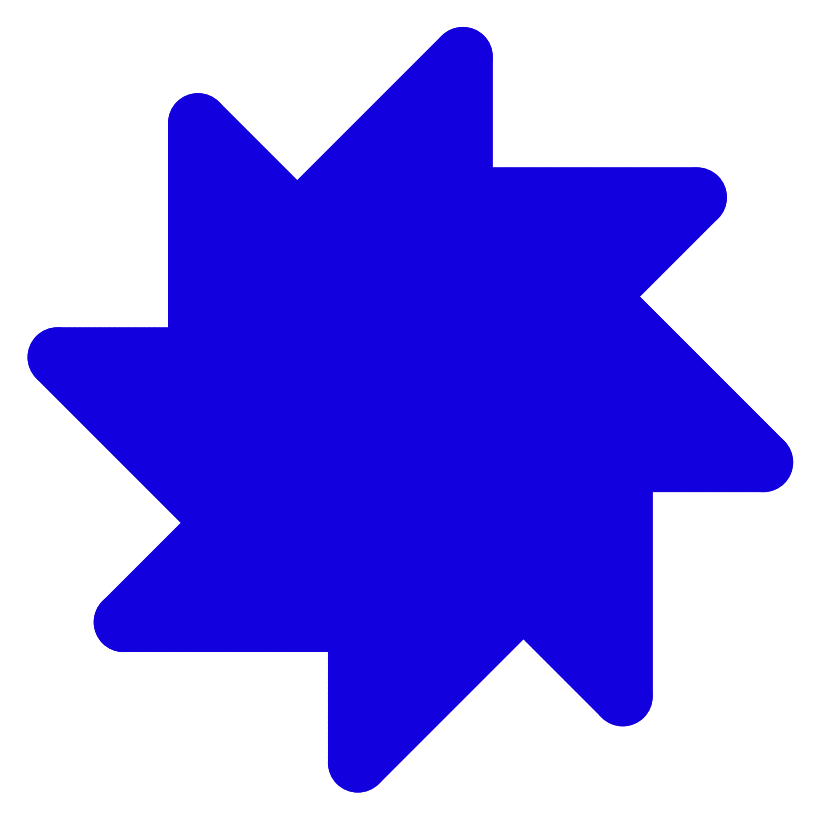}}}} &

          \multicolumn{2}{c}{\raisebox{-.5\totalheight}{\includegraphics[width=0.08\textwidth]{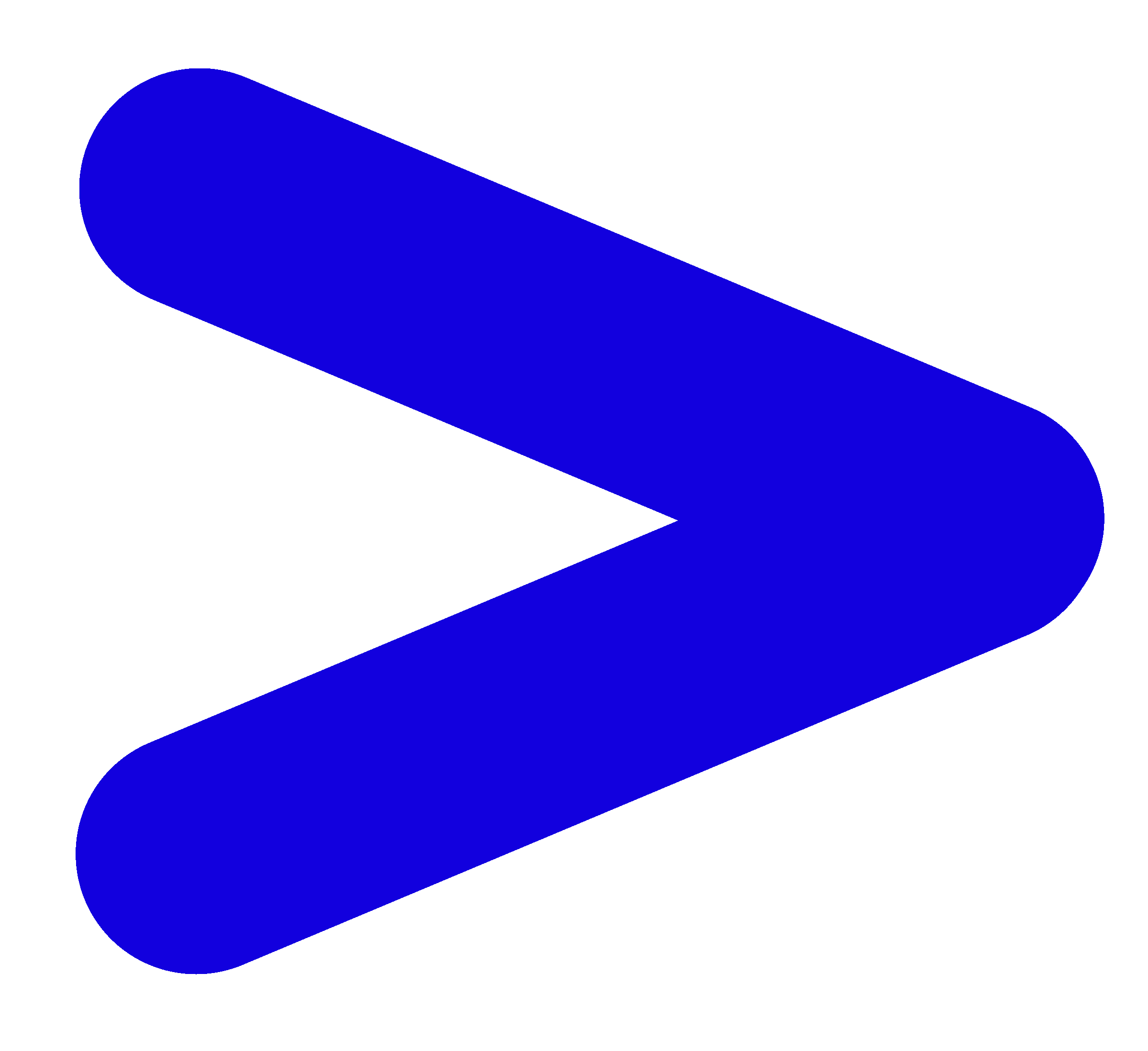}}} \\
     \cmidrule(lr){2-3} \cmidrule(lr){4-5} \cmidrule(lr){6-7} \cmidrule(lr){8-9}
     \makecell[l]{\textbf{Bath chirality}}&
     \makecell{achiral\\$\omega=0$} & \makecell{chiral\\$\omega\ne0$} &
     \makecell{achiral\\$\omega=0$} & \makecell{chiral\\$\omega\ne0$} &
     \makecell{achiral\\$\omega=0$} & \makecell{chiral\\$\omega\ne0$} &
     \makecell{achiral\\$\omega=0$} & \makecell{chiral\\$\omega\ne0$}  \\\midrule
    \makecell[l]{\textbf{Odd dynamics} \\\quad $D_\perp$, $\mu_\perp$, $\lambda_\perp$, $\zeta_\perp$} & \xmark & \cmark & \xmark & \cmark & \cmark & \cmark & \xmark & \cmark \\
    \makecell[l]{\textbf{Net torque} $\Gammaavg$} & \xmark & \xmark & \xmark & \cmark & \cmark & \cmark & \xmark & \cmark \\
    \makecell[l]{\textbf{Net force} $\bFavg$} & \xmark & \xmark & \xmark & \xmark & \xmark & \xmark & \cmark & \cmark \\
    \makecell[l]{\textbf{Cross couplings}\\\quad $\bzeta_{\bP L}$, $\bzeta_{L \bP}$, $\blambda_{\bP L}$, $\blambda_{L \bP}$} & \xmark & \xmark & \xmark & \xmark & \xmark & \xmark & \xmark & \cmark \\\bottomrule
  \end{tabular}
  \caption{Consequences of a passive object's symmetry group on its dynamics in the adiabatic limit. The symbols \xmark\ and \cmark\ denote zero or nonzero values, respectively, of the terms in the first column. (Objects breaking both $C_n$ and $\Pi$ symmetries contain all terms regardless of the bath chirality.)}
  \label{tab:object-symmetries}
\end{table*}

\section{How object symmetries impact transport parameters}\label{section:tracer-symmetries}

We now discuss the impact of the object shape on its dynamics, explaining under what circumstances certain terms in Eq.~\eqref{eq:tracerdynamics} vanish due to symmetry, as summarized in Table~\ref{tab:object-symmetries}.
A circular object (i.e.~ a disk) is isotropic, being invariant under rotation by any angle. These rotations form the special orthogonal group $SO(2)$.
The steady-state distribution of the bath is isotropic surrounding a disk, implying that $\langle \bF \rangle_\text{b}=\mathbf{0}$.
Moreover, in the absence of tangential contact forces (which are not considered in our model), the bath exerts no torque on a disk, and thus its angular variables are irrelevant. Its motion is then entirely described by $\bR$ and $\bP$, 
with the $2 \times 2$ friction and force correlation matrices
\begin{align}
        \label{eq:zeta_2x2}
        \bzeta_{\bP\bP} &= \int_0^{\infty}  \dd\tau\, \langle \bF(\tau)\otimes \bnabla_\bR \ln \rhob(0) \rangle_{\text{b}}\,, \\
        \label{eq:lambda_2x2}
        \blambda_{\bP\bP} &= \int_0^{\infty} \dd\tau\,  \langle \bF(\tau)\otimes \bF(0) \rangle_{\text{b}}\,.
\end{align}
The most general isotropic form of these matrices is
\begin{equation}\label{eq:parallel_plus_perp}
    \begin{split}
      \bzeta_{\bP\bP} &= \zeta_\parallel \id + \zeta_\perp \bA\,, \\
      \blambda_{\bP\bP} &=  \lambda_\parallel \id + \lambda_\perp \bA\,,
    \end{split}
\end{equation}
with $\id = \begin{bmatrix} 1 & 0 \\ 0 & 1 \end{bmatrix}$
and $\bA = \begin{bmatrix} 0 & -1 \\ 1 & 0 \end{bmatrix}$.
Here, $\zeta_\perp$ is the odd friction, causing the object to deflect in the direction perpendicular to its motion, and $\lambda_\perp$ is the odd momentum diffusivity.
These can be seen from Eqs.~\eqref{eq:zeta_2x2} and~\eqref{eq:lambda_2x2} to require breaking time-reversal and parity symmetries, as satisfied by the chiral active bath. By the same symmetry considerations, objects exhibiting nonzero $\zeta_\perp$ also exhibit odd diffusivity $D_\perp$ and odd mobility $\mu_\perp$, as described in the following two sections.

For other shapes, the angular variables $\Theta$ and $L$ are also needed to fully characterize the state of the object. We specifically identify two relevant symmetries. The first is an $n$-fold rotational symmetry, denoted by $C_n$, and meaning the object is invariant to rotations of angles $\frac{2\pi}{n}$, with $n\geq 2$ an integer. The second is parity (i.e.~reflection) symmetry, which we denote by $\Pi$. Examples of objects obeying or breaking these symmetries are portrayed in Table~\ref{tab:object-symmetries}, together with a characterization of their dynamics.

Objects in $C_n$ exhibit odd dynamics in a chiral active bath. In the adiabatic limit, they have fully independent translational and rotational degrees of freedom: $\bzeta_{\bP L}$, $\bzeta_{L \bP}$, $\blambda_{\bP L}$, and $\blambda_{L \bP}$ vanish by symmetry. Such objects have no polarity, and thus do not behave as translational ratchets, exhibiting $\bFavg = 0$, but do receive a nonvanishing torque $\Gammaavg \neq 0$ from a chiral active bath. Moreover, when a $C_n$ symmetric object is chiral (i.e.~belongs to $C_n$ but not $\Pi$), it can exhibit odd dynamics and a net torque whether or not the bath is chiral. Thus, to achieve odd dynamics and ratchet torques, the breaking of parity symmetry can be supplied either by the chirality of the object or by the chirality of the active bath.

Objects breaking $C_n$ symmetry exhibit a net force $\bFavg$ due to their polarity, whether or not the bath is chiral. If they preserve the $\Pi$ symmetry and the bath is achiral, the decoupling of rotational and translational degrees of freedom survives the breaking of $C_n$ symmetry in the adiabatic limit. This is, for instance, the case of a wedge in a bath of active Brownian particles.

Finally, objects with neither $C_n$ nor $\Pi$ symmetries are expected to exhibit the most general form of the dynamics, with odd dynamics, a net torque and force, and all cross-couplings, regardless of the chirality of the bath.

To establish these symmetry claims, we observe that for an object with $C_n$ symmetry, the steady-state bath probability distribution respects
\begin{equation}\label{eq:cn-symmetry-bath}
\rhob\left(\br^N | \bR, \Theta + \frac{2\pi i}{n}\right) = \rhob(\br^N | \bR, \Theta)\,,
\end{equation}
with $i \in \{0,1,\dots, n-1\}$.
Here, we have defined the marginalized steady-state bath probability density $\rhob(\br^N | \bR, \Theta) = \int \dd\theta^N\ \rhob(\br^N, \theta^N | \bR, \Theta)$, since the orientations $\theta^N$ do not affect bath-object interactions in our model, and can thus be ignored in the following treatment.
What follows, however, straightforwardly generalizes to cases where the orientations $\theta^N$ do affect the interaction force and torque.

With these considerations in mind, the ratchet force then evaluates to
\begin{align}\label{eq:mean-force}
        &\bFavg = \int \dd\br^N\ \rhob(\br^N | \bR, \Theta) \bF(\br^N, \bR, \Theta)\notag\\
        &= \frac{1}{n} \sum_{i=0}^{n-1} \int \dd\br^N\ \rhob\left(\br^N | \bR, \Theta+\frac{2\pi i}{n}\right) \bF\left(\br^N, \bR, \Theta+\frac{2\pi i}{n} \right)\notag\\
        &= \frac{1}{n} \sum_{i=0}^{n-1} \mathcal{R}_\frac{2\pi i}{n} \int \dd\br^N\ \rhob(\br^N | \bR, \Theta) \bF(\br^N, \bR, \Theta)\notag\\
        &= \frac{1}{n} \sum_{i=0}^{n-1} \mathcal{R}_\frac{2\pi i}{n} \bFavg = \mathbf{0}\,.
\end{align}
The second line follows from the isotropy of the bath and holds for an arbitrarily-shaped object.
On the third line, we have used the identity $\bF( \br^N, \bR, \Theta+\frac{2\pi i}{n}) = \mathcal{R}_\frac{2\pi i}{n} \bF(\br^N, \bR, \Theta)$, which holds for any object, where we define the set of $n$ rotation operators $\{\mathcal{R}_\frac{2\pi i}{n}\}$ with $i \in \{0,1,\dots, n-1\}$.
Finally, inserting the identity from Eq.~\eqref{eq:cn-symmetry-bath}, the ratchet force is seen to vanish for $C_n$ symmetric objects.

Unlike the force $\bF$, the torque $\Gamma$ is invariant under rotations by $\frac{2\pi i}{n}$ with respect to the bath; that is,  $\Gamma(\br^N, \bR, \Theta + \frac{2\pi i}{n}) = \Gamma(\br^N)$. Thus, while $C_n$ objects experience no net force, they do receive a net torque if either the bath or the object is chiral.

\begin{figure}[t]
    \centering
    \includegraphics[width=.49\textwidth]{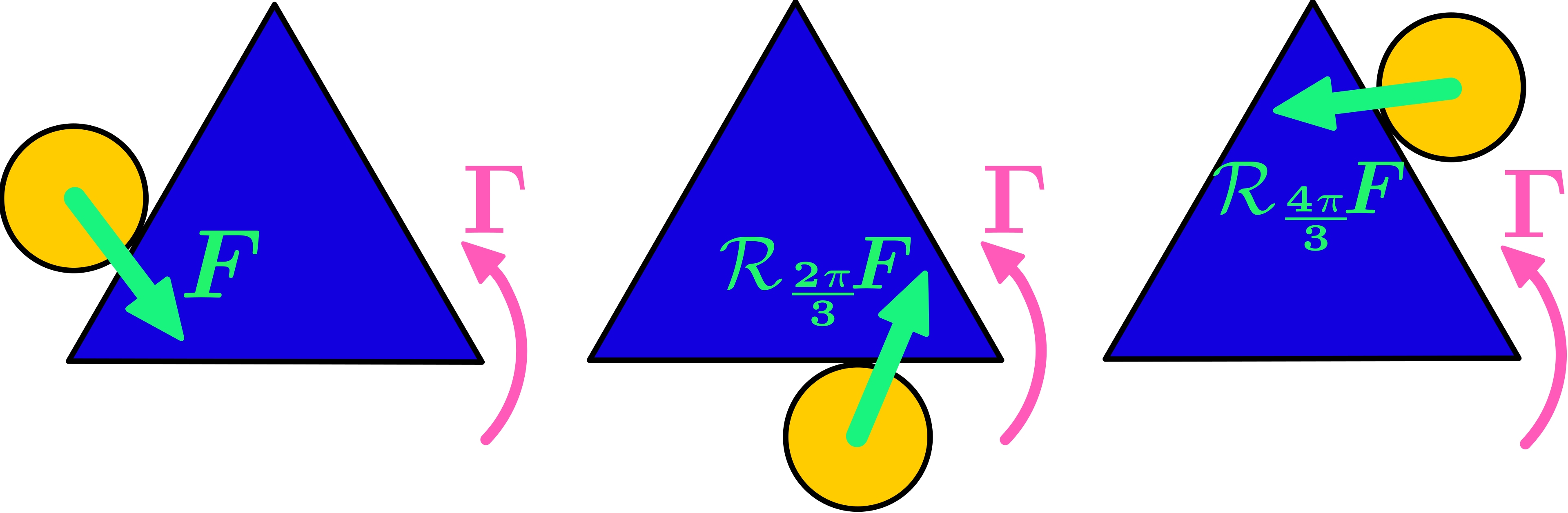}
    \caption{Three equally probable configurations of an object with $C_3$ symmetry (blue) interacting with a bath particle (yellow), related by rotations of $\frac{2\pi}{3}$. Averaging over the three configurations yields a net torque but no net force. The same principle also applies to fluctuations of the torque and force, so that the force-torque time correlations vanish and, consequently, translational and rotational degrees of freedom decouple from one another.}
    \label{fig:rotational-symmetry}
\end{figure}

By the same principle, dynamical correlations between the fluctuating force and torque are seen to vanish at all times as
\begin{equation}\label{eq:decoupling}
        \langle \delta\bF(t) \delta\Gamma(0) \rangle_\text{b}
        = \frac{1}{n} \sum_{i=0}^{n-1} \mathcal{R}_\frac{2\pi i}{n} \langle \delta\bF(t) \delta\Gamma(0) \rangle_\text{b} = \bm{0}\,,
\end{equation}
because, as in Eq.~\eqref{eq:mean-force}, fluctuations in the force vanish statistically when summing over rotations. These considerations are illustrated in Fig.~\ref{fig:rotational-symmetry}.

As a consequence,  the Green-Kubo relations in Eq.~\eqref{eq:green-kubo} implies that $\blambda_{\bP L} = \bm{0}$ and $\blambda_{L \bP} = \mathbf{0}^\mathrm{T}$.
An identical argument for the correlation functions $\langle \delta\bF(\tau)\partial_\Theta\ln \rho_\text{b} (0)\rangle_\text{b}$ and $\langle \delta\Gamma(\tau)\bnabla_\bR^\mathrm{T}\ln \rho_\text{b}(0)\rangle_\text{b}$ appearing in Eq.~\eqref{eq:kubo-agarwal} leads to the same result for the frictional cross couplings, i.e.~$\bzeta_{\bP L} = \mathbf{0}$ and $\bzeta_{L \bP} = \mathbf{0}^\mathrm{T}$. For similar reasons, these cross-couplings also vanish for achiral objects. In this case, reflecting the object over its symmetry axis preserves the fluctuation in the force but changes the sign of the torque, so that the correlator vanishes statistically when summing over the original configuration and the (statistically identical) mirror image.

This leaves the transport matrices $\bzeta(\Theta)$ and $\blambda(\Theta)$ with the block structure
\begin{equation}
    \begin{split}
        \bzeta(\Theta) &= \begin{bmatrix} \bzeta_{\bP\bP}(\Theta) & \mathbf{0} \\ \mathbf{0}^\mathrm{T} &\zeta_{LL} \end{bmatrix}\,, \\
        \blambda(\Theta) &= \begin{bmatrix} \blambda_{\bP\bP}(\Theta) & \mathbf{0} \\ \mathbf{0}^\mathrm{T} & \lambda_{LL} \end{bmatrix}\,.
    \end{split}
\end{equation}
As such, the translational motion becomes independent of the rotational motion.

Finally, we discuss the effect of inverting the chirality of the bath $\omega_0 \to -\omega_0$ on the ratchet effects and the transport matrices. Achiral objects, which can be superimposed on their mirror image, experience no net torque when the active bath is achiral ($\omega_0 = 0$), and thus their direction of rotation is determined solely by the bath chirality.
That is,
\begin{equation}\label{eq:omega0-symmetry-ratchet}
    \Gammaavg \xrightarrow{\omega_0 \rightarrow -\omega_0} -\Gammaavg\,.
\end{equation}
If the object has both parity and rotational symmetries, inverting the bath chirality leads necessarily to a sign change in the off-diagonal elements of the friction and noise correlation matrices
\begin{equation}\label{eq:omega0-symmetry-transport}
    \bzeta_{\bP\bP} \xrightarrow{\omega_0 \rightarrow -\omega_0} \bzeta_{\bP\bP}^\mathrm{T} \,, \quad\quad
    \blambda_{\bP\bP} \xrightarrow{\omega_0 \rightarrow -\omega_0} \blambda_{\bP\bP}^\mathrm{T}\,.
\end{equation}
These relations are invoked in Section~\ref{section:entropy-production-and-hidden-TRS} to investigate the time-reversal symmetry behavior of rotationally-symmetric objects.

\section{Diffusion, mobility, and Einstein relations}\label{section:diffusion-mobility-einstein}

\subsection{Effective temperatures of rotationally-symmetric objects}\label{section:effective-temperatures}
In this section, we show that an effective equilibrium steady state exists for isotropic or $C_n$ objects, such as a disk or a rod, in the adiabatic limit.

For a disk in a chiral active bath, the Fokker-Planck equation~\eqref{eq:FPrhoT} reduces to
\begin{align}\label{eq:kramers-disk}
    \begin{split}
    &\p_t \rhoo(\bR, \bP, t) = \big[-M^{-1}\bP \cdot \bnabla_\bR \\
    &+ \bnabla_\bP\cdot (\zeta_\parallel\id
    + \zeta_\perp\bA) M^{-1} \bP + \lambda_\parallel\bnabla_\bP^2\big]\rhoo(\bR, \bP)\,,
    \end{split}
\end{align}
where $\lambda_\perp$ plays no role due to its antisymmetry. In the presence of periodic boundary conditions, the steady state is uniform in $\bR$ and takes the Boltzmann form:
\begin{equation}
    \rhooss \propto \ee^{-\frac{1}{2 M\Tefftrans}|\bP|^2}\,,
\end{equation}
with the translational effective temperature given by:
\begin{equation}\label{eq:temp-transl}
    \Tefftrans \equiv \frac{\lambda_\parallel}{\zeta_\parallel} = \frac{1}{2M}\big\langle |\bP|^2 \big\rangle\,.
\end{equation}
Here, and in the following, $\langle\ldots\rangle$ indicate an average over $\rhoo$.

We next consider, more broadly, all objects with $C_n$ rotational symmetry, for which the corresponding Fokker-Planck equation is
    \begin{align}
        &\p_t \rhoo(\bR, \bP, \Theta, L, t) = \nonumber\\
        &\bigg[-\frac{\bP}{M} \cdot \bnabla_\bR
        + \bnabla_\bP\cdot \frac{\bzeta_{\bP\bP}(\Theta)}{M} \bP + \bnabla_\bP \cdot \blambda_{\bP\bP}(\Theta)\bnabla_\bP\nonumber\\
        &- \frac{L}{I} \p_{\Theta} -\langle\Gamma\rangle_\text{b}\p_L + \p_L \frac{\zeta_{LL}}{I} L + \lambda_{LL}\p_L^2\bigg]\rhoo\,.\label{eq:fp_rod}
\end{align}
In the steady state, the independence of translational and rotational degrees of freedom make $\rhooss$  factorize. The second and third lines of Eq.~\eqref{eq:fp_rod} vanish separately and $\rhooss$ is given by the product of two Boltzmann distributions:
\begin{equation}\label{eq:rod-boltzmann}
    \rhooss(\bR, \bP, \Theta, L) \propto \ee^{-\frac{1}{2I\Teffrot}(L - I\Omega)^2} \times \ee^{-\frac{1}{2M\Tefftrans}|\bP|^2}\,.
\end{equation}
The rotational effective temperature $\Teffrot$ is defined by
\begin{equation}\label{eq:temp-rot}
    \Teffrot \equiv \frac{\lambda_{LL}}{\zeta_{LL}} = I^{-1}\big\langle |L - I\Omega|^2\big\rangle\,,
\end{equation}
where we have introduced the steady state angular velocity $\Omega$, defined by
\begin{equation}
    \Omega = \frac{\langle \Gamma\rangle_\text{b}}{\zeta_{LL}}\,.
\end{equation}
The steady state solution in Eq.~\eqref{eq:rod-boltzmann} requires the symmetric part of the friction and noise correlation matrices $\bzeta_{\bP\bP}(\Theta)$ and $\blambda_{\bP\bP}(\Theta)$ to obey the fluctuation-dissipation relation
\begin{equation}
    \Tefftrans \left(\bzeta(\Theta) + \bzeta(\Theta)^\mathrm{T}\right) = \left(\blambda(\Theta) + \blambda(\Theta)^\mathrm{T}\right).
\end{equation}

For objects with a $C_n$ symmetry, the decoupling of translational degrees of freedom from the rotational ones thus gives rise to two independent effective  temperatures. In Section~\ref{section:confined-density-and-flux}, we confirm the thermodynamic role of these temperatures by considering the dynamics of objects trapped in external potentials.

\begin{figure}[t]
    \centering
    \includegraphics[width=.5\textwidth]{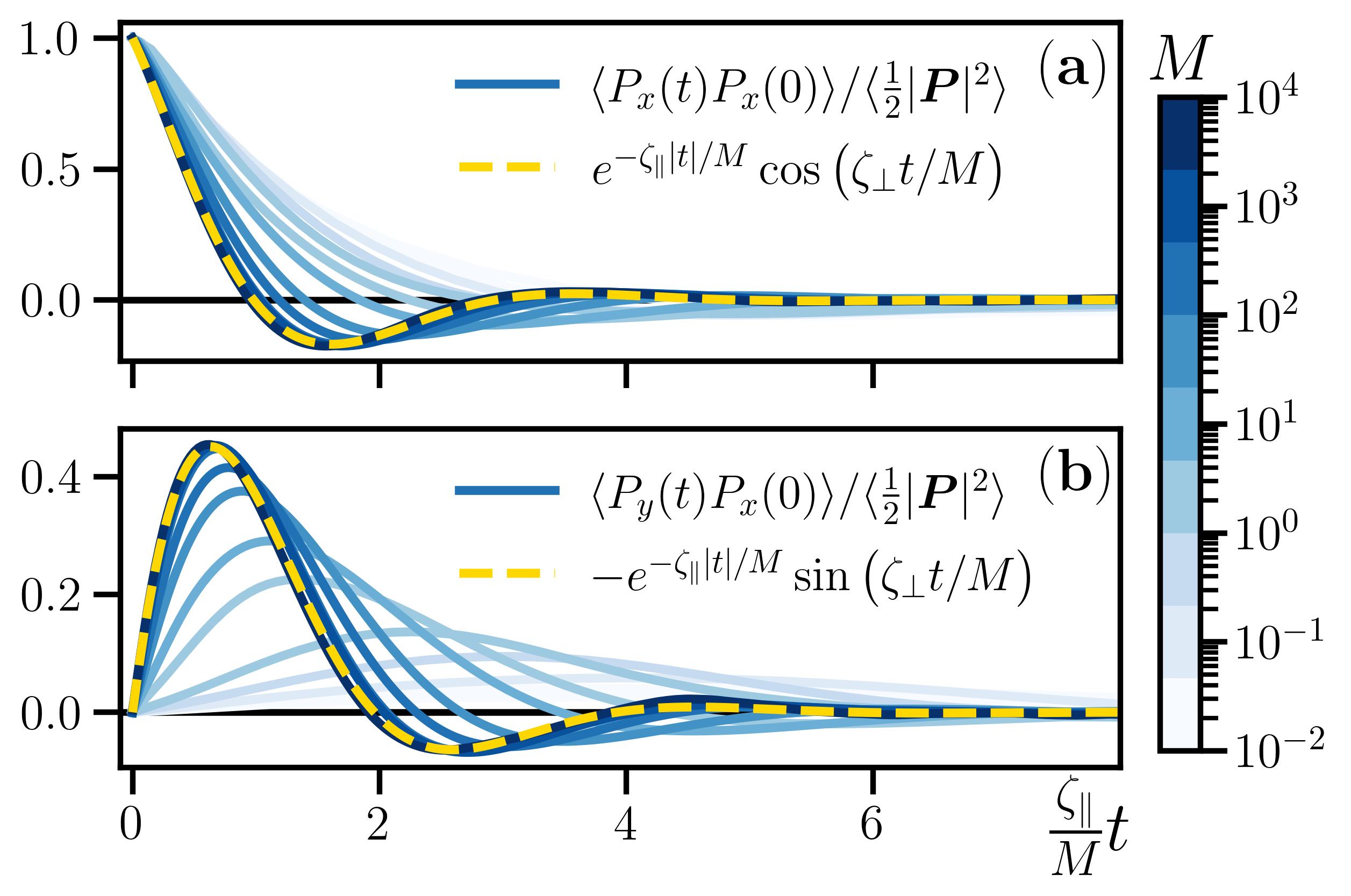}
    \caption{\textbf{Momentum autocorrelations} of the disk. Both the even \textbf{(a)} and odd \textbf{(b)} parts  converge to the analytical solution (dashed line) at large mass.
    \label{fig:autocorrelations}
    }
\end{figure}

\subsection{Disk diffusivity and mobility}
The translational diffusivity matrix of the passive object can be computed from the Green-Kubo relation
\begin{equation}\label{eq:translational-diffusivity}
\mathbf{D}_\mathrm{transl} = \frac{1}{M^2}\int_0^\infty
\dd t\ \langle \bP(t) \otimes \bP(0) \rangle\,.
\end{equation}
For disk, isotropy imposes $\mathbf{D}_\mathrm{transl} = D_\parallel \id + D_\perp \bA$, where $D_\perp$ is the odd diffusivity~\cite{Hargus2021}.
The Langevin equation for the disk reduces from Eq.~\eqref{eq:tracerdynamics} to
\begin{equation}\label{eq:disk-langevin}
    \dot\bR = M^{-1} \bP\,, \hspace{4mm} \dot\bP = -M^{-1}\bzeta_{\bP\bP} \bP + \bxi_\bP\,,
\end{equation}
which is solved by
\begin{equation}\label{eq:P_tracer}
    \bP(t) = \ee^{-\frac{1}{M}\bzeta_{\bP\bP} t}\bP(0) + \int_0^{t}\dd \tau\ \ee^{-\frac{1}{M}\bzeta_{\bP\bP} (t-\tau)}\bxi_\bP(\tau)\,.
\end{equation}
The disk momentum autocorrelation function appearing in Eq.~\eqref{eq:translational-diffusivity} is then
\begin{equation}\label{eq:autocorrelation}
    \begin{split}
        \langle \bP(t) \otimes \bP(0)\rangle = \ee^{-\frac{1}{M}\bzeta_{\bP\bP} t} \left\langle \bP(0) \otimes \bP(0) \right\rangle \\
        + \int_0^t \dd \tau\ \ee^{-\frac{1}{M}\bzeta_{\bP\bP} (t-\tau)}\left\langle \bxi_\bP(\tau) \otimes \bP(0) \right\rangle \\
        = M\Tefftrans e^{-\frac{\zeta_\parallel}{M} t} {\begin{bmatrix} \cos(\frac{\zeta_\perp}{M}t) & \sin(\frac{\zeta_\perp}{M}t) \\ -\sin(\frac{\zeta_\perp}{M}t) & \cos(\frac{\zeta_\perp}{M}t) \end{bmatrix}}
    \end{split}
\end{equation}
for all $t > 0$. On the final line we have identified $\Tefftrans$ using Eq.~\eqref{eq:temp-transl}.
We also exploit the structure of the friction matrix $\bzeta$ given by Eq.~\eqref{eq:parallel_plus_perp} together with the useful identity
\begin{equation}
    \exp\left({-\bA t}\right) = \begin{bmatrix} \cos(t) & \sin(t) \\ -\sin(t) & \cos(t) \end{bmatrix}\,.
\end{equation}

The momentum autocorrelation decays through dampened oscillations, and its behavior is entirely characterized by the friction matrix $\bzeta_{\bP\bP}$ and the object mass. The even friction coefficient $\zeta_\parallel$ determines the rate of the envelope decay, while the odd friction coefficient $\zeta_\perp$ sets the frequency of the oscillations. In Figure~\ref{fig:autocorrelations} we use numerical simulations (numerical details available in companion letter~\cite{companionPRL,github,lammps}) to show that as the adiabatic limit is approached, the momentum autocorrelation functions converge towards the prediction of Eq.~\eqref{eq:autocorrelation}.

Inserting the result of Eq.~\eqref{eq:autocorrelation} into Eq.~\eqref{eq:translational-diffusivity}, we obtain the following expression for the diffusivity matrix:
\begin{equation}\label{eq:calculation_D_disk}
    \begin{split}
        \bD_{\text{transl}} &= \frac{\Tefftrans}{M}\int_0^\infty \dd t\ e^{-\frac{1}{M}\bzeta_{\bP\bP}t}\\
        &= \Tefftrans \bzeta_{\bP\bP}^{-1}
        = \frac{\Tefftrans}{\zeta_{\parallel}^2 + \zeta_{\perp}^2}
        \begin{bmatrix} \zeta_{\parallel} & \zeta_{\perp} \\ -\zeta_{\perp} & \zeta_\parallel \end{bmatrix}.
    \end{split}
\end{equation}

\begin{figure}[t]
    \centering
    \includegraphics[width=.48\textwidth]{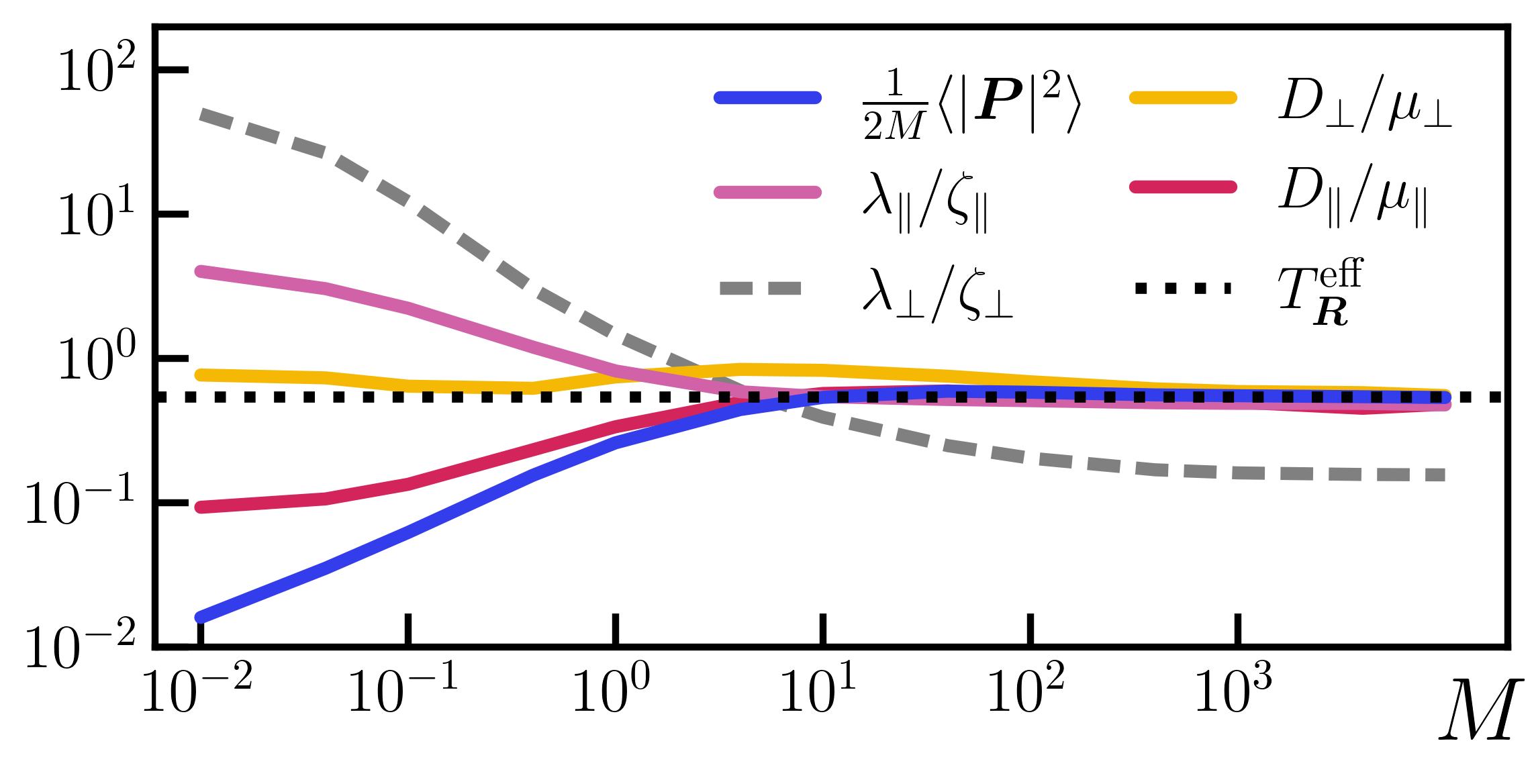}
    \caption{\textbf{Even and odd thermometers} ($\ell_p=10$, $\ell_g=2$). The effective temperature $\Tefftrans$ of a disk can be approximated using both even and odd Einstein relations (Eq.~\eqref{eq:einstein-full}), and the even FDT (Eq.~\eqref{eq:2FDT-full}). All of these measurements agree in at large $M$ but differ for smaller masses. In the adiabatic limit, however, the ratio $\lambda_\perp / \zeta_\perp$ still does not measure the effective temperature, due to the absence of an odd FDT for $\lambda_\perp$. This mismatch is an odd signature of the bath nonequilibrium nature.}
    \label{fig:disk-adiabaticity-2}
\end{figure}

We now turn to the computation of the mobility tensor. When pulling the disk through the chiral active bath with a weak external force $\bF^\mathrm{ext}$, the evolution equation becomes
\begin{equation}
    \dot\bP = -M^{-1}\bzeta_{\bP\bP} \bP + \bF^\mathrm{ext} + \bxi_\bP\,.
\end{equation}
Averaging over the steady state in the presence of $\bF^\mathrm{ext}$, denoted by $\langle \cdot \rangle_{\bF^\mathrm{ext}}$, the disk mobility $\bmu_\text{transl}$ is defined by
\begin{equation}\label{eq:mobility-relation}
    M^{-1}\langle \bP \rangle_{\bF^\mathrm{ext}} = \bzeta_{\bP\bP}^{-1} \bF^\mathrm{ext} \equiv \bmu_\mathrm{transl} \bF^\mathrm{ext}\,.
\end{equation}
Finally, a comparison with the result of Eq.~\eqref{eq:calculation_D_disk} yields the Einstein relation
\begin{equation}\label{eq:einstein-full}
    \mathbf{D}_\mathrm{transl} = \Tefftrans \bmu_\mathrm{transl}\,,
\end{equation}
holding for both the even and odd parts. Figure~\ref{fig:disk-adiabaticity-2} shows that, in the adiabatic limit, Eqs.~\eqref{eq:temp-transl} and~\eqref{eq:einstein-full} indeed allow defining the same translational effective temperature.
By convention, our numerical simulations set $\omega_0 < 0$, so that the bath particles rotate clockwise. In response, the disk then acquires counterclockwise random motion with $D_\perp > 0$, $\mu_\perp > 0$, and $\zeta_\perp < 0$.

\subsection{Diffusion and mobility of $C_n$ objects}
Let us now turn to the relation between diffusivity and mobility matrices for objects with $C_n$ symmetry. For such objects, the $3\times3$ diffusion and mobility matrices assume the block form
\begin{equation}
    \bD = \begin{bmatrix} \bD_{\text{transl}} & \mathbf{0} \\ \mathbf{0}^\mathrm{T} & D_{\text{rot}} \end{bmatrix}\,, \quad \quad
        \bmu_\text{rot} = \begin{bmatrix} \bmu_\text{transl} & \mathbf{0} \\ \mathbf{0}^\mathrm{T} & \mu_{\text{rot}}
        \end{bmatrix}
    \,.
\end{equation}
Here, $\bD_{\text{transl}}$ and $\bmu_\text{transl}$ are the $2\times 2$ translational diffusion and mobility matrices defined in the previous subsection, while the scalar $D_{\text{rot}}$ and $\mu_{\text{rot}}$ are the rotational diffusion and mobility constants.
Despite the angular dependence of the friction coefficient, the long-timescale translational motion of $C_n$ objects is formally identical to that of the disk.
This can be seen by integrating Eq.~\eqref{eq:fp_rod} over $\Theta$ and $L$, which 
recovers Eq.~\eqref{eq:kramers-disk}, but with $\bzeta$ and $\blambda$ replaced by $\bar{\bzeta}$ and $\bar{\blambda}$, defined by
\begin{align}
    \bar{\bzeta} \rhooss(\bR,\bP) \equiv \iint \dd\Theta \dd L\ \bzeta_{\bP\bP}(\Theta)\rhooss(\bR,\bP,\Theta,L)\,,\\
        \bar{\blambda} \rhooss(\bR,\bP) \equiv \iint \dd \Theta \dd L\ \blambda_{\bP\bP}(\Theta)\rhooss(\bR,\bP,\Theta,L)\,.
\end{align}
The translational motion of $C_n$ objects thus obeys the Einstein relation 
\begin{equation}
    \bD_{\text{transl}} = \Tefftrans \bmu_\text{transl}\,,
\end{equation} 
analogous to Eq.~\eqref{eq:einstein-full}, with $\bmu_\text{transl} = \bar{\bzeta}_{\bP\bP}^{-1}$.

Meanwhile, the rotational diffusion coefficient is
\begin{equation}\label{eq:D_rot_def}
        D_{\text{rot}} = \frac{1}{I^2}\int_0^{\infty} \dd t\ \left\langle [L(t) - I\Omega][L(0) -I\Omega]\right\rangle,
\end{equation}
where we recall that $\Omega = \frac{\Gammaavg}{\zeta_{LL}}$ is the average angular velocity in steady state. The equation of motion of the angular momentum is
\begin{equation}\label{eq:rod-langevin}
    \dot L= \Gammaavg - I^{-1}\zeta_{LL}L + \xi_L(t)\,,
\end{equation}
which is solved by
\begin{equation}
    \begin{split}
    L(t) - I\Omega &= \ee^{-\frac{\zeta_{LL}}{I}t}\left[L(0) - I\Omega \right] \\
    &+ \sqrt{2\Teffrot\zeta_{LL}}\int_0^t \dd \tau\ \ee^{-\frac{\zeta_{LL}}{I}(t-\tau)} \xi_L(\tau)\,.
    \end{split}
\end{equation}
From Eq.~\eqref{eq:D_rot_def}, the rotational diffusivity is then
\begin{equation}\label{eq:D_rot}
        D_{\text{rot}} = \frac{1}{I \zeta_{LL}} \langle|L(0) - I\Omega|^2 \rangle = \frac{\Teffrot}{\zeta_{LL}}\,,
\end{equation}
where $\Teffrot$ is defined in Eq.~\eqref{eq:temp-rot}.

\if Following a calculation similar to the one performed for the disk, the translational mobility tensor reads
\begin{equation}
    \begin{split}
    \bmu_{\text{transl}} &= \frac{1}{2\Tefftrans}(\bD_{\text{transl}} + \bD_{\text{transl}}^T)\\
    &\left(\bm{\delta} + \frac{1}{4}\langle (\bzeta_{\bP\bP}^T - \bzeta_{\bP\bP})(\bzeta_{\bP\bP} + \bzeta_{\bP\bP}^{T})^{-1}\rangle\right)\,.
    \end{split}
\end{equation}
Note that an Einstein relation links the symmetric part of the translational mobility and diffusivity tensors as
\begin{equation}    \Tefftrans(\bmu_{\text{transl}}^T + \bmu_{\text{transl}}) = \bD_{\text{transl}}^T + \bD_{\text{transl}}\,.
\end{equation}
When $\bm{F}^{\mathrm{ext}}$ is weak, the probability distribution is expected to remain uniform in $\Theta$.
Averaging over the object orientation then yields $\bD_{\text{transl}} = \Tefftrans \bmu_{\text{transl}}$.
\fi
Finally, analogous to Eq.~\eqref{eq:mobility-relation}, the rotational mobility $\mu_{\text{rot}}$ in response to an external torque $\Gamma^\mathrm{ext}$ is defined by
\begin{equation}\label{eq:mu_rot}
    I^{-1}\langle L \rangle_{\Gamma^\mathrm{ext}} = \zeta_{LL}^{-1} \Gamma^\mathrm{ext} \equiv \mu_{\text{rot}} \Gamma^\mathrm{ext}\,,
\end{equation}
where $\langle\ldots\rangle_{\Gamma^\text{ext}}$ is an average over the steady state in the presence of a weak external torque $\Gamma^\text{ext}$. Comparing Eq. \eqref{eq:D_rot} with Eq. \eqref{eq:mu_rot}yieldis the Einstein relation
\begin{equation}
    D_{\text{rot}} = \Teffrot \mu_{\text{rot}}\,,
\end{equation}
involving the effective rotational temperature $\Teffrot$.

Our calculations thus predict that rotational and translational degrees of freedom obey Einstein relations, albeit with different temperatures, in line with the simluation results presented in~\cite{companionPRL}.

\section{Density and flux in a confining potential}\label{section:confined-density-and-flux}
To test whether the effective temperatures defined in the previous section indeed play a thermodynamic role, we now study the steady state of objects whose translational and rotational degrees of freedom are confined by an external potential.

\subsection{Translational confinement}

When the passive disk is confined by an external potential $U(\bR)$, Eq.~\eqref{eq:kramers-disk} becomes
\begin{equation}\label{eq:kramers-disk-confined}
\begin{split}
    &\p_t \rhoo(\bR, \bP, t) = \big[-M^{-1}\bP \cdot \bnabla_\bR \\
    &+ M^{-1}\bnabla_\bP\cdot\bzeta_{\bP\bP}\cdot \bP +\lambda_\parallel\bnabla_\bP^2 + \bnabla_\bP \cdot (\bnabla_\bR U) \big] \rhoo\,.
    \end{split}
\end{equation}
Eq.~\eqref{eq:kramers-disk-confined} is solved in steady state by the Boltzmann distribution
\begin{equation}\label{eq:boltzmann-2}
    \rhooss(\bR,\bP) \propto \ee^{-\big(\frac{|\bP|^2}{2M} + U(\bR)\big)/\Tefftrans}\,,
\end{equation}
which confirms the interpretation of $\Tefftrans$ as a temperature.

\begin{figure*}[t]
    \centering
    \includegraphics[width=.9\textwidth]{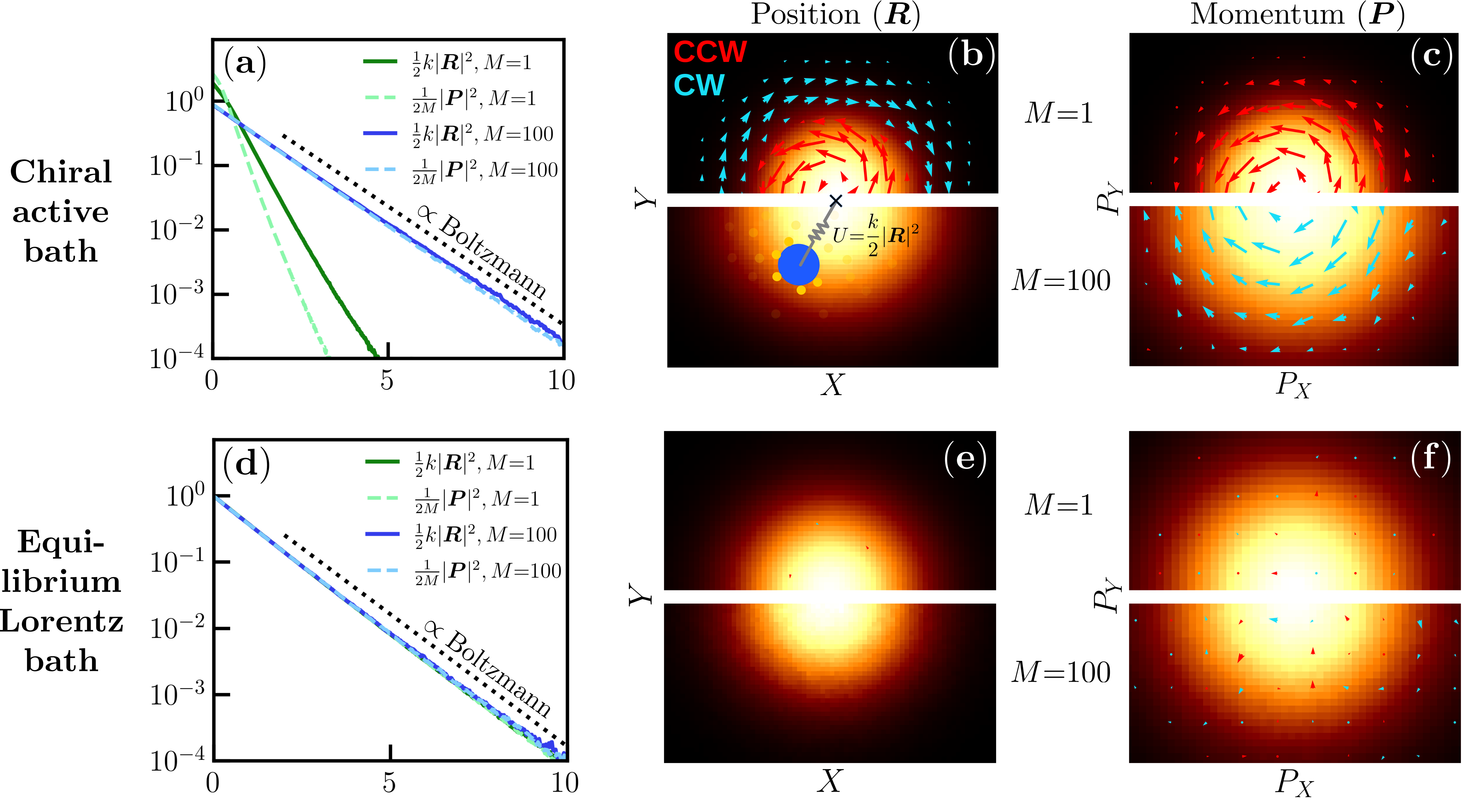}
    \caption{
    \textbf{Confined disk.}
    \textbf{Top row: Chiral active bath (Eq.~\eqref{eq:cABP-EOM})}. Adapted from companion Letter~\cite{companionPRL}. \textbf{(a)} Distribution of the potential energy $\frac{1}{2}k|\bR|^2$ (solid lines) and kinetic energy $\frac{1}{2M}|\bP|^2$ (dashed lines) of the disk. In the adiabatic limit (blue lines, $M=100$) these are Boltzmann distributed with the same effective temperature.
    \textbf{(b)} Position-space probability $\rho_\bR$ (heatmap) and associated steady-state flux $\bm{J}_\bR^{\text{ss}}$ (arrows) of a disk in the potential $U(\bR) = \frac{k}{2}|\bR|^2$ (cartoon overlay). Circulating currents are permitted at steady state (top) but vanish at large $M$ (bottom) due to Eq.~\eqref{eq:einstein-full}. \textbf{(c)} In contrast, the momentum-space density $\rho_\bP$ exhibits steady-state currents $\bm{J}_\bP^{\text{ss}}$ that persist even in the adiabatic limit due to the absence of an odd 2FDT.
    \textbf{Bottom row: Equilibrium Lorentz bath (Eq.~\eqref{eq:Lorentz-bath})}. \textbf{(d)} Regardless of the disk mass, the dynamics admit a Boltzmann solution with the disk and bath in thermal equilibrium. \textbf{(e) \& (f)} Unlike the chiral active bath, both the odd Einstein relation and odd FDT hold regardless of the object mass. Odd driving forces therefore cancel and currents vanish in both position and momentum spaces.
    Simulation parameters:  $\ell_p=10$, $\ell_g=5$ $T=1$, $m=1$, $f_\text{Lorentz}=1$, $\gamma=1$. $X\,,Y \in [-2,2]$, in (b) and (e). $P_X\,,P_Y \in [-2,2]$ in (c) and (f).
    \label{fig:disk-lorentz-bath}
    }
\end{figure*}

We now define the number density $\rho_\bR(\bR, t) = \int \dd\bP\ \rhoo(\bR,\bP,t)$. Integrating Eq.~\eqref{eq:kramers-disk-confined} over $\bP$ then yields the continuity equation
\begin{equation}
    \p_t \rho_\bR = - \bnabla_\bR \cdot \bm{J}_\bR\,,
\end{equation}
where $\bm{J}_\bR(\bR, t) = M^{-1}\int \dd\bP\ \bP \rhoo(\bR,\bP,t)$.
Furthermore, taking the first moment of Eq.~\eqref{eq:kramers-disk-confined} with respect to $\bP$ yields
\begin{equation}
    \p_t \bm{J}_\bR = -\bnabla_\bR \cdot \bm{\sigma} - M^{-1}\bzeta_{\bP\bP}\cdot\bm{J}_\bR - M^{-1}\rho_\bR \bnabla_\bR U\,,
\end{equation}
where $\bm{\sigma} = \frac{1}{M^2} \int \dd\bP\ \bP \otimes \bP \rhoo(\bR, \bP, t)$.
In the steady state, $\bm{\sigma} = M^{-1}\Tefftrans \rho(\bR)\id$
and the flux becomes
\begin{align}\label{eq:steady-state-flux}
    \begin{split}
    \bm{J}_\bR^\mathrm{ss}(\bR) &= -\mathbf{D} \bnabla_\bR \rho_\bR - \rho_\bR \bm{\mu} \bnabla_\bR U \\
    &= \left[\Tefftrans \bmu_{\text{transl}} - \mathbf{D}_{\text{transl}}\right]\bnabla_\bR \rho_\bR = \bm{0}\,,
    \end{split}
\end{align}
where the equality to zero in the final line results from the Einstein relation in Eq.~\eqref{eq:einstein-full}.
Thus, in the adiabatic limit a disk exhibits no circulation in position space in a confining potential: the current vanishes due to an exact cancellation between the flux driven by diffusion and that driven by the mobility, as observed in Fig.~\subfigref{fig:disk-lorentz-bath}{b}.

A nonzero net circulation appears however in momentum space. Defining the momentum density $\rho_\bP(\bP, t) = \int \dd\bR\ \rhoo(\bR,\bP,t)$ and marginalizing Eq.~\eqref{eq:kramers-disk-confined} with respect to $\bR$ yields the continuity equation in momentum space
\begin{equation}
    \partial_t \rho_\bP = - \bnabla_\bP \cdot \bm{J}_\bP\,,
\end{equation}
where the momentum-space flux is defined by
\begin{equation}
    \bm{J}_\bP = -[M^{-1} \bzeta_{\bP\bP}\bP + \blambda_{\bP\bP}\bnabla_\bP] \rho_\bP - \int \dd\bR\ (\bnabla_\bR U) \rhoo\,.
\end{equation}
Inserting $\rhooss$ from Eq.~\eqref{eq:boltzmann-2} yields
\begin{equation}
    \int \dd\bR (\bnabla_\bR U) \rhoo(\bR,\bP,t) = -\Tefftrans \int \dd\bR \bnabla_\bR \rhooss = \mathbf{0}\,,
\end{equation} and finally
\begin{equation}\label{eq:Jss-momentum-even-and-odd}
        \bm{J}_\bP^\mathrm{ss} = \left[\Tefftrans \bzeta_{\bP\bP} - \blambda_{\bP\bP} \right] \bnabla_\bP \rho_\bP\,.
\end{equation}
The steady state condition $\bnabla_\bP \cdot \bm{J}_\bP^\mathrm{ss} = 0$ imposes a fluctuation-dissipation theorem
\begin{equation}\label{eq:2FDT-full}
\lambda_\parallel = \Tefftrans \zeta_\parallel\,.
\end{equation}
This leaves the steady-state momentum flux as:
\begin{equation}\label{eq:Jss-momentum}
    \bm{J}_\bP^\mathrm{ss} = \left[\Tefftrans \zeta_\perp - \lambda_\perp \right] \bA \bnabla_\bP \rho_\bP\,.
\end{equation}
Note that, while the Einstein relation~\eqref{eq:einstein-full} holds for both the odd and even parts of $\bD_{\rm transl}$, the FDT~\eqref{eq:2FDT-full} holds only for the even part: $\lambda_\perp/\zeta_\perp$ is in general not given by $\Tefftrans$, even in the adiabatic limit, as shown in Fig.~\ref{fig:disk-adiabaticity-2}. An important consequence is that Eq.~\eqref{eq:Jss-momentum} predicts the mismatch between $\lambda_\perp$ and $\zeta_\perp$ to drive a steady-state circulation $\bm{J}_\bP^\mathrm{ss}$. This is confirmed in Fig.~\subfigref{fig:disk-lorentz-bath}{c}, even in the adiabatic limit.

\subsection{Rotational confinement}
We next impose a confining external potential $U(\bR, \Theta)$ on a $C_n$ object such as a rod. So long as this potential decomposes as $U(\bR, \Theta) = u(\bR)+w(\Theta)$, the translational and rotational degrees of freedom remain independent. As long as the potential traps the object within a finite range $[\Theta_\text{min}, \Theta_\text{max}]$, the rotational degrees of freedom admit the steady-state distribution
\begin{equation}\label{eq:confined-rod-boltzmann}
        \rhooss \propto \exp\bigg\{-\frac{I^{-1}\big(L - I\Omega\big)^2 + \big(w(\Theta) - \Gammaavg\Theta\big)\big]}{2\Teffrot}\bigg\} \,.
\end{equation}
This validates the interpretation of $\Teffrot$ as an effective temperature. Note, however, that Eq.~\eqref{eq:confined-rod-boltzmann} breaks down for a periodic, nonconfining potential $w(\theta)$. This shows that, for $C_n$ objects, the equilibrium regime for the angular degrees of freedom is more fragile than for the translational degrees of freedom, due to the non-vanishing average torque $\Gammaavg$.

\subsection{Comparison with an equilibrium Lorentz bath}
The mismatch between $\zeta_\perp$ and $\lambda_\perp$ that drives momentum circulation in Eq.~\eqref{eq:Jss-momentum} has no equivalent among even transport coefficients. It is a signature of the nonequilibrium nature of the chiral active bath and it would not be present in a passive chiral bath. To show this, we consider  bath particles subject to an equilibrium Langevin dynamics with a Lorentz force orthogonal to their direction of motion:
\begin{align}\label{eq:Lorentz-bath}
    \begin{split}
        \dot\br_i &= \bm{v}_i\,,\\
        m \dot{\bm{v}}_i &= \bF_{i} + f_\text{Lorentz} \bA \bm{v} - \gamma \bm{v} + \sqrt{2 T \gamma} \beeta_i\,.
    \end{split}
\end{align}
Here, $m$ is the mass of a bath particle, $\gamma$ is the friction, $T$ is the  temperature, and $\beeta_i(t)$ are Gaussian white noises satisfying $\langle \beeta_i(t) \otimes \beeta_j(t') \rangle = \delta_{ij} \id \delta(t-t')$.
Equation~\eqref{eq:Lorentz-bath} can be thought as describing thermalized charged particles moving in the plane normal to a magnetic field.
As in Eqs.~\eqref{eq:newton} and~\eqref{eq:cABP-EOM}, bath particles interact with the passive object through a potential $V^\mathrm{int}$ (i.e.~$\bF = -\bnabla_\bR V^\mathrm{int}(\br^N, \bR) = -\sum_{i=1}^N \bF_i$).
$f_\text{Lorentz}$ is a Lorentz force acting perpendicular to the motion of a bath particle, responsible for breaking parity symmetries. $f_\text{Lorentz}$ also breaks time-reversal symmetry if its sign is not flipped under time reversal~\cite{obyrneTimeIrreversibilityActive2022}, which results in nonzero $\zeta_\perp$ and $\lambda_\perp$. The steady state distribution is, however, given by the equilibrium Boltzmann weight
\begin{equation}
\rhob \propto e^{-V^\mathrm{int}(\br^N, \bR)/T}\,.
\end{equation}

Equation~\eqref{eq:kubo-agarwal} then evaluates to
\begin{align}\label{eq:2FDT-lorentz}
    \begin{split}
        \bzeta_{\bP \bP} &= \int_0^{\infty} \dd s\ \langle \delta\bF(\tau) \otimes \bnabla_\bR \ln \rhob (0)\rangle_\text{b}\\
        &= \int_0^{\infty} \dd s\ \langle \delta\bF(\tau) \otimes \delta\bF(0)\rangle_\text{b} / T \\
        &= \blambda_{\bP \bP} / T \,,
    \end{split}
\end{align}
where the final equality is made through identification with Eq.~\eqref{eq:green-kubo}.
Thus, we recover
a FDT for both the even and odd parts of $\blambda_{\bP\bP}$ and $\bzeta_{\bP\bP}$.
As a result, the momentum-space flux in Eq.~\eqref{eq:Jss-momentum} is seen to vanish, even when $\zeta_\perp$ and $\lambda_\perp$ are individually nonzero, as shown numerically in Fig.~\subfigref{fig:disk-lorentz-bath}{f}.
This stands in contrast to the case of a chiral active bath, where such circulation persists, indicating the non-equilibrium nature of the bath.

\section{Time-reversal symmetry and comparison with a charged particle in a magnetic field}\label{section:entropy-production-and-hidden-TRS}
In this Section, we address the irreversibility of the trajectories of the object. We focus on the disk, whose symmetry properties allow for analytical progress and we show that the entropy production contains a contribution that diverges in the adiabatic limit. We then observe that a hidden symmetry exists, corresponding to simultaneous reversal of time and the bath chirality, under which the entropy production of the disk vanishes. This symmetrized entropy production is also found to vanish in the contribution from rotational motion of objects with both $C_n$ and $\Pi$ symmetries. These findings rationalize the existence of equilibrium-like steady states for such objects, as discussed in the previous Sections.
 
\subsection{Entropy production rate}
The irreversibility of the object's dynamics is quantified by the entropy production rate
\begin{equation}\label{eq:entropy_production}
    \sigma \equiv \lim_{t \to \infty} \frac{1}{t}\left\langle \ln \frac{\mathscr{P}[\big\{\bGamma(\tau) \big\} \lvert \bGamma(0), \omega_0]}{\mathscr{P}[\big\{\tilde{\bGamma}(t-\tau) \big\} \lvert \tilde{\bGamma}(t), \omega_0]}\right\rangle_\text{path}\,,
\end{equation}
where $\big\{ \bGamma(\tau) \big\} = \big\{\bR(\tau), \Theta(\tau), \bP(\tau), L(\tau)\big\}$ is a trajectory through phase space for $\tau \in [0,t]$ and $\big\{ \tilde{\bGamma}(t-\tau) \big\} = \big\{\bR(t - \tau), \Theta(t - \tau), -\bP(t - \tau), -L(t - \tau)\big\}$ is the time-reversed trajectory.
The quantity $\mathscr{P}[\big\{\bGamma(\tau) \big\} \lvert \bGamma(0), \omega_0]$ is the probability to observe a trajectory $\big\{\bGamma(\tau) \big\}$ beginning at $\bGamma(0)$ at time $\tau=0$ in a chiral active bath with the rotational frequency $\omega_0$. 
The brackets 
\begin{equation}
    \langle\ldots\rangle_\text{path} \equiv \int \mathcal{D}\bGamma(\tau) \ldots \mathscr{P}[\{\bGamma(\tau)|\bGamma(0),\omega_0\}]
\end{equation}
denote the average over the ensemble of paths $\{\bGamma(\tau)\}$. 
As discussed at the beginning of Sec. \ref{sec:noise}, path probabilities and entropy production must be computed at finite $\epsilon$, and the adiabatic limit $\epsilon\to0$ must be taken only afterwards. This procedure is implemented using the non-Markovian Gaussian process $\bxi_\epsilon$, which recovers the statistics of $\bxi$ in the adiabatic limit, as described by Eq.~\eqref{eq:markovian_from_nonmarkovian}. We also use the generalized coordinates system introduced in Eq.~\eqref{eq:tracerdynamics-vectorial}. The probability density of a trajectory that solves Eq.~\eqref{eq:tracerdynamics-vectorial} is given by
\begin{equation}
    \mathscr{P}[\{\bGamma(\tau) \} \lvert \bGamma(0), \omega_0] \propto  \exp\left[-S[\{\bGamma(\tau)\}\lvert \bGamma(0), \omega_0] \right]\,,
\end{equation}
where the action $S[\{\bGamma(\tau)\}\lvert \bGamma(0), \omega_0]$ is defined as
\begin{widetext}
\begin{equation}\label{eq:S_def}
    S\{\bGamma(\tau) \} \lvert \bGamma(0), \omega_0] \equiv \frac{1}{2} \int_0^t \dd \tau\int_0^t \dd \tau' \left( \dot\bW(\tau) - \langle\bG\rangle_\text{b} + \tilde{\bzeta} \cdot \bW(\tau)\right) \cdot \bT_\epsilon(\tau,\tau') \left( \dot\bW(\tau') - \langle\bG\rangle_\text{b} + \tilde{\bzeta}\cdot\bW(\tau')\right)\,.
\end{equation}
\end{widetext}
The operator $\bT_\epsilon(\tau,\tau')$ is the inverse of the noise-noise correlation function at finite $\epsilon$ given by Eq.~\eqref{eq:markovian_from_nonmarkovian}, such that
\begin{equation}\label{eq:T_def}
    \int \dd \tau \langle \bxi_\epsilon(t) \otimes \bxi_\epsilon(\tau)\rangle \bT_\epsilon(\tau,t') = \id\delta(t-t')\,.
\end{equation}
Finding an explicit expression of $\bT_\epsilon$ for arbitrary object shapes is challenging, because the correlation function $\langle\bxi_\epsilon(t) \otimes \bxi_\epsilon(\tau)\rangle$ depends on the evolution of the object orientation. 
To allow for analytical progress, we focus on the disk and, as mentioned in Section~\ref{sec:current}, we introduce a regularized explicit form for the correlations of $\bxi_\epsilon(t)$. We use the following memory kernel that decays over a time scaled by $\epsilon^2$:
\begin{equation}\label{eq:correlation}
    \langle \bxi_\epsilon(t) \otimes \bxi_\epsilon(t')\rangle =\begin{cases} \frac{1}{\epsilon^2}\ee^{-\blambda_{\bP\bP}^{-1}(t-t')/\epsilon^2} &\text{ if $t\geq t'$}\\
    \frac{1}{\epsilon^2}\ee^{-(\blambda_{\bP\bP}^{-1})^\mathrm{T}(t'-t)/\epsilon^2} &\text{ if $t< t'$}\,
    \end{cases}
\end{equation}
which indeed recovers Eq.~\eqref{eq:noise} as $\epsilon \to 0$. 

In the case of the disk, the operator $\bT_{\epsilon}(\tau,\tau')$ is a $2\times2$ operator matrix which depends only on the time difference $\tau-\tau'$. We denote this operator by $\bT_{\epsilon}^{\bP\bP}(\tau-\tau')$. Using the expression of the noise-noise correlation given by Eq.~\eqref{eq:correlation} we find, after some algebra detailed in Appendix \ref{app:algebra_T_PP} the following expression:
\begin{equation}\label{eq:T_PP}
    \begin{split}
        \bT^{\bP\bP}_\epsilon(t) &= \frac{1}{2}[ (\blambda_{\bP\bP,S})^{-1} \\
        &+ \epsilon^2 \left((\blambda_{\bP\bP,S})^{-1}\blambda_{\bP\bP}- \blambda_{\bP\bP}^\mathrm{T}(\blambda_{\bP\bP,S})^{-1} \right)\p_t \\
        &+ \epsilon^4 \blambda_{\bP\bP}(\blambda_{\bP\bP,S})^{-1}\blambda_{\bP\bP}^\mathrm{T}\p_t^2]\delta(t)\\
        &= \frac{1}{2\lambda_\parallel}[\id + 2\epsilon^2\lambda_\perp\bA\p_t + \epsilon^4(\lambda_\parallel^2 + \lambda_\perp^2)\id\p^2_t ]\\
        &\equiv \frac{1}{2}[\bT^{(0)} + \epsilon^2\bT^{(1)}\p_t + \epsilon^4\bT^{(2)}\p^2_t]\delta(t)\,.
    \end{split}
\end{equation}
where $\blambda_{\bP\bP,S} \equiv \frac{1}{2}\left[ \blambda_{\bP\bP} + \blambda^\mathrm{T}_{\bP\bP}\right]$ is the symmetric part of the matrix $\blambda_{\bP\bP}$. In the second equality we used the explicit form of $\blambda_{\bP\bP}$ for the disk. The last line defines three matrices $\bT^{(n)}$ ($n\in\{0,1,2\}$) multiplying the operators $\epsilon^{2n}\p_t^n\delta(t)$. The symmetric matrices $\bT^{(0)}$ and $\epsilon^4\bT^{(2)}\p_t^2$ are even under time reversal, while the antisymmetric matrix $\epsilon^2\bT^{(1)}\p_t$ is odd under time reversal. 

The path probability for a disk is given by
\begin{equation}
    \begin{split}
        &\mathscr{P}[\{\bP(\tau) \} \lvert \bGamma(0), \omega_0] \propto \exp\Biggl[-\frac{1}{2}\int_0^t \dd \tau\int_0^t \dd \tau'\ \bigg(\dot\bP(\tau) \\
        &+ \frac{1}{M}\bzeta_{\bP\bP} \bP(\tau) \bigg) \cdot \bT^{\bP\bP}_\epsilon(\tau-\tau')  
         \bigg(\dot\bP(\tau') + \frac{1}{M}\bzeta_{\bP\bP} \bP(\tau')\bigg)\Biggr]\,.
    \end{split}
\end{equation}
Using the expression of  $\bT^{\bP\bP}_\epsilon$ given by Eq.~\eqref{eq:T_PP} and its symmetry properties under time reversal, we obtain
\begin{equation}\label{eq:sigma_disk}
    \begin{split}
        \sigma_\text{disk} &= -\lim_{t\to\infty} \frac{1}{t}\int_0^t \dd \tau \ \Biggl\langle  \dot\bP \cdot\bT^{(0)}\frac{\bzeta_{\bP\bP}}{M}\bP \\
        &+ \epsilon^4 \dot\bP\cdot\bT^{(2)}\frac{\bzeta_{\bP\bP}}{M}\ddot\bP \\
        &+ \frac{\epsilon^2}{2}\left[ \dot\bP \cdot\bT^{(1)} \ddot\bP +  \frac{\bzeta_{\bP\bP}}{M}\bP\cdot \frac{\bzeta_{\bP\bP}}{M}\dot\bP\right] \Biggr\rangle_\text{path}\,.
    \end{split}
\end{equation}
To identify the leading contribution to Eq.~\eqref{eq:sigma_disk} in the adiabatic limit, we observe that the integrand contains correlations among time derivatives of various orders of the disk momentum. Their contribution can be estimated, using the effective Langevin equation Eq.~\eqref{eq:tracerdynamics} and the noise correlations in Eq.~\eqref{eq:correlation}, as 
\begin{equation}
    \begin{split}
        &\left\langle \frac{\dd^n\bP(\tau)}{\dd \tau^n} \otimes \frac{\dd^m \bP(\tau)}{\dd\tau^m}\, \right\rangle_\text{\tiny{path}} \\
        &\sim \left\langle\frac{\dd^{n-1}\bxi_\epsilon(\tau)}{\dd \tau^{n-1}}\otimes \frac{\dd^{m-1}\bxi_\epsilon(\tau)}{\dd \tau^{m-1}} \right\rangle 
        \sim \epsilon^{-2(n+m-1)}\,.
    \end{split}
\end{equation}
Using this scaling, we see that the leading order to the entropy production $\sigma_\text{disk}$ is given by the term $\epsilon^2\langle\dot\bP\bT^{(1)}\ddot\bP\rangle$, which is of order $O(\epsilon^{-2})$, while all the other terms are of higher order in $\epsilon$. The entropy production $\sigma_\text{disk}$ is thus, to leading order in $\epsilon$,
\begin{align}
    \begin{split}
        &\sigma_\text{disk} \approx -\lim_{t\to+\infty} \frac{\epsilon^2}{2t}\int_0^t \dd\tau\,\left\langle \dot\bP(\tau)\cdot\bT^{(1)}\ddot\bP(\tau) \right\rangle_\text{path}  \\
        &= \lim_{t\to+\infty} \frac{\epsilon^2}{2t}\int_0^t\dd\tau\, \Tr\left[\bT^{(1)}\left\langle \bxi_\epsilon(\tau) \otimes \dot\bxi_\epsilon(\tau)\right\rangle \right]\\
        &= \lim_{t\to+\infty}\frac{1}{2 t}\int_0^t   \dd\tau\, \Tr \left[\bT^{(1)}\left(\p_{\tau'} \ee^{-\blambda_{\bP\bP}^{-1}(\tau-\tau')/\epsilon^2}\right)\biggr\rvert_{\tau'=\tau}\right] \\
        &= \epsilon^{-2}\lim_{t\to+\infty}\frac{1}{2t}\int_0^t\dd\tau\, \Tr\left[\bT^{(1)}\blambda_{\bP\bP}^{-1}\right] \\
        &=\frac{1}{2\epsilon^2}\Tr\left[ \bT^{(1)}\blambda_{\bP\bP}^{-1}\right]\\
        &= \frac{2\lambda_\perp^2}{\epsilon^2\lambda_\parallel(\lambda_\parallel^2 + \lambda_\perp^2)}\,. 
    \end{split}
\end{align}
In the second equality, we made use of the fact that $\bT^{(1)}$ is antisymmetric. This expression clearly shows that the adiabatic limit being singular is a consequence of the chirality of the bath. Interestingly, the leading term in $\sigma_{\text{disk}}$ depends only on the properties of the force-force correlation matrix $\blambda_{\bP\bP}$, and not on the friction matrix $\bzeta_{\bP\bP}$. It can be shown that treating the friction matrix $\bzeta_{\bP\bP}$ as the adiabatic limit of a non Markovian process does not change the singular contribution to the entropy production rate.

\subsection{Hidden time-reversal symmetry for the disk}

In this section, we show that the disk obeys  a hidden time-reversal symmetry under which the entropy production vanishes, when the chirality of the bath is flipped upon time reversal. 
To do so, we  consider the \textit{symmetrized} entropy production rate
\begin{equation}\label{eq:entropy_symm}
    \sigma_s \equiv \lim_{t \to \infty} \frac{1}{t}\left\langle \ln \frac{\mathscr{P}[\big\{\bGamma(\tau) \big\} \lvert \bGamma(0), \omega_0]}{\mathscr{P}[\big\{\tilde{\bGamma}(t-\tau) \big\} \lvert \tilde{\bGamma}(t), -\omega_0]}\right\rangle_\text{path}\,,
\end{equation}
where $\omega_0$ has been flipped in the time-reversed trajectories. 

Let us compute the symmetrized entropy production rate for the disk, $\sigma_{s,\text{disk}}$. We recall that inverting the bath chirality has the effect of transposing the force-force correlation and friction matrices $\blambda_{\bP\bP}$, $\bzeta_{\bP\bP}$, see Eq.~\eqref{eq:omega0-symmetry-transport}. Using this fact, we can see that the operator $\bT^{\bP\bP}_\epsilon$ defined in Eq.~\eqref{eq:T_PP} is even under simultaneous reversal of time and the chirality of the bath. The symmetrized entropy production of the disk is thus
\begin{widetext}
\begin{equation}\label{eq:entropy-symm-disk}
    \begin{split}
        \sigma_{s,\text{disk}} &= -\lim_{t\to\infty}\frac{1}{2t }\int_0^t \dd \tau \Biggl\langle \left(\dot\bP(\tau) + \frac{1}{M}\bzeta_{\bP\bP}\bP(\tau)\right)\cdot\bT^{\bP\bP}_\epsilon \left(\dot\bP(\tau) + \frac{1}{M}\bzeta_{\bP\bP}\bP(\tau)\right)\\
        &\hspace{2.7cm}-\left(\dot\bP(\tau) - \frac{1}{M}\bzeta^\mathrm{T}_{\bP\bP}\bP(\tau)\right)\cdot \bT^{\bP\bP}_\epsilon \left(\dot\bP(\tau) - \frac{1}{M}\bzeta^\mathrm{T}_{\bP\bP}\bP(\tau)\right)\Biggr\rangle_\text{path}\\
        &= -\lim_{t\to +\infty} \frac{1}{Mt}\int_0^t \dd\tau\, \Biggl\langle\bzeta_{\bP\bP,S}\bP\cdot \bT^{\bP\bP}_\epsilon\dot\bP +  \dot\bP\cdot\bT^{\bP\bP}_\epsilon\bzeta_{\bP\bP,S}\bP \\
        &\hspace{2.7cm}+ \frac{1}{2M}\left( \bzeta_{\bP\bP}^\mathrm{T}\bP\cdot\bT^{\bP\bP}_\epsilon\bzeta_{\bP\bP}^\mathrm{T}\bP - \bzeta_{\bP\bP}\bP\cdot\bT^{\bP\bP}_\epsilon\bzeta_{\bP\bP}\bP\right)\Biggr\rangle_\text{path}\hspace{-3mm}, 
    \end{split}
\end{equation}
\end{widetext}
where $\bzeta_{\bP\bP,S} \equiv \frac{1}{2}(\bzeta_{\bP\bP} + \bzeta_{\bP\bP}^\mathrm{T})$. In Appendix \ref{app:sigma_disk}, we show that all terms in the integrand entering Eq.~\eqref{eq:entropy-symm-disk} either vanish by symmetry or contribute boundary terms that vanish as $1/t$,  so that
\begin{equation}\label{eq:entropy-symm-P}
    \sigma_{s,\text{disk}} = 0\,.
\end{equation}

For $C_n$-symmetric objects, the friction and noise-correlation matrices depend on the object orientation and evolve in time. As a consequence, computing the entropy production rate in a controlled way as the adiabatic limit is approached becomes challenging. We can nevertheless show that the contribution to the entropy production due to the net rotational motion of objects with both $C_n$ and $\Pi$ symmetries once again vanishes when simultaneously reversing time and the bath chirality. The noise correlation function $\langle\bxi_\epsilon(\tau) \otimes \bxi_\epsilon(\tau')\rangle$ for such objects has a block matrix structure, with a $2\times2$ block associated with the translational motion and a $1\times 1$ block $\langle\xi^L_\epsilon(\tau) \otimes \xi^L_\epsilon(\tau')\rangle$ for the rotational motion. As for the disk, we choose for the latter an exponential decay:
\begin{equation}\label{eq:corr_xi_epsilon_L}
    \langle\xi^L_\epsilon(\tau) \otimes \xi^L_\epsilon(\tau')\rangle \equiv \frac{1}{\epsilon^2}\ee^{-\lambda_{LL}|\tau-\tau'|/\epsilon^2}\,.
\end{equation} 
The operator $\bT_\epsilon(t)$ defined in Eq.~\eqref{eq:T_def} has a block matrix structure as well, with a $2\times2$ block associated with translational motion and a $1\times 1$ block $T_\epsilon^{LL}(t)$ associated with the rotational one. The latter can be explicitly computed using Eq.~\eqref{eq:T_def} and Eq.~\eqref{eq:corr_xi_epsilon_L}, yielding,
\begin{equation}\label{eq:T_LL}
    T_\epsilon^{LL}(t) = \frac{1}{2\lambda_{LL}}\left( 1 + \epsilon^2 \lambda_{LL}\p^2_t\right)\delta(t)\,.
\end{equation} 
The action defined in Eq.~\eqref{eq:S_def} thus decomposes into two contributions
\begin{equation}
    \begin{split}
        S[\{\bGamma(\tau)\}|\bGamma(0),\omega_0] &= S_\text{transl}[\{\bGamma(\tau)\}|\bGamma(0),\omega_0] \\
        &+ S_\text{rot}[\{\bGamma(\tau)\}|\bGamma(0),\omega_0]\,,
    \end{split}
\end{equation}
where the rotational contribution $S_\text{rot}$ is defined as
\begin{widetext}
    \begin{equation}
        S_\text{rot}[L(\tau)|L(0),\omega_0] \equiv \frac{1}{2}\int_0^t\int_0^t \dd \tau\dd \tau' \left(\dot L(\tau) - \Gammaavg + \frac{\zeta_{LL}}{I}L(\tau)\right)T_\epsilon^{LL}(\tau-\tau')\left(\dot L(\tau') - \Gammaavg + \frac{\zeta_{LL}}{I}L(\tau')\right)\,.
    \end{equation}
\end{widetext}
The symmetrized entropy production rate can be decomposed accordingly,
\begin{equation}
    \sigma_s \equiv \sigma_{s,\text{transl}} + \sigma_{s,\text{rot}}\,.
\end{equation}

While an analytical computation of the translational contribution $\sigma_{s,\text{transl}}$ is not straightforward (for reasons described following Eq.~\eqref{eq:T_def}), the rotational contribution is readily evaluated as
\begin{equation}\label{eq:entropy_symm_L}
    \begin{split}
        \sigma_{s,\text{rot}} &=- \lim_{t \to+\infty} \frac{1}{t}\bigl( S_\text{rot}[L(\tau)|L(0),\omega_0] \\
        &\hspace{1.8cm}- S_\text{rot}[\tilde L(t-\tau)|L(t),-\omega_0] \bigr)\\
        &=\lim_{t\to+\infty} \frac{1}{\lambda_{LL} t}\int_0^t\dd\tau\,  \Biggl\langle\Gammaavg (\dot L + \epsilon^2\lambda_{LL}\dddot L) \\
        &\hspace{2cm}- \frac{\zeta_{LL}}{I}(\dot L L + \epsilon^2 \lambda_{LL}\ddot L \dot L) \Biggr\rangle_\text{path}\\
        &= 0\,.
    \end{split}
\end{equation}
In the second equality we have used that $\Gammaavg$ changes sign under reversal of the bath chirality (see Sec.~\ref{section:tracer-symmetries}) and that the operator $T_\epsilon^{LL}$ given by Eq.~\eqref{eq:T_LL} is symmetric upon reversal of time and chirality of the bath. The final equality to zero then follows from the fact that all terms in the integrand contribute boundary terms that vanish as $1/t$. Note that since the noise correlations in Eq.~\eqref{eq:corr_xi_epsilon_L} are time-reversal symmetric, the computation of $\sigma_{s,\text{rot}}$ can equally be done by taking $\epsilon=0$ directly in Eq.~\eqref{eq:T_LL}.

The hidden symmetries in Eq.~\eqref{eq:entropy-symm-P} and Eq.~\eqref{eq:entropy_symm_L} can be understood through analogy with the motion of a charged particle in an external magnetic field. Just as that system is invariant under simultaneous reversal of time and the magnetic field~\cite{nakajimaQuantumTheoryTransport1958,obyrneTimeIrreversibilityActive2022}, the dynamics of these rotationally-symmetric objects are invariant under simultaneous reversal of time and the bath chirality. The presence of such a symmetry rationalizes the finding, discussed in Sec. \ref{section:effective-temperatures} that the stationary distribution of the disk has a Boltzmann form.

\if As discussed in the next section, both the freely rotating and fully confined case leads to a vanishing entropy-production rate upon flipping $t\to -t$ and $\omega_0\to-\omega_0$
Unlike the freely rotating case discussed in Section~\ref{section:diffusion-mobility-einstein}, a net rotational flux in an angular potential would leads to a finite entropy production, breaking the effective equilibrium of Eq.~\eqref{eq:confined-rod-boltzmann}.\fi

\section{Multipole description of far-field current}\label{section:multipole}

\subsection{Dynamics of the chiral active bath}
In the previous sections, we developed an adiabatic theory for the motion of passive objects by integrating out the bath degrees of freedom. This allowed to us to determine the impact of the object shape on its dynamics. We now complement this perspective with the impact that the object shape has on the bath itself.
The dynamics of the bath are given by Eq.~\eqref{eq:cABP-EOM}, which corresponds to the Fokker-Planck equation
\begin{equation}\label{eq:cABP-fokker-planck}
    \partial_t \psi = -\bnabla \cdot \big[ v_0 \psi \bu(\theta)  - \mu \psi \bnabla V \big]
    - \partial_\theta \big[ \omega_0 \psi - D_r \partial_\theta \psi \big]\,.
\end{equation}
Here, $\psi(\br, \theta, t)$ is the probability of any of the non-interacting bath particles existing at position $\br$ with orientation $\theta$ at time $t$. We define $\mu = \gamma^{-1}$ as the mobility of the bath particle and $v_0 = \mu f_0$ as its self-propulsion speed.
The first three angular moments of $\psi(\br, \theta, t)$ define the density, polar order, and nematic order, respectively, as
\begin{align}
\rho(\br, t) &= \int \dd\theta\ \psi(\br, \theta, t)\,,\\
\bm{m}(\br, t) &=\int \dd\theta\ \psi(\br, \theta, t) \bu(\theta)\,,\\
\mathbf{Q}(\br, t) &= \int \dd\theta\ \psi(\br, \theta, t) \left(\bu(\theta)\otimes \bu(\theta) - \frac{1}{2} \id \right)\,.
\end{align}
From Eq.~\eqref{eq:cABP-fokker-planck}, we then obtain evolution equations for the density and polar fields as:
\begin{align}
\label{eq:non-interacting-moment-1}
\partial_t \rho &= -\bnabla \cdot \big[ v_0 \bm{m} - \mu \rho \bnabla V \big]\,, \\
\label{eq:non-interacting-moment-2}
\begin{split}
    \partial_t \bm{m} &= -\bnabla \cdot \big[ v_0 (\mathbf{Q} + \frac{1}{2}\rho\id) -\mu \bm{m} \otimes \bnabla V \big] \\
    &\hspace{9mm} + \big[ \omega_0 \bA - D_r \id \big] \bm{m} \,.
\end{split}
\end{align}

\begin{figure}[t!]
    \includegraphics[width=.49\textwidth]{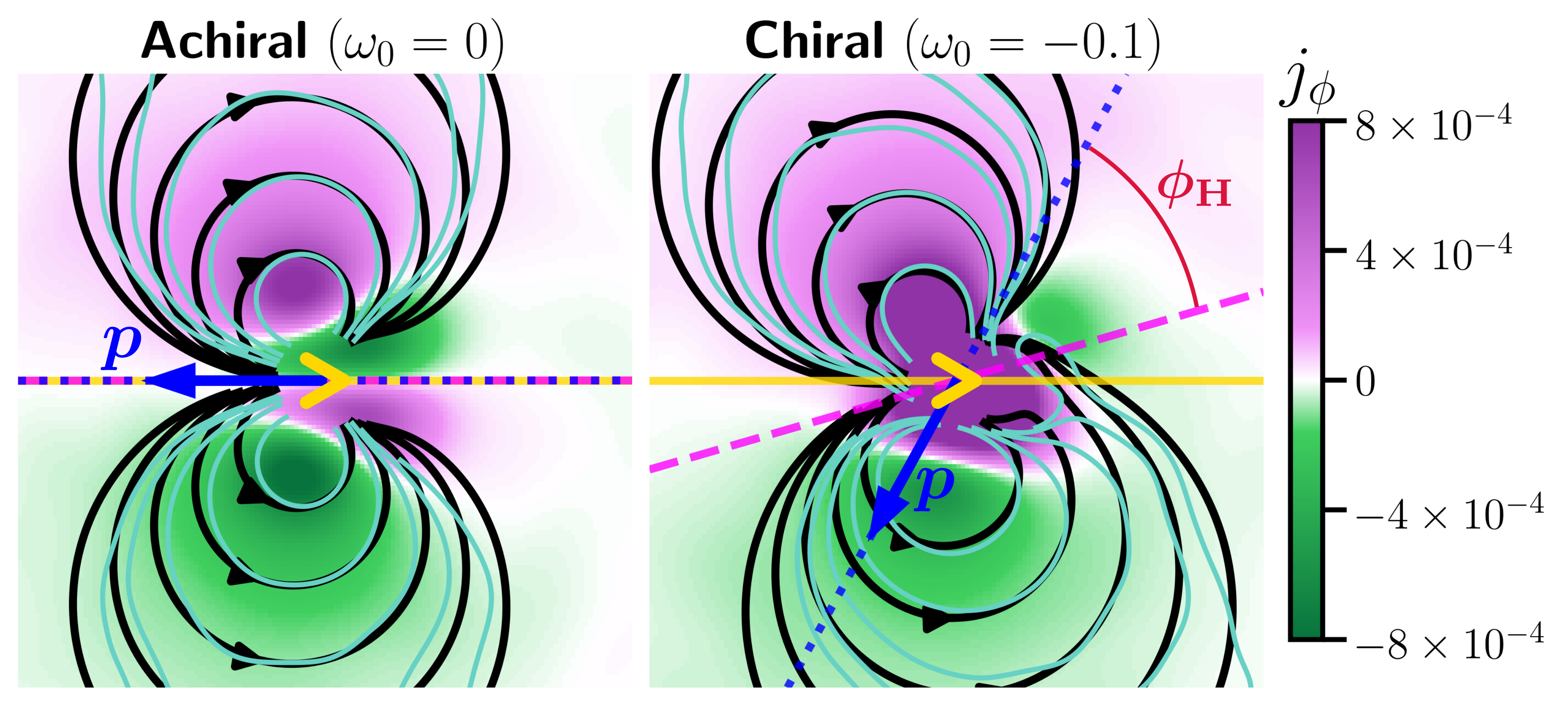}
    \caption{\textbf{Flux of the active bath} in steady state (blue streamlines) induced by a wedge, \textit{vs.}\ multipole prediction (black streamlines), and heatmap of the rotational flux $j_\phi = \br\times\bm{j}/|\br|$. In an achiral bath \textbf{(left)} the dipole moment $\bm{p}$ (blue arrow and dotted line) lies on the symmetry axis of the object (yellow line), which coincides with the symmetry axis of the bath flux (dashed magenta line). In chiral bath \textbf{(right)} $j_\phi$ exhibits a large net circulation near the object, and $\bm{p}$ is rotated with respect to the symmetry axis of the object. The flux field is then further rotated with respect to $\bm{p}$ by the Hall angle $\phi_\text{Hall} = \arctan(D^\mathrm{b}_\perp / D^\mathrm{b}_\parallel) = \pi/4$ for $\ell_p=10$, $\ell_g=10$. (Figure adapted from the companion Letter~\protect\cite{companionPRL}.)
    \label{fig:multipole-flux}
    }
\end{figure}
Recalling the definition $\bu(\theta) = [\cos(\theta),
  \sin(\theta)]^\text{T}$ and differentiating yields the identities $\partial_\theta \bu = -\bA \bu$ and $\partial^2_\theta \bu = - \bu$.
Equation~\eqref{eq:non-interacting-moment-1} is a continuity equation $\partial_t \rho = -\bm{\nabla} \cdot \bm{j}$, where the bath current is $\bm{j}\equiv v_0 \bm{m} - \mu \rho \bnabla V$. 
Solving for the steady-state value of $\bm{m}$ in Eq.~\eqref{eq:non-interacting-moment-2} and inserting the result into Eq.~\eqref{eq:non-interacting-moment-1} then yields for the steady-state current
\begin{equation}\label{eq:flux-noninteracting}
    \bm{j} =  - \mu \rho \bm{\nabla} V -\mathbf{D}^\text{b} \bm{\nabla} \rho - \bm{\nabla}\cdot \big[2 \mathbf{D}^\text{b} (\mathbf{Q} - \frac{\mu}{v_0} \bm{m}\otimes \bm{\nabla} V) \big]\,.
\end{equation}
The three terms on the right-hand side account for the current driven directly by the bath-object interaction potential $V$, the diffusive flux, and the divergence of a stress-like term due to orientational order near the object.
$\mathbf{D}^\text{b}$ is the diffusivity of the bath particles defined by
\begin{equation}
    \mathbf{D}^\text{b} = \frac{v_0^2}{2(D_r^2 + \omega_0^2)} (D_r \id + \omega_0 \bA) \equiv D^\text{b}_\parallel \id + D^\text{b}_\perp \bA\,,
\end{equation}
with $D^\text{b}_\parallel$ and $D^\text{b}_\perp$ the even and odd diffusivity of the bath particles, respectively.

\subsection{Multipole expansion}

Placing a passive object into an active bath induces density modulations and currents in the bath which can extend far away from the object~\cite{granekInclusionsBoundariesDisorder2023}. Their large-scale structures has been shown to be well captured within a multipole description~\cite{baekGenericLongrangeInteractions2018,Granek2020}. Here we extend this approach to characterize the bath density and current fields when the bath is also chiral, revealing how the bath chirality induces large-scale modifications of the far-field flow and density fields. 

Far from the object, the bath flux is well described by its diffusive approximation
\begin{equation}\label{eq:far-field-flux}
    \bm{j}_D = - \mathbf{D^\text{b}} \bm{\nabla} \rho\,.
\end{equation}
Close to the object, it deviates from this expression by an amount $\bm{\delta j} = \bm{j} - \bm{j}_D$. Requiring that $\bm{\nabla} \cdot \bm{j} = 0$ in the steady state leads to the Poisson equation
\begin{equation}\label{eq:poisson-eqn}
    D_\parallel^\mathrm{b} \nabla^2 \rho = \bm{\nabla} \cdot \bm{\delta j}\,,
\end{equation}
which admits the solution
\begin{equation}\label{eq:density-greens-function}
    \rho(\br) = \rho_0 + \frac{1}{2 \pi D_\parallel^\mathrm{b}} \int \dd\bm{r'} \ln | \br - \bm{r'}|\ \bm{\nabla'} \cdot \bm{\delta j}(\bm{r'})\,.
\end{equation}
Here, $\bm{\nabla'}$ denotes the gradient with respect to $\bm{r'}$ and $\ln | \br - \bm{r'}|$ is the Green's function for the Laplacian in two dimensions, which can be expanded when $|\br| \gg |\bm{r'}|$ as
\begin{align}
    \begin{split}
        &\ln | \br - \bm{r'}| = \frac{1}{2} \ln \big[(\br-\bm{r'}) \cdot (\br-\bm{r'}) \big] \\
        &= \ln |\br| - \frac{\br \cdot \bm{r'}}{r^2} - \bm{r'} \cdot \frac{2\br\otimes \br - r^2 \id}{2r^4} \cdot  \bm{r'} + \mathcal{O}(r^{-3})\,.
    \end{split}
\end{align}
Inserting this expansion into Eq.~\eqref{eq:density-greens-function} yields
\begin{widetext}
\begin{align}\label{eq:multipole-expansion_initial}
    \begin{split}
    \rho(\br) &= \rho_0 + \frac{1}{2 \pi D_\parallel^\mathrm{b}}\int \dd\bm{r'}\ \bigg[ \ln |\br| - \frac{\br \cdot \bm{r'}}{r^2} - \bm{r'} \cdot \frac{2\br\otimes \br - r^2 \id}{2r^4} \cdot  \bm{r'} + \mathcal{O}(r^{-3}) \bigg] \bm{\nabla'} \cdot \bm{\delta j}\\
    &= \rho_0 + \frac{1}{2 \pi D_\parallel^\mathrm{b}}\int \dd\bm{r'}\ \bigg[ \frac{\br}{r^2} + \frac{2\br\otimes \br - r^2 \id}{r^4} \cdot  \bm{r'} + \mathcal{O}(r^{-3}) \bigg] \cdot \bm{\delta j} \\
    &= \rho_0 + \frac{1}{2 \pi D_\parallel^\mathrm{b}}\int \dd\bm{r'}\ \bigg[ \frac{\br}{r^2} + \frac{2\br\otimes \br - r^2 \id}{r^4} \cdot  \bm{r'} + \mathcal{O}(r^{-3}) \bigg] \cdot \bigg[ -\mu \rho \bm{\nabla'} V - 2 \bm{\nabla'} \cdot \bigg( \mathbf{D}^\mathrm{b} \big[ \mathbf{Q} - \frac{\mu}{v_0} \bm{m} \otimes \bm{\nabla'} V \big] \bigg) \bigg] \\
    &= \rho_0 + \frac{1}{2 \pi D_\parallel^\mathrm{b}}\int \dd\bm{r'}\ -\frac{\br}{r^2} \cdot \mu \rho \bm{\nabla'} V - \frac{2\br\otimes \br - r^2 \id}{r^4} :  \bigg[\bm{r'} \otimes \mu \rho \bm{\nabla'} V + \frac{2 \mu}{v_0} \mathbf{D}^\mathrm{b}  \bm{m} \otimes \bm{\nabla'} V \bigg] + \mathcal{O}(r^{-3})\,,
    \end{split}
\end{align}
\end{widetext}
where we have applied the divergence theorem in the second and the fourth equalities, and used that
the nematic tensor $\mathbf{Q}$ enters only at order $\mathcal{O}(r^{-3})$.
We use $\mathbf{A}:\mathbf{B}$ to indicate the double contraction of matrices $\mathbf{A}$ and $\mathbf{B}$.
Finally, evaluating the integral in Eq.~\eqref{eq:multipole-expansion_initial} yields the multipole expansion
\begin{equation}\label{eq:multipole-expansion}
    \rho(\br) = \rho_0 + \frac{\mu}{2 \pi D_\parallel^\mathrm{b}}\bigg(\frac{\br \cdot \bm{p}}{r^2} + \frac{\br \cdot \mathbf{q} \cdot \br}{2r^4}\bigg) + \mathcal{O}(r^{-3})\,,
\end{equation}
where the dipole $\bm{p}$ and quadrupole $\mathbf{q}$ moments are defined as
\begin{align}
    \label{eq:dipole}
    \bm{p} &= -\int \dd\br\ \rho(\br) \bm{\nabla} V\,, \\
    \label{eq:quadrupole}
    \begin{split}
    \mathbf{q} &= -2\int \dd\br\ \bigg\{\rho(\br) \br \otimes \bm{\nabla} V
    + \frac{2}{v_0} \mathbf{D}^\mathrm{b}\bm{m}(\br)\otimes \bm{\nabla} V\\
    &\hspace{4mm}- \frac{1}{2} \bigg[ \rho(\br) \br \cdot \bm{\nabla} V + \frac{2}{v_0} \big(\mathbf{D}^\mathrm{b} \bm{m}(\br)\big) \cdot \bm{\nabla} V \bigg] \id \bigg\}\,.
    \end{split}
\end{align}
The far-field bath flux induced by the object is then obtained as
\begin{equation}\label{eq:multipole-flux}
\begin{split}
    \bm{j} = - \mathbf{D}^\mathrm{b} \bm{\nabla}\rho
    = -\frac{\mu}{2\pi D_\parallel^\mathrm{b}} \mathbf{D}^\mathrm{b} \bigg[\frac{r^2 \bm{p} - 2(\br\cdot\bm{p})\br}{r^4} \\+ \frac{r^2(\mathbf{q} + \mathbf{q}^T)\cdot \br  - 4(\br\cdot\mathbf{q}\cdot\br)\br}{r^6} \bigg] + \mathcal{O}(r^{-4})\,.
\end{split}
\end{equation}
Equations~\eqref{eq:multipole-expansion}-\eqref{eq:multipole-flux} constitute the main result of this section. They are shown in Fig.~\ref{fig:multipole-flux} to predict the bath flux far from an object. Note that the integrands in Eqs.~\eqref{eq:dipole}-\eqref{eq:quadrupole} vanish everywhere except at the object, where $\bm{\nabla} V \ne \mathbf{0}$.
In this sense, the multipole expansion of~\eqref{eq:multipole-expansion} predicts the far-field density modulation induced by the object from measurements local to the object. This is analogous to how a multipole expansion in classical electrostatics approximates the electric field far from a localized charge distribution. Let us now discuss the properties of the flux and density modulations and relate them to the bath chirality.

\begin{figure}[t]
    \centering
    \includegraphics[width=.48\textwidth]{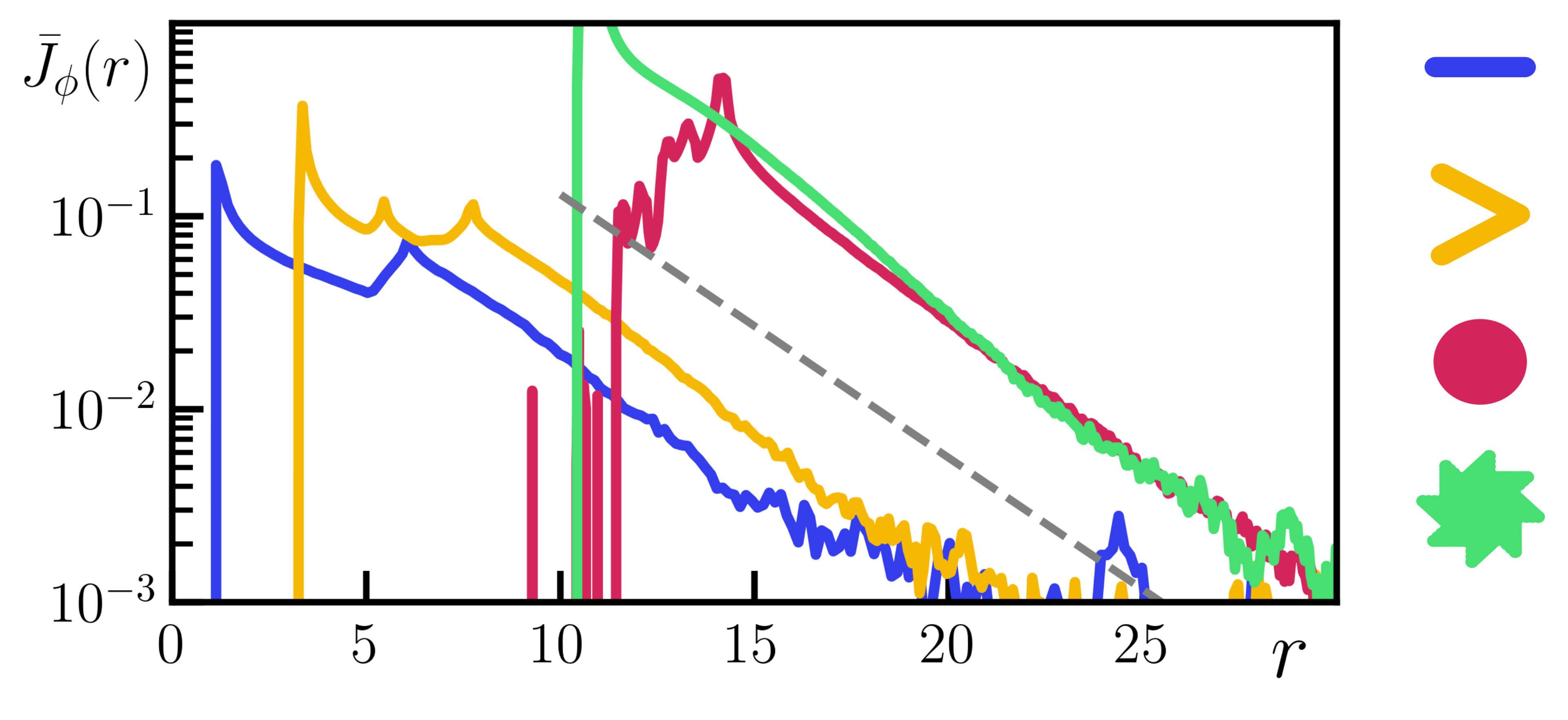}
    \caption{Profiles of the rotational current $\bar{j_\phi}(r)$ for the four object shapes shown at right, with corresponding colors. For the disk, rod and wedge, the bath is chiral ($\ell_p=10$, $\ell_g=10$) while for the chiral gear (green) the bath is achiral ($\ell_p=10$, $\ell_g=\infty$). In all cases, $\bar{j}_\phi(r)$ decays rapidly far from the object, with exponential tails $\bar{j}_\phi(r) \propto e^{-\pi r / \ell_p}$ (dashed line).
    \label{fig:Jtheta-exp-decay}
    }
\end{figure}

On the one hand, some characteristics are common to chiral and achiral bath. For instance, Eq.~\eqref{eq:dipole} shows the dipole moment to equal the total force exerted by the object on the bath and Eq.~\eqref{eq:flux-noninteracting} shows that it measures the total bath current~\cite{baekGenericLongrangeInteractions2018,Granek2020,granekInclusionsBoundariesDisorder2023}:
\begin{equation}\label{eq:dipole-ratchet}
     \bm{p} = -\bFavg = \gamma \int \dd\br\ \bm{j}(\br) \,.
\end{equation}

On the other hand, Fig.~\ref{fig:multipole-flux} shows important differences between chiral and achiral baths.
First, in a chiral bath the axis of ${\bm p}$ does not coincide with the symmetry axis of the object.
Then, from Eq.~\eqref{eq:far-field-flux} the far-field flux in a chiral bath has an odd diffusive contribution orthogonal to $\bnabla \rho$. This rotates the flux axis with respect to $\bm{p}$ by a ``Hall angle'' $\phi_\text{H} = \arctan(D^\mathrm{b}_\perp / D^\mathrm{b}_\parallel)$, as follows from the polar decomposition 
\begin{equation}
    \mathbf{D}^\text{b} = \sqrt{(D^\mathrm{b}_\parallel)^2 + (D^\mathrm{b}_\perp)^2} \begin{bmatrix} \cos(\phi_\text{H}) & \sin(\phi_\text{H}) \\ -\sin(\phi_\text{H}) & \cos(\phi_\text{H})
        \end{bmatrix}\,.
\end{equation}

Taking the curl of Eq.~\eqref{eq:far-field-flux} yields $\bnabla \times \bm{j}_D = D_\perp \nabla^2 \rho$, showing that in general a non-vanishing $D_\perp$ allows for a rotational diffusive flux. However, in the far field, $\nabla^2\rho$ vanishes  due to Eq.~\eqref{eq:poisson-eqn}, so that the far-field flux remains irrotational. 
In contrast, the near-field flux contains a significant rotational component, which is related to the antisymmetric part of the quadrupole moment $q^\mathrm{A} = -\frac{1}{2}\bA:\mathbf{q}$ through
\begin{align}
\label{eq:circulation-quadrupole}
        \int \dd\br\ \br\times\bm{j}
         &= -\mu\Gammaavg \notag\\
         &+ \frac{2}{f_0} \int \dd\br \big(D^\mathrm{b}_\perp \bm{m} \cdot \bnabla V
         + D^\mathrm{b}_\parallel \bm{m} \times \bnabla V\big) \notag\\
         &= \mu q^\mathrm{A} \,.
\end{align}
This expression is the angular analog to Eq.~\eqref{eq:dipole-ratchet}.
Here, the two-dimensional cross-product is a scalar defined as $\bm{a} \times \bm{b} = (\bA\bm{a})\cdot\bm{b}$.
The first equality follows from inserting Eq.~\eqref{eq:flux-noninteracting} for $\bm{j}$, and the second equality from the definition of $\mathbf{q}$ in Eq.~\eqref{eq:quadrupole}.
We have invoked the divergence theorem to show that
\begin{equation}
    \int \dd\br\ \br \times \bm{\nabla} \rho(\br) = -\int \dd\br\ \rho(\br) \bm{\nabla} \times \br = 0\,,
\end{equation}
and
\begin{align}
    \begin{split}
        &\int \dd\br\ \br \times \bm{\nabla} \cdot (\mathbf{D}^\mathrm{b} \mathbf{Q})
        = \int \dd\br\ (\mathbf{D}^\mathrm{b} \mathbf{Q}) : \bA\\
        = &\int \dd\br\ D_\parallel^\mathrm{b} (Q_{xy}-Q_{yx}) + D_\perp^\mathrm{b}(Q_{xx} + Q_{yy}) = 0\,,
    \end{split}
\end{align}
because $\bm{\nabla} \times \br = 0$, $Q_{xy} = Q_{yx}$, and $Q_{xx} = - Q_{yy}$.
Note that while $q^\mathrm{A}$ is related to the circulating currents by Eq.~\eqref{eq:circulation-quadrupole}, it drops out of the multipole expansion~\eqref{eq:multipole-expansion} due to its antisymmetric nature.

To characterize these circulating currents for different object shapes, we measured the net circulation\footnote{Note that the radial component of the flux $j_r \equiv \br\cdot\bm{j}(\br, \phi)/|\br|$ vanishes for all $r$ upon averaging over $\phi$; that is, $\int_0^{2\pi}d\phi\ j_r(r) = 0$, due to the steady-state continuity condition.} around the object as $\bar{j}_\phi(r) \equiv \int_0^{2\pi}d\phi\ {j}_\phi$, where $j_\phi(r) \equiv \br\times\bm{j}/|\br|$ and polar coordinates have been introduced as $x = r\cos(\phi)$ and $y = r\sin(\phi)$.
In Fig.~\ref{fig:Jtheta-exp-decay} we observe that $\bar{j}_\phi$ vanishes exponentially away from the object, regardless of its size, at a rate set by the persistence length $\ell_p$. Note that all considerations above on the circulating flux directly extends to the case of a chiral object in an achiral active bath, as illustrated using an asymmetric gear in Fig.~\ref{fig:Jtheta-exp-decay}.

\section{Conclusion}
We have constructed a detailed theory for the dynamical properties of a passive object immersed in a chiral active bath. Our derivations show how the chiral nature of the bath endows the resulting Langevin dynamics with unusual properties that are reflected in the non-standard relationship between the Langevin and Fokker-Planck equations. This derivation provides a microscopic grounding of many observations presented in our companion Letter~\cite{companionPRL}. 
In particular, we show that broken symmetries in the bath intersect with broken symmetries of the object to give rise to odd transport properties and rotational ratchet currents. 
The corresponding odd diffusivity and odd mobility are connected by an Einstein relation, but no equivalent odd fluctuation-dissipation theorem exists between the noise correlations and the friction. 
The effective equilibrium regimes reported in our companion letter are here rationalized by the existence of hidden time-reversal symmetries of the object dynamics when the bath chirality is flipped upon time reversal.
Finally, we develop a multipole expansion to predict the far-field structure and flows induced in the chiral bath and we connect these effects to the ratchet force and torque exerted on the object.

A number of intriguing directions are suggested by these results, including investigating interactions between passive objects mediated by a chiral active bath, with consequences for tunable self-assembly. Another direction of interest would be to consider the limit of large bath before the adiabatic limit is taken. Then, hydrodynamics modes lead to long-time tails~\cite{VanBeijeren1982} that can have important consequences in active systems~\cite{Granek2022}. How they would manifest in chiral active bath remains an interesting open question. On a more mathematical side, odd Langevin dynamics have now arisen in two different contexts: for a passive tracers in a chiral bath, here, and for a Brownian charged particle in a magnetic field in Ref~\cite{chun2018emergence}. Much remains to be done to fully characterize the stochastic calculus emerging from these equations and its physical consequences (at $\epsilon =0^+$).

\vspace{5mm}
\noindent\textbf{\textit{Acknowledgements.}}
 CH, FG, JT and FvW acknowledge the financial support of the ANR THEMA AAPG2020 grant.

\appendix
\addcontentsline{toc}{section}{Appendices}

\section{Statistics of the projected force}\label{app:random_force}

In this Appendix we show that the statistics of the projected force $\bF^+(\tau) = \mU[\mLB^\dagger + \epsilon\mLo^{*\dagger}](\tau,0)\bF(0)$ matches the statistics of the force $\bF_0$, exerted by the bath on a fixed object. By iteratively applying the operator identity in Eq.~\eqref{eq:operator_identity},  with $\mA =\mLB^\dagger$ and $\mB = \epsilon \mQ \mLo^{*\dagger}$, to $\bF^+(\tau)$, we construct the Dyson series 
\begin{widetext}
\begin{equation}\label{eq:F+_dyson}
        \bF^+(\tau) = \bF_0(\tau) + \sum_{n=1}^{+\infty} \epsilon^n \int_0^\tau \dd\tau_1\, \int_0^{\tau_1}\dd\tau_2\, \ldots\int_0^{\tau_{n-1}}\dd\tau_n\, \mO(\tau,\tau_1)\mO(\tau_1,\tau_2)\ldots\mO(\tau_{n-1},\tau_n)\bF_0(\tau_n)\,,
\end{equation}
where we defined an operator $\mO(t_1,t_2)\equiv \mU[\mLB](t_1,t_2)\mQ\mLo$ and $\bF_0(\tau) \equiv \mU[\mLB](\tau,0)\bF(0)$ is the force exerted on the object by the bath, after evolving the latter for a time  $\tau$ with the object held fixed in its configuration at time $0$. 

Averaging Eq.~\eqref{eq:F+_dyson} over the bath degrees of freedom and using the triangle inequality we obtain
\begin{equation}
    \lvert \langle\bF^+(\tau)\rangle_\text{b} - \langle \bF_0(\tau)\rangle_\text{b}\rvert \leq \sum_{n=1}^{+\infty} \epsilon^n c_n\,,
\end{equation}
with the coefficients $c_n$ given by
\begin{equation} 
    c_n \equiv \lim_{s\to+\infty}\int_0^s \dd\tau_1\, \int_0^{\tau_1}\dd\tau_2\, \ldots\int_0^{\tau_{n-1}}\dd\tau_n\, \bigl\lvert\left\langle\mO(s,\tau_1)\mO(\tau_1,\tau_2)\ldots\mO(\tau_{n-1},\tau_n)\bF_0(\tau_n)\right\rangle_\text{b}\bigr\rvert\,.
\end{equation}
\end{widetext}
We now use that the bath dynamics for a fixed object is ergodic, and assume that the decay of correlation functions of physical observables happens sufficiently fast, so that all the coefficients $c_n$ are finite. 
Note that this is expected to hold when the adiabatic limit is taken before the large-bath limit, so that the bath hydrodynamic modes also have time to relax~\cite{VanBeijeren1982,van1986brownian}.  We also assume that the series $\sum_{n=1}^{+\infty}\epsilon^n c_n$ converges. Using the stationarity of the bath dynamics we then have, to lowest order in $\epsilon$,
\begin{equation}
    \begin{split}
        \langle \bF^+(t)\rangle_\text{b} &= \langle \bF_0(t)\rangle_\text{b} + O(\epsilon)= \langle\bF_0\rangle_\text{b} + O(\epsilon)\,.
    \end{split}
\end{equation}
The average value $\bF^+(t)$ matches the average, steady-state force exerted on the object by the bath when the object is held fixed. As discussed in the main text, the average force $\langle\bF_0\rangle_\text{b}$ can be different from $\mathbf{0}$. We thus introduce the fluctuations of the force and the torque on the object as
\begin{equation}
    \delta\bF_0 \equiv\bF_0-\left\langle\bF_0\right\rangle_\text{b},
\end{equation} 
and analogous fluctuating quantities for $\delta\bF^+$. 

The two-point correlation function of $\delta\bF^+$ can also be expanded in terms of the small adiabaticity parameter $\epsilon$. From Eq.~\eqref{eq:F+_dyson}, using $\delta\bF^+(0)=\delta\bF_0(0)$, we obtain
\begin{equation}
    \lvert \langle \delta\bF^+(0)\otimes  \delta\bF^+(\tau) \rangle_\text{b} - \langle \delta \bF_0(0) \otimes \delta \bF_0(\tau) \rangle_\text{b}\rvert \leq \sum_{n=1}^{+\infty} d_n \epsilon^n\,, 
\end{equation}
with the coefficient $d_n$ given by
\begin{widetext}
\begin{equation}
    d_n \equiv \lim_{s\to+\infty}\int_0^s \dd\tau_1\, \int_0^{\tau_1}\dd\tau_2\, \ldots\int_0^{\tau_{n-1}}\dd\tau_n\, \bigl\lvert\left\langle\delta\bF_0(0)\mO(s,\tau_1)\mO(\tau_1,\tau_2)\ldots\mO(\tau_{n-1},\tau_n)\delta \bF_0(\tau_n)\right\rangle_\text{b}\bigr\rvert\,.
\end{equation}
\end{widetext}
If all the coefficients $d_n$ are finite, which follows again from our assumption of an ergodic bath of finite size we obtain
\begin{equation}
    \langle \delta \bF^+(0) \otimes \delta\bF^+(\tau)\rangle_\text{b} = \langle\delta \bF_0(0) \otimes \delta\bF_0(\tau)\rangle_\text{b} + O(\epsilon)\,.
\end{equation}  
The proof for higher moments of $\delta\bF^+$ proceeds in a similar fashion~\cite{mazur1970molecular}. The statistics of the projected force $\bF^+$ and of $\bF_0$ are thus identical in the adiabatic limit. A similar derivation can be performed for the projected torque.

\section{The random force $\xi$ is Gaussian}\label{app:force_is_gaussian}
In this Appendix we show that the random force $\bxi(t)$ defined in Eq.~\eqref{eq:noise_rescaled}, with two-point correlations given by Eq.~\eqref{eq:corr_xi_gen}, has Gaussian statistics in the adiabatic limit. To do so, we consider the moment generating function $\psi_{\bxi}[\bg(\tau)]$, defined as
\begin{align}\label{eq:cgf}
    \psi_{\bxi}[\bg(\tau)] \equiv \left\langle\ee^{\epsilon^{-1}\int_0^{+\infty}\dd\tau\, \bg(\tau)\cdot \bxi(\epsilon^{-2}\tau)}\right\rangle_\text{b}\,.
\end{align}
The statistics of $\epsilon^{-1}\bxi$ can be obtained from the knowledge of $\psi_{\bxi}$ by taking its functional derivatives with respect to the auxiliary field $\bg(\tau)$ and setting the auxiliary field $\bg(\tau)=\mathbf{0}$. An expansion in powers of $\epsilon$ yields
\begin{widetext}
\begin{align}
    \begin{split}
        \psi_{\bxi}(\bg(\tau)) &=  1 + \epsilon\int_0^{+\infty} \dd \tau \bg(\epsilon^{2} \tau)\cdot \langle \bxi(\tau)\rangle_\text{b} + \frac{\epsilon^{2}}{2}\int_0^{+\infty}\int_0^{+\infty}\dd \tau \dd \tau' \bg(\epsilon^{2} \tau)\cdot \langle\bxi(\tau)\otimes\bxi(\tau')\rangle_\text{b}\bg(\epsilon^{2} \tau') +\mathcal{O}(\epsilon^3)\\
        &\approx  1   + \frac{\epsilon^{2}}{2}\int_0^{+\infty}\int_0^{+\infty}\dd \tau \dd \tau' \bg(\epsilon^{2} \tau)\cdot \langle\bxi(\tau)\otimes\bxi(\tau')\rangle_\text{b}\bg(\epsilon^{2} \tau') \\
        &\approx \exp\left[{\frac{\epsilon^2}{2}}\int_0^{+\infty}\int_0^{+\infty}\dd \tau \dd \tau' \bg(\epsilon^{2} \tau)\cdot \langle\bxi(\tau)\otimes\bxi(\tau')\rangle_\text{b}\bg(\epsilon^{2} \tau')\right]\\
        &= \exp\left[{\frac{1}{2}}\int_0^{+\infty}\int_0^{+\infty}\dd \tau \dd \tau' \bg(\tau)\cdot \langle\epsilon^{-1}\bxi(\epsilon^{-2}\tau)\otimes\epsilon^{-1}\bxi(\epsilon^{-2}\tau')\rangle_\text{b}\bg( \tau')\right]\,.
    \end{split}
\end{align}
\end{widetext}
In the first line, we expanded the exponential in Eq.~\eqref{eq:cgf}, changed the integration variable, and made use of the fact that
\begin{align}
\begin{split}
    &\epsilon^{-n}\int_0^{+\infty}\dd \tau_1\int_0^{+\infty}\dd \tau_2\ldots \int_0^{+\infty} \dd \tau_n \langle \Pi_{i=1}^n \xi_{\alpha_i}(\epsilon^{-2} \tau_i)\rangle_\text{b} \\
    &= \epsilon^{n}\int_0^{+\infty}\dd \tau_1\int_0^{+\infty}\dd \tau_2\ldots \int_0^{+\infty} \dd \tau_n \langle \Pi_{i=1}^n \xi_{\alpha_i}(\tau_i)\rangle_\text{b} \\ 
    &=O(\epsilon^{n })\,.
\end{split}
\end{align}

In the second line, we neglected the $O(\epsilon^3)$ terms and used the fact that $\langle \bxi\rangle_\text{b} =\mathbf{0}$. In the third line, we have reverted the Taylor expansion to an exponential, neglecting again $O(\epsilon^3)$ terms. Finally in the last line, by means of a change of variable, we have obtained the cumulant generating function for a Gaussian process, with correlations given by Eq.~\eqref{eq:corr_xi_gen}. This concludes the demonstration that the random force fluctuations $\epsilon^{-1}\bxi(\epsilon^{-2} \tau)$ has Gaussian statistics. 

\section{Derivation of Eq.~\eqref{eq:dynmomrescaled} and Eq.~\eqref{eq:dynandmomrescaled}}\label{app:PLF+}
The term $\mP\mLo^{*\dagger}\bF^+(\epsilon^{-2}\tau)$  in Eq.~\eqref{eq:Pdot_effective_Langevin_F+_as_F0} reads
\begin{equation}
    \begin{split}                       &\mP\mLo^{*\dagger}\bF^+(\epsilon^{-2}\tau) = \left\langle \left[\frac{\bP^*}{M^*}\cdot\bnabla_{\bR} + \bF\cdot\bnabla_{\bP^*}\right]\bF^+(\epsilon^{-2}\tau)\right\rangle_\text{b} \\
    &+ \left\langle\left[\frac{L}{I}\p_{\Theta} + \Gamma \p_{L^*} \right] \bF^+(\epsilon^{-2}\tau) \right\rangle_\text{b}\\
    &= -\frac{\bP^*}{M^*}\cdot \left\langle \left(\bnabla_{\bR} \ln \rhob\right) \otimes \bF^+(\epsilon^{-2}\tau) \right\rangle_\text{b} \\
    &+ \bnabla_{\bP^*} \cdot \left\langle \bF(0)\otimes  \bF^+(\epsilon^{-2}\tau)\right\rangle_\text{b}\\
    &-\frac{L^*}{I^*}\left\langle (\p_{\Theta} \ln \rhob) \bF^+(\epsilon^{-2}\tau) \right\rangle_\text{b} +\p_{L^*} \left\langle \Gamma \bF^+(\epsilon^{-2}\tau)\right\rangle_\text{b} \\
    &+ \frac{L^*}{I^*}\p_{\Theta} \left\langle \bF^+(\epsilon^{-2}\tau)\right\rangle_\text{b}\,.
    \end{split}
\end{equation}
To pass from the first to the second equality, we used the fact that, due to translational invariance, $\bnabla_{\bR} \langle \bF^+\rangle_\text{b} = \mathbf{0}$, which implies that $\langle \bnabla_{\bR} \bF^+ \rangle_\text{b} = -\langle (\bnabla_{\bR} \ln \rhob)\bF^+ \rangle_\text{b}$. On the other hand, since the object has an arbitrary shape, rotational invariance can be broken, and thus $\langle \p_{\Theta} \bF^+\rangle_\text{b} = \p_{\Theta} \langle \bF^+ \rangle_\text{b} - \langle(\p_{\Theta} \ln \rhob) \bF^+\rangle_\text{b}$.

As shown in the previous section, in the adiabatic limit we can replace $\bF^+(\epsilon^{-2}\tau)\approx \langle \bF_0 \rangle_\text{b} + \delta\bF_0(\epsilon^{-2}\tau)$. The correlation functions obtained through this approximation are independent of the angular or linear momentum of the object, so the terms proportional to $\bnabla_{\bP^*}$ and $\partial_{L^*}$ vanish. Terms like $\langle \bnabla_{\bR} \ln \rhob \rangle_\text{b}\otimes \langle\bF \rangle_\text{b}$ and $\langle\p_{\Theta} \ln \rhob \rangle_\text{b}\langle\bF \rangle_\text{b}$ vanish too since, because of the normalization of $\rhob$, we have 
\begin{align}\label{eq:normalization-identity}
    \begin{split}
        \langle \bnabla_{\bR} \ln \rhob \rangle_\text{b} &= \bnabla_{\bR} \int \dd\br^N \dd\theta^N\ \rhob(\br^N,\theta^N|\bR, \Theta)
        \\&= \mathbf{0}\,,\\
        \langle \partial_\Theta \ln \rhob \rangle_\text{b} &= \partial_\Theta \int \dd\br^N \dd\theta^N\ \rhob(\br^N,\theta^N|\bR, \Theta)
        \\&= 0\,.\\
    \end{split}
\end{align}
We are thus left with
\begin{align}
    \begin{split}           &\mP\mLo^{*\dagger}\bF^+(\epsilon^{-2}\tau) = \\&-\langle\delta\bF_0(\epsilon^{-2}\tau)\otimes\bnabla_{\bR} \ln\rhob(0)\rangle_\text{b}\cdot \frac{\bP^*}{M^*} \\
    &- \langle\delta \bF_0(\epsilon^{-2}\tau) \p_{\Theta}\ln\rhob(0)\rangle_\text{b}\frac{L^*}{I^*}
        + \frac{L^*}{I^*}\p_{\Theta} \left\langle \bF_0\right\rangle_\text{b}\,. 
    \end{split}
\end{align}
Using the fact that $\frac{L^*(\tau)}{I^*}=\epsilon^{-1}\dot\Theta(\epsilon^{-2}\tau) = \epsilon\dot\Theta^*(\tau)$, the last term is rewritten as $\epsilon\frac{\dd}{\dd \tau} \langle \bF_0\rangle_\text{b}(\Theta^*(\tau))\rvert_{\tau=0}$. Here, $\Theta^{*}(t^*) \equiv \Theta(t^*/\epsilon^{2})$ is the orientation of the object measured in the rescaled time. When inserted in the integral in Eq.~\eqref{eq:integral_P}, this term yields
\begin{equation}\label{eq:integration}
    \begin{split}
        &\epsilon^{-1}\int_0^{t^*}\dd\tau\, \mU[\mLB^\dagger + \mLo^{*\dagger}](\epsilon^{-2}t^*,\epsilon^{-2}\tau) \frac{\dd}{\dd\tau}\langle \bF_0\rangle_\text{b}\\ &= \epsilon^{-1}\left[\langle\bF_0\rangle_\text{b}(\Theta^*(t^*)) -\langle\bF_0\rangle_\text{b}(\Theta^*(0)) \right]\,,
    \end{split}
\end{equation}
where we used the fact that rescaled variables, like $\bP^*(t^*)$, are evolved from time $0$ to time $t^*$ by the operator $\mU[\mLB^\dagger + \mLo^{*\dagger}](\epsilon^{-2} t^*,0)$ 
\footnote{This can be seen by taking the time derivative of $\bP^*(t^*)$ and using the equation of motion for $\bP(\epsilon^{-2}t)$, which gives $$\frac{\dd \bP^*(t^*)}{\dd t^*} = \epsilon^{-2}(\mLB^\dagger(\epsilon^{-2}t^*) + \mLo^\dagger)\bP^{*}(t^*)\,.$$}. The second term in Eq.~\eqref{eq:integration} cancels out the constant force appearing on the right-hand side of Eq.~\eqref{eq:Pdot_effective_Langevin_F+_as_F0}. The effective equation for $\bP^*$ then becomes 
\begin{equation}\label{eq:app_Pdot_eff_Langevin_intermediate}
    \begin{split}
        &\dot\bP^*(t^*) = \epsilon^{-1}\langle \bF_0 \rangle_\text{b}(\Theta^*(t^*)) + \epsilon^{-1}\delta\bF_0(\epsilon^{-2}t^*) \\
        &- \epsilon^{-2}\int_0^{t^*}\dd\tau\,  \langle\delta\bF_0(\epsilon^{-2} \tau)\otimes\bnabla_{\bR} \ln\rhob(0)\rangle_\text{b}\cdot \frac{\bP^*(t^*-\tau)}{M^*}\\
        &-\epsilon^{-2}\int_0^{t^*}\dd\tau\, \langle\delta \bF_0(\epsilon^{-2}\tau) \p_{\Theta^*}\ln\rhob(0)\rangle_\text{b}\frac{L^*(t^*-\tau)}{I^*}\, .
    \end{split}
\end{equation}
In the adiabatic limit the first memory term becomes
\begin{equation}\label{eq:app_zeta_PP}
    \begin{split}
        &\lim_{\epsilon\to0} \epsilon^{-2}\int_0^{t^*}\dd\tau\,  \langle\delta\bF_0(\epsilon^{-2} \tau)\otimes\bnabla_{\bR} \ln\rhob(0)\rangle_\text{b}\cdot \frac{\bP^*(t^*-\tau)}{M^*} \\
        &= \lim_{\epsilon\to0} \int_0^{t^*/\epsilon^2}\dd\tau\,  \langle\delta\bF_0( \tau)\otimes\bnabla_{\bR} \ln\rhob(0)\rangle_\text{b}\cdot \frac{\bP^*(t^*-\epsilon^2\tau)}{M^*} \\
        &= \left[\int_0^{+\infty}\dd  \tau  \langle\delta\bF_0( \tau)\otimes\bnabla_{\bR} \ln\rhob(0)\rangle_\text{b} \right]\cdot \frac{\bP^*(t^*)}{M^*} \\
        &\equiv \bzeta_{\bP\bP}(\Theta^*(t^*))\cdot \frac{\bP^*(t^*)}{M^*}\,.
    \end{split}
\end{equation}
The last line defines the linear momentum friction matrix $\bzeta_{\bP\bP}(\Theta^*(t^*))$. In the adiabatic limit it thus satisfies an Agarwal-like formula~\cite{agarwal1972}
\begin{equation}
    \bzeta_{\bP\bP}(\Theta^*(t^*)) = \int_0^{+\infty} \dd\tau\, \langle \delta\bF_0( \tau)\otimes\bnabla_{\bR} \ln\rhob(0)\rangle_\text{b}\,,
\end{equation}
computed by letting the bath evolve while the object is held in fixed position $\bR^*(t^*)$ and orientation $\Theta^*(t^*)$. An analogous manipulation shows that
\begin{equation}\label{eq:app_zetaPL}
    \begin{split}       &\lim_{\epsilon\to0}\epsilon^{-2}\int_0^{t^*}\dd\tau\, \langle\delta \bF_0(\epsilon^{-2}\tau) \p_{\Theta^*}\ln\rhob(0)\rangle_\text{b}\frac{L^*(t^*-\tau)}{I^*} \\
    &= \left[ \int_0^{+\infty}\dd\tau\, \langle\delta \bF_0(\tau) \p_{\Theta^*}\ln\rhob(0)\rangle_\text{b}\right]\frac{L^*(t^*)}{I^*}\\
    &\equiv \bzeta_{\bP L}\frac{L^*(t^*)}{I^*}
    \end{split}
\end{equation}
where the friction coefficients $\bzeta_{\bP L}$ couple the linear momentum with the angular one. Plugging Eq.~\eqref{eq:app_zeta_PP} and Eq.~\eqref{eq:app_zetaPL} into Eq.~\eqref{eq:app_Pdot_eff_Langevin_intermediate}, and introducing the rescaled noise $\bxi_\bP^*(t^*) \equiv \epsilon^{-1}\delta\bF_0(\epsilon^{-2}t^*)$, we obtain Eq.~\eqref{eq:dynmomrescaled} of the main text. 

A similar analysis can be carried out to study the memory term in the equation of the angular momentum. Through a similar analysis, we can show that the projected random torque reads
\begin{equation}
    \begin{split}              &\mP\mLo^{*\dagger}\Gamma^+(\epsilon^{-2}\tau) = \\&-\langle\delta\Gamma_0(\epsilon^{-2}\tau)(\bnabla_{\bR}\ln\rhob(0))\rangle_\text{b}\cdot \frac{\bP^*}{M^*} \\
    &-\langle\delta\Gamma_0(\epsilon^{-2}\tau)(\p_{\Theta}\ln\rhob(0))\rangle_\text{b}\frac{L^*}{I^*}\,.
    \end{split}
\end{equation}
Note that, in this case, there is no extra angular ratchet term, since $\p_{\Theta^*}\langle\Gamma_0\rangle_\text{b}=0$. The memory terms in Eq.~\eqref{eq:Pdot_effective_Langevin_F+_as_F0} can be studied as in the derivation for the linear momentum, leading to the following dynamics for $L^*$:
\begin{equation}\label{eq:app_dynL}
    \dot L^* = \langle \Gamma_0 \rangle_\text{b} - \bzeta_{L\bP}\cdot\frac{\bP^*}{M^*} - \zeta_{LL}\frac{L^*}{I^*} + \epsilon^{-1}\delta\Gamma_0(\epsilon^{-2}t^*)
\end{equation}
where the friction coefficients $\bzeta_{L\bP}$ and $\zeta_{LL}$ respectively read
\begin{equation}
    \begin{split}
        \zeta_{LL} &\equiv \int_0^{+\infty}\dd \tau \langle \delta\Gamma_0(\tau)\p_{\Theta^*}\ln\rhob(0)\rangle_\text{b}\,, \\
        \bzeta_{L\bP}^\mathrm{T} &\equiv \int_0^{+\infty}\dd \tau\langle \delta\Gamma_0(\tau)\bnabla_{\bR} \ln\rhob(0)\rangle_\text{b}\,.
    \end{split}
\end{equation}
Upon introducing the rescaled noise $\xi_L^*(t^*)\equiv \epsilon^{-1}\delta\Gamma_0(\epsilon^{-2}t^*)$ into Eq.~\eqref{eq:app_dynL}, we obtain Eq.~\eqref{eq:dynandmomrescaled} of the main text.

\section{Derivation of Eq.~\eqref{eq:Novikov_to_result}}\label{app:novikov}

In this Appendix we derive Eq.~\eqref{eq:Novikov_to_result} starting from  Eq.~\eqref{eq:novikov}. We have
\begin{align}\label{eq:lim_xi_delta}
    \begin{split}
        &\lim_{\epsilon\to0} \langle \bxi_\epsilon(t)\delta(\bnu - \bnu(t))\rangle = -\lim_{\epsilon\to 0}\int_0^t \dd s \langle\bxi_\epsilon(t) \otimes \bxi_\epsilon(s)\rangle\\
        &\cdot \bnabla_{\bW}\left\langle\frac{\delta\bW(t)}{\delta\bxi_\epsilon(s)}\delta(\bnu - \bnu(t))\right\rangle\,\\
        &= - \lim_{\epsilon\to0}\int_0^t \dd s\blambda \delta_+(t-s)\cdot \bnabla_{\bW}\left\langle\frac{\delta\bW(t)}{\delta\bxi_\epsilon(s)}\delta(\bnu - \bnu(t))\right\rangle\\
        &= -\lim_{\epsilon\to0}\lim_{s\to t_-}\blambda\cdot\bnabla_\bW\left\langle\frac{\delta\bW(t)}{\delta\bxi_\epsilon(s)}\delta(\bnu - \bnu(t))\right\rangle\,.
    \end{split}
\end{align}
In the second equality, we used the fact that, in the adiabatic limit, the correlations of the noise $\bxi_\epsilon$ recover the correlations of $\bxi$, given in Eq.~\eqref{eq:noise}. The functional derivative $\frac{\delta\bW(t)}{\delta\bxi_\epsilon(s)}$ can be computed expressing $\bW(t)$ as $\bW(t)=\int_0^t\dd\tau\, \dot\bW(\tau) + \bW(0)$ and using the effective Langevin equation~\eqref{eq:tracerdynamics-vectorial}. We obtain
\begin{equation}
    \begin{split}
    \frac{\delta\bW(t)}{\delta\bxi_\epsilon(s)} &= \int_0^t \dd \tau\left[ \frac{\delta\langle\bG\rangle_\text{b}}{\delta\bxi_\epsilon(s)} - \frac{\delta \tilde\bzeta \bW(\tau)}{\delta\bxi_\epsilon(s)}\right]
    +\int_0^t\dd\tau\, \frac{\delta\bxi_\epsilon(\tau)}{\delta \bxi_\epsilon(s)}\\
    &= \int_s^t \dd\tau\,\left[ \frac{\delta\langle\bG\rangle_\text{b}}{\delta\bxi_\epsilon(s)} - \frac{\delta \tilde\bzeta \bW(\tau)}{\delta\bxi_\epsilon(s)}\right] + \id\,,
    \end{split}
\end{equation}
where the second equality follows from causality. Taking the limit $s\to t_-$ on both sides, we get
\begin{equation}\label{eq:func_derivative}
    \lim_{s\to t_-} \frac{\delta\bW(t)}{\delta\bxi_\epsilon(s)} = \id\,.
\end{equation}
Substituting Eq.~\eqref{eq:func_derivative} into Eq.~\eqref{eq:lim_xi_delta}, we conclude that
\begin{equation}
    \lim_{\epsilon\to 0} \langle\bxi_\epsilon(t)\delta(\bnu - \bnu(t)) = -\blambda\cdot\bnabla_\bW\langle\delta(\bnu-\bnu(t))\rangle\,,
\end{equation}
which is Eq.~\eqref{eq:Novikov_to_result} of the main text.
\section{Derivation of Eq.~~\eqref{eq:T_PP}}\label{app:algebra_T_PP}

We start from the definition of $\bT_\epsilon^{\bP\bP}$ as the inverse of the noise-noise correlation matrix given by Eq.~~\eqref{eq:correlation}
\begin{equation}\label{eq:inverse_relation}
    \int_{-\infty}^{+\infty}\dd \tau\langle \bxi_\epsilon(t) \otimes \bxi_\epsilon(\tau)\rangle\bT_\epsilon^{\bP\bP}(\tau) = \id \delta(t)\,.
\end{equation}
It is convenient to rewrite its expression using the matrix $\bM \equiv \blambda_{\bP\bP}^{-1}$, so that the noise-noise correlations read
\begin{equation}
    \langle \bxi_\epsilon(t) \otimes \bxi_\epsilon(t')\rangle =\begin{cases} \frac{1}{\epsilon^2}\ee^{-\bM(t-t')/\epsilon^2} &\text{ if $t\geq t'$}\\
    \frac{1}{\epsilon^2}\ee^{-\bM^\mathrm{T}(t'-t)/\epsilon^2} &\text{ if $t< t'$}\,
    \end{cases}
\end{equation}
We then apply a Fourier transform $\mF[\bO](\omega) \equiv \int_{-\infty}^{+\infty}\dd t \bO(t) \ee^{-i\omega t}$ to both sides of Eq.~\eqref{eq:inverse_relation} and use the convolution theorem to obtain
\begin{equation}
    \begin{split}           \mF&[\bT_\epsilon^{\bP\bP}](\omega) = \left[ \left[ \bM + i\epsilon^2\omega\id  \right]^{-1} + \left[\bM^{T} - i\epsilon^2\omega\id \right]^{-1}\right]^{-1} \\
        &= [\bM^\mathrm{T} - i\epsilon^2 \omega\id](\bM + \bM^\mathrm{T})^{-1}[\bM + i\epsilon^2 \omega\id]\\
        &= [\id - i\epsilon^2 \omega(\bM^{T})^{-1}](\bM^{-1} + (\bM^T)^{-1})^{-1}\\
        &\cdot[\id + i\epsilon^2 \omega\bM^{-1}]\\
        &= \frac{1}{2}[\id - i\epsilon^{2}\omega \blambda^\mathrm{T}_{\bP\bP}](\blambda_{\bP\bP,S})^{-1}[\id + i\epsilon^2\omega \blambda_{\bP\bP}]\\
        &= \frac{1}{2}\bigl[ (\blambda_{\bP\bP,S})^{-1} \\
        &+ i\epsilon^2\omega((\blambda_{\bP\bP,S})^{-1}\blambda_{\bP\bP}- \blambda_{\bP\bP}^\mathrm{T}(\blambda_{\bP\bP,S})^{-1}  ) \\
        &+ \epsilon^4\omega^2 \blambda_{\bP\bP}^\mathrm{T}(\blambda_{\bP\bP,S})^{-1}\blambda_{\bP\bP}\bigr]\,,
    \end{split}
\end{equation}
where we made use of the matrix identity $[\bA + \bB]^{-1}= \bB^{-1}[\bB^{-1} + \bA^{-1}]^{-1}\bA^{-1}$, and we introduced the symmetrized matrix $\blambda_{\bP\bP,S} \equiv \frac{1}{2}(\blambda_{\bP\bP} + \blambda_{\bP\bP}^\mathrm{T})$. Taking an inverse Fourier transform we obtain Eq.~\eqref{eq:T_def} of the main text.

\section{Derivation of $\sigma_\text{s,disk}=0$}\label{app:sigma_disk}
In this Appendix we show that the symmetrized entropy production rate of the disk, $\sigma_{s,\text{disk}}$, is zero. Its expression is given by Eq.~\eqref{eq:entropy-symm-disk}, which we report again here
\begin{equation}
    \begin{split}
        &\sigma_{s,\text{disk}} =  -\lim_{t\to +\infty} \frac{1}{Mt}\int_0^t \dd\tau\, \Biggl\langle\bzeta_{\bP\bP,S}\bP \cdot \bT^{\bP\bP}_\epsilon\dot\bP \\ 
        &+  \dot\bP\cdot \bT^{\bP\bP}_\epsilon\bzeta_{\bP\bP,S}\bP 
        + \frac{1}{2M}( \bzeta_{\bP\bP}^\mathrm{T}\bP\cdot \bT^{\bP\bP}_\epsilon\bzeta_{\bP\bP}^\mathrm{T}\bP \\&- \bzeta_{\bP\bP}\bP\cdot \bT^{\bP\bP}_\epsilon\bzeta_{\bP\bP}\bP)\Biggr\rangle_\text{path}\,.
    \end{split}
\end{equation}
The first two terms under the integral contribute
\begin{equation}\label{eq:appscal1}
    \begin{split}
        &\int_0^t\dd\tau\, \big\langle\bzeta_{\bP\bP,S}\bP \cdot \bT^{\bP\bP}_\epsilon\dot\bP +  \dot\bP\cdot\bT^{\bP\bP}_\epsilon\bzeta_{\bP\bP,S}\bP\big\rangle_\text{path}\\
        &= \int_0^t\dd\tau\, \frac{\dd}{\dd\tau}\big\langle\bzeta_{\bP\bP,S}(\bP\cdot\bT^{(0)}\bP-\epsilon^4\dot\bP\cdot\bT^{(2)}\dot\bP(\tau))\big\rangle_\text{path}\,,
    \end{split}
\end{equation}
where we made use of the symmetry properties of $\bT^{\bP\bP}_\epsilon$, of integration by parts and we have eventually recognized a total time derivative\footnote{In Eqs.~\eqref{eq:appscal1} and~\eqref{eq:appscal2}, despite the parenthesis, the matrix-vector product has priority over the scalar product}. This integral thus contributes finite boundary terms that have a vanishing contribution in the entropy-production rate:
\begin{equation}\label{eq:appscal2}
    \lim_{t\to+\infty} \frac{1}{t}\Big[\frac{\dd}{\dd\tau}\big\langle\bzeta_{\bP\bP,S}(\bP\cdot \bT^{(0)}\bP-\epsilon^4\dot\bP\cdot \bT^{(2)}\dot\bP(\tau))\big\rangle_\text{path}\Bigl]_0^t=0\,.
\end{equation}

Let us now consider the second contribution, involving the term
\begin{equation}\label{eq:second_term}
    \int_0^t\dd\tau\,\left\langle\bzeta_{\bP\bP}^\mathrm{T}\bP\cdot\bT^{\bP\bP}_\epsilon\bzeta_{\bP\bP}^\mathrm{T}\bP - \bzeta_{\bP\bP}\bP\cdot\bT^{\bP\bP}_\epsilon\bzeta_{\bP\bP}\bP\right\rangle_\text{path}\,.
\end{equation}
From the definition of $\bT_\epsilon^{\bP\bP}$, we need to compute three different terms, respectively, proportional to $\bT^{(0)}$, $\bT^{(1)}$ and $\bT^{(2)}$. The contribution proportional to $\bT^{(0)}$ is
\begin{align}
        &\int_0^t\dd\tau\, \Biggl\langle\bzeta_{\bP\bP}^\mathrm{T}\bP\cdot\bT^{(0)}\bzeta_{\bP\bP}^\mathrm{T}\bP - \bzeta_{\bP\bP}\bP\cdot\bT^{(0)}\bzeta_{\bP\bP}\bP\Biggr\rangle_\text{path}\notag\\
        &= \int_0^t \dd\tau\, \Tr\Biggl[\biggl(\bzeta_{\bP\bP}\bT^{(0)}\bzeta_{\bP\bP}^\mathrm{T} \notag\\
        &\hspace{1.8cm}- \bzeta_{\bP\bP}^\mathrm{T}\bT^{(0)}\bzeta_{\bP\bP}\Biggr) 
        \langle\bP \otimes \bP\rangle_\text{path}\Biggr] = 0\,.
\end{align}
In the last equality we used the fact that, since $\bT^{(0)}$ is symmetric, the integrand is the trace of the product of an antisymmetric matrix and a symmetric matrix, which thus evaluates to $0$. A similar argument can be used for the term in Eq.~\eqref{eq:second_term} proportional to $\bT^{(2)}$, after an integration by parts. Finally, the contribution proportional to $\bT^{(1)}$ reads 
\begin{align}
    \begin{split}
        &\frac{\epsilon^2}{t}\int_0^t\dd\tau\, \langle\bzeta_{\bP\bP}^\mathrm{T}\bP\cdot\bT^{(1)}\bzeta_{\bP\bP}^\mathrm{T}\dot\bP - \bzeta_{\bP\bP}\bP\cdot\bT^{(1)}\bzeta_{\bP\bP}\dot\bP\rangle_\text{path}\\
        &= \frac{\epsilon^2}{t}\int_0^t \dd\tau\, \Tr\biggl[\big(\bzeta_{\bP\bP}\bT^{(1)}\bzeta_{\bP\bP}^\mathrm{T} \\
        &\hspace{2.1cm}- \bzeta_{\bP\bP}^\mathrm{T}\bT^{(1)}\bzeta_{\bP\bP}\big)
        \langle\dot\bP \otimes \bP\rangle_\text{path}\biggr]\\
        &= \frac{\epsilon^2}{t}\int_0^t \dd\tau\, \Tr\biggl[\bigg\{\bzeta_{\bP\bP}\bT^{(1)}\bzeta_{\bP\bP}^\mathrm{T} \\
        &\hspace{2cm}+ \left(\bzeta_{\bP\bP}\bT^{(1)}\bzeta_{\bP\bP}^\mathrm{T}\right)^{\mathrm{T}}\bigg\}
        \langle\dot\bP \otimes \bP\rangle_\text{path}\biggr]\\
        &= \frac{\epsilon^2}{t}\int_0^t \dd\tau\, \frac{\dd}{\dd\tau}\Tr\biggl[\bigg\{\bzeta_{\bP\bP}\bT^{(1)}\bzeta_{\bP\bP}^\mathrm{T} \\
        &\hspace{1.5cm} + \left(\bzeta_{\bP\bP}\bT^{(1)}\bzeta_{\bP\bP}^\mathrm{T}\right)^{\mathrm{T}}\bigg\}
        \langle\bP(\tau) \otimes \bP(\tau)\rangle_\text{path}\biggr]\\
    \end{split}
\end{align}
In the second equality, we used the fact that $\bT^{(1)}$, defined in Eq.~\eqref{eq:T_PP} is an antisymmetric matrix and thus $\big(\bzeta_{\bP\bP}\bT^{(1)}\bzeta_{\bP\bP}^\mathrm{T}\big)^\mathrm{T} = -\big(\bzeta_{\bP\bP}^\mathrm{T}\bT^{(1)}\bzeta_{\bP\bP}\big)$. Thus, only the symmetric part of $\langle\dot\bP \otimes \bP\rangle_\text{path}$ contributes to the trace, leading to the final equality. Finally, we recognized the time derivative of a bounded quantity which does not contribute to the entropy production rate since:
\begin{align}
    \begin{split}
        \lim_{t\to+\infty}\frac{1}{t}&\bigl[ \bigl\langle\bzeta_{\bP\bP}^\mathrm{T}\bP\cdot\bT^{(1)}\bzeta_{\bP\bP}^\mathrm{T}\bP \\
        &- \bzeta_{\bP\bP}\bP\cdot\bT^{(1)}\bzeta_{\bP\bP}\bP\bigr\rangle_\text{path}\bigr]_0^t = 0\,.
    \end{split}
\end{align}
We thus conclude that $\sigma_{s,\text{disk}}=0$, which is the result of Eq.~\eqref{eq:entropy-symm-P} in the main text.
\addtocontents{toc}{\string\tocdepth@munge}  

\newpage
\bibliography{ref}

\begin{thebibliography}{84}%
\makeatletter
\providecommand \@ifxundefined [1]{%
 \@ifx{#1\undefined}
}%
\providecommand \@ifnum [1]{%
 \ifnum #1\expandafter \@firstoftwo
 \else \expandafter \@secondoftwo
 \fi
}%
\providecommand \@ifx [1]{%
 \ifx #1\expandafter \@firstoftwo
 \else \expandafter \@secondoftwo
 \fi
}%
\providecommand \natexlab [1]{#1}%
\providecommand \enquote  [1]{``#1''}%
\providecommand \bibnamefont  [1]{#1}%
\providecommand \bibfnamefont [1]{#1}%
\providecommand \citenamefont [1]{#1}%
\providecommand \href@noop [0]{\@secondoftwo}%
\providecommand \href [0]{\begingroup \@sanitize@url \@href}%
\providecommand \@href[1]{\@@startlink{#1}\@@href}%
\providecommand \@@href[1]{\endgroup#1\@@endlink}%
\providecommand \@sanitize@url [0]{\catcode `\\12\catcode `\$12\catcode `\&12\catcode `\#12\catcode `\^12\catcode `\_12\catcode `\%12\relax}%
\providecommand \@@startlink[1]{}%
\providecommand \@@endlink[0]{}%
\providecommand \url  [0]{\begingroup\@sanitize@url \@url }%
\providecommand \@url [1]{\endgroup\@href {#1}{\urlprefix }}%
\providecommand \urlprefix  [0]{URL }%
\providecommand \Eprint [0]{\href }%
\providecommand \doibase [0]{https://doi.org/}%
\providecommand \selectlanguage [0]{\@gobble}%
\providecommand \bibinfo  [0]{\@secondoftwo}%
\providecommand \bibfield  [0]{\@secondoftwo}%
\providecommand \translation [1]{[#1]}%
\providecommand \BibitemOpen [0]{}%
\providecommand \bibitemStop [0]{}%
\providecommand \bibitemNoStop [0]{.\EOS\space}%
\providecommand \EOS [0]{\spacefactor3000\relax}%
\providecommand \BibitemShut  [1]{\csname bibitem#1\endcsname}%
\let\auto@bib@innerbib\@empty
\bibitem [{\citenamefont {Einstein}(1905)}]{Einstein1905}%
  \BibitemOpen
  \bibfield  {author} {\bibinfo {author} {\bibfnamefont {A.}~\bibnamefont {Einstein}},\ }\bibfield  {title} {\bibinfo {title} {Investigations on the {{Theory}} of the {{Brownian Movement}}},\ }\bibfield  {journal} {\bibinfo  {journal} {Physics Bulletin}\ }\textbf {\bibinfo {volume} {7}},\ \href {https://doi.org/10.1088/0031-9112/7/10/012} {10.1088/0031-9112/7/10/012} (\bibinfo {year} {1905})\BibitemShut {NoStop}%
\bibitem [{\citenamefont {Smoluchowski}(1906)}]{Smoluchowski1906}%
  \BibitemOpen
  \bibfield  {author} {\bibinfo {author} {\bibfnamefont {M.}~\bibnamefont {Smoluchowski}},\ }\bibfield  {title} {\bibinfo {title} {{On the Kinetic Theory of the Brownian Molecular Motion and of Suspensions}},\ }\href@noop {} {\bibfield  {journal} {\bibinfo  {journal} {Annals of physics}\ }\textbf {\bibinfo {volume} {326}},\ \bibinfo {pages} {756} (\bibinfo {year} {1906})}\BibitemShut {NoStop}%
\bibitem [{\citenamefont {Langevin}(1908)}]{Langevin1908}%
  \BibitemOpen
  \bibfield  {author} {\bibinfo {author} {\bibfnamefont {P.}~\bibnamefont {Langevin}},\ }\bibfield  {title} {\bibinfo {title} {{On the Theory of Brownian Motion}},\ }\href@noop {} {\bibfield  {journal} {\bibinfo  {journal} {C. R. Acad. Sci.}\ }\textbf {\bibinfo {volume} {146}} (\bibinfo {year} {1908})}\BibitemShut {NoStop}%
\bibitem [{\citenamefont {Nakajima}(1958)}]{nakajimaQuantumTheoryTransport1958}%
  \BibitemOpen
  \bibfield  {author} {\bibinfo {author} {\bibfnamefont {S.}~\bibnamefont {Nakajima}},\ }\bibfield  {title} {\bibinfo {title} {On {{Quantum Theory}} of {{Transport Phenomena}}},\ }\bibfield  {journal} {\bibinfo  {journal} {Progress of Theoretical Physics}\ }\textbf {\bibinfo {volume} {20}},\ \href {https://doi.org/10.1002/andp.19844960610} {10.1002/andp.19844960610} (\bibinfo {year} {1958})\BibitemShut {NoStop}%
\bibitem [{\citenamefont {Zwanzig}(1960)}]{zwanzigEnsembleMethodTheory1960}%
  \BibitemOpen
  \bibfield  {author} {\bibinfo {author} {\bibfnamefont {R.}~\bibnamefont {Zwanzig}},\ }\bibfield  {title} {\bibinfo {title} {Ensemble {{Method}} in the {{Theory}} of {{Irreversibility}}},\ }\href {https://doi.org/10.1063/1.1731409} {\bibfield  {journal} {\bibinfo  {journal} {The Journal of Chemical Physics}\ }\textbf {\bibinfo {volume} {33}},\ \bibinfo {pages} {1338} (\bibinfo {year} {1960})}\BibitemShut {NoStop}%
\bibitem [{\citenamefont {Mori}(1965)}]{moriTransportCollectiveMotion1965}%
  \BibitemOpen
  \bibfield  {author} {\bibinfo {author} {\bibfnamefont {H.}~\bibnamefont {Mori}},\ }\bibfield  {title} {\bibinfo {title} {Transport, {{Collective Motion}}, and {{Brownian Motion}}},\ }\href@noop {} {\bibfield  {journal} {\bibinfo  {journal} {Progress of Theoretical Physics}\ }\textbf {\bibinfo {volume} {33}},\ \bibinfo {pages} {423} (\bibinfo {year} {1965})}\BibitemShut {NoStop}%
\bibitem [{\citenamefont {Van~Kampen}\ and\ \citenamefont {Oppenheim}(1986)}]{van1986brownian}%
  \BibitemOpen
  \bibfield  {author} {\bibinfo {author} {\bibfnamefont {N.}~\bibnamefont {Van~Kampen}}\ and\ \bibinfo {author} {\bibfnamefont {I.}~\bibnamefont {Oppenheim}},\ }\bibfield  {title} {\bibinfo {title} {Brownian motion as a problem of eliminating fast variables},\ }\href@noop {} {\bibfield  {journal} {\bibinfo  {journal} {Physica A: Statistical Mechanics and its Applications}\ }\textbf {\bibinfo {volume} {138}},\ \bibinfo {pages} {231} (\bibinfo {year} {1986})}\BibitemShut {NoStop}%
\bibitem [{\citenamefont {Zwanzig}(2001)}]{zwanzigNonequilibriumStatisticalMechanics2001}%
  \BibitemOpen
  \bibfield  {author} {\bibinfo {author} {\bibfnamefont {R.}~\bibnamefont {Zwanzig}},\ }\href@noop {} {\emph {\bibinfo {title} {Nonequilibrium {{Statistical Mechanics}}}}}\ (\bibinfo  {publisher} {Oxford University Press},\ \bibinfo {address} {Oxford, New York},\ \bibinfo {year} {2001})\BibitemShut {NoStop}%
\bibitem [{\citenamefont {Van~Kampen}(1981)}]{VanKampen1981}%
  \BibitemOpen
  \bibfield  {author} {\bibinfo {author} {\bibfnamefont {N.~G.}\ \bibnamefont {Van~Kampen}},\ }\href {https://doi.org/10.1063/1.2915501} {\emph {\bibinfo {title} {Stochastic {{Processes}} in {{Physics}} and {{Chemistry}}}}},\ Vol.~\bibinfo {volume} {36}\ (\bibinfo {year} {1981})\BibitemShut {NoStop}%
\bibitem [{\citenamefont {Kubo}\ \emph {et~al.}(1991)\citenamefont {Kubo}, \citenamefont {Toda},\ and\ \citenamefont {Hashitsume}}]{kuboStatisticalPhysicsII1991}%
  \BibitemOpen
  \bibfield  {author} {\bibinfo {author} {\bibfnamefont {R.}~\bibnamefont {Kubo}}, \bibinfo {author} {\bibfnamefont {M.}~\bibnamefont {Toda}},\ and\ \bibinfo {author} {\bibfnamefont {N.}~\bibnamefont {Hashitsume}},\ }\href {https://doi.org/10.1007/978-3-642-58244-8} {\emph {\bibinfo {title} {Statistical {{Physics II}}}}},\ edited by\ \bibinfo {editor} {\bibfnamefont {M.}~\bibnamefont {Cardona}}, \bibinfo {editor} {\bibfnamefont {P.}~\bibnamefont {Fulde}}, \bibinfo {editor} {\bibfnamefont {K.}~\bibnamefont {Von~Klitzing}}, \bibinfo {editor} {\bibfnamefont {H.-J.}\ \bibnamefont {Queisser}},\ and\ \bibinfo {editor} {\bibfnamefont {H.~K.~V.}\ \bibnamefont {Lotsch}},\ \bibinfo {series} {Springer {{Series}} in {{Solid-State Sciences}}}, Vol.~\bibinfo {volume} {31}\ (\bibinfo  {publisher} {Springer Berlin Heidelberg},\ \bibinfo {address} {Berlin, Heidelberg},\ \bibinfo {year} {1991})\BibitemShut {NoStop}%
\bibitem [{\citenamefont {Agarwal}(1972)}]{agarwal1972}%
  \BibitemOpen
  \bibfield  {author} {\bibinfo {author} {\bibfnamefont {G.~S.}\ \bibnamefont {Agarwal}},\ }\bibfield  {title} {\bibinfo {title} {Fluctuation-dissipation theorems for systems in non-thermal equilibrium and applications},\ }\href {https://doi.org/10.1007/BF01391621} {\bibfield  {journal} {\bibinfo  {journal} {Zeitschrift f{\"u}r Physik A Hadrons and nuclei}\ }\textbf {\bibinfo {volume} {252}},\ \bibinfo {pages} {25} (\bibinfo {year} {1972})}\BibitemShut {NoStop}%
\bibitem [{\citenamefont {Speck}\ and\ \citenamefont {Seifert}(2006)}]{Speck2006}%
  \BibitemOpen
  \bibfield  {author} {\bibinfo {author} {\bibfnamefont {T.}~\bibnamefont {Speck}}\ and\ \bibinfo {author} {\bibfnamefont {U.}~\bibnamefont {Seifert}},\ }\bibfield  {title} {\bibinfo {title} {Restoring a fluctuation-dissipation theorem in a nonequilibrium steady state},\ }\href {https://doi.org/10.1209/epl/i2005-10549-4} {\bibfield  {journal} {\bibinfo  {journal} {Europhysics Letters}\ }\textbf {\bibinfo {volume} {74}},\ \bibinfo {pages} {391} (\bibinfo {year} {2006})}\BibitemShut {NoStop}%
\bibitem [{\citenamefont {Seifert}\ and\ \citenamefont {Speck}(2010)}]{Seifert2010}%
  \BibitemOpen
  \bibfield  {author} {\bibinfo {author} {\bibfnamefont {U.}~\bibnamefont {Seifert}}\ and\ \bibinfo {author} {\bibfnamefont {T.}~\bibnamefont {Speck}},\ }\bibfield  {title} {\bibinfo {title} {Fluctuation-dissipation theorem in nonequilibrium steady states},\ }\href@noop {} {\bibfield  {journal} {\bibinfo  {journal} {Europhysics Letters}\ }\textbf {\bibinfo {volume} {89}} (\bibinfo {year} {2010})}\BibitemShut {NoStop}%
\bibitem [{\citenamefont {Baiesi}\ \emph {et~al.}(2009)\citenamefont {Baiesi}, \citenamefont {Maes},\ and\ \citenamefont {Wynants}}]{Baiesi2009}%
  \BibitemOpen
  \bibfield  {author} {\bibinfo {author} {\bibfnamefont {M.}~\bibnamefont {Baiesi}}, \bibinfo {author} {\bibfnamefont {C.}~\bibnamefont {Maes}},\ and\ \bibinfo {author} {\bibfnamefont {B.}~\bibnamefont {Wynants}},\ }\bibfield  {title} {\bibinfo {title} {Fluctuations and response of nonequilibrium states},\ }\href {https://doi.org/10.1103/PhysRevLett.103.010602} {\bibfield  {journal} {\bibinfo  {journal} {Physical Review Letters}\ }\textbf {\bibinfo {volume} {103}},\ \bibinfo {pages} {010602} (\bibinfo {year} {2009})},\ \Eprint {https://arxiv.org/abs/0902.3955} {arXiv:0902.3955} \BibitemShut {NoStop}%
\bibitem [{\citenamefont {Baiesi}\ and\ \citenamefont {Maes}(2013)}]{Baiesi2013}%
  \BibitemOpen
  \bibfield  {author} {\bibinfo {author} {\bibfnamefont {M.}~\bibnamefont {Baiesi}}\ and\ \bibinfo {author} {\bibfnamefont {C.}~\bibnamefont {Maes}},\ }\bibfield  {title} {\bibinfo {title} {An update on the nonequilibrium linear response},\ }\bibfield  {journal} {\bibinfo  {journal} {New Journal of Physics}\ }\textbf {\bibinfo {volume} {15}},\ \href {https://doi.org/10.1088/1367-2630/15/1/013004} {10.1088/1367-2630/15/1/013004} (\bibinfo {year} {2013}),\ \Eprint {https://arxiv.org/abs/1205.4157} {arXiv:1205.4157} \BibitemShut {NoStop}%
\bibitem [{\citenamefont {Bertini}\ \emph {et~al.}(2015)\citenamefont {Bertini}, \citenamefont {De~Sole}, \citenamefont {Gabrielli}, \citenamefont {{Jona-Lasinio}},\ and\ \citenamefont {Landim}}]{bertiniMacroscopicFluctuationTheory2015}%
  \BibitemOpen
  \bibfield  {author} {\bibinfo {author} {\bibfnamefont {L.}~\bibnamefont {Bertini}}, \bibinfo {author} {\bibfnamefont {A.}~\bibnamefont {De~Sole}}, \bibinfo {author} {\bibfnamefont {D.}~\bibnamefont {Gabrielli}}, \bibinfo {author} {\bibfnamefont {G.}~\bibnamefont {{Jona-Lasinio}}},\ and\ \bibinfo {author} {\bibfnamefont {C.}~\bibnamefont {Landim}},\ }\bibfield  {title} {\bibinfo {title} {Macroscopic fluctuation theory},\ }\href {https://doi.org/10.1103/RevModPhys.87.593} {\bibfield  {journal} {\bibinfo  {journal} {Reviews of Modern Physics}\ }\textbf {\bibinfo {volume} {87}},\ \bibinfo {pages} {593} (\bibinfo {year} {2015})}\BibitemShut {NoStop}%
\bibitem [{\citenamefont {Mason}\ and\ \citenamefont {Weitz}(1995)}]{MasonWeitz1995}%
  \BibitemOpen
  \bibfield  {author} {\bibinfo {author} {\bibfnamefont {T.~G.}\ \bibnamefont {Mason}}\ and\ \bibinfo {author} {\bibfnamefont {D.~A.}\ \bibnamefont {Weitz}},\ }\bibfield  {title} {\bibinfo {title} {Optical {{Measurements}} of {{Frequency-Dependent Linear Viscoelastic Moduli}} of {{Complex Fluids}}},\ }\href {https://doi.org/10.1103/PhysRevLett.74.1250} {\bibfield  {journal} {\bibinfo  {journal} {Physical Review Letters}\ }\textbf {\bibinfo {volume} {74}},\ \bibinfo {pages} {1250} (\bibinfo {year} {1995})}\BibitemShut {NoStop}%
\bibitem [{\citenamefont {Crocker}\ \emph {et~al.}(2000)\citenamefont {Crocker}, \citenamefont {Valentine}, \citenamefont {Weeks}, \citenamefont {Gisler}, \citenamefont {Kaplan}, \citenamefont {Yodh},\ and\ \citenamefont {Weitz}}]{CrockerWeitz2000}%
  \BibitemOpen
  \bibfield  {author} {\bibinfo {author} {\bibfnamefont {J.~C.}\ \bibnamefont {Crocker}}, \bibinfo {author} {\bibfnamefont {M.~T.}\ \bibnamefont {Valentine}}, \bibinfo {author} {\bibfnamefont {E.~R.}\ \bibnamefont {Weeks}}, \bibinfo {author} {\bibfnamefont {T.}~\bibnamefont {Gisler}}, \bibinfo {author} {\bibfnamefont {P.~D.}\ \bibnamefont {Kaplan}}, \bibinfo {author} {\bibfnamefont {A.~G.}\ \bibnamefont {Yodh}},\ and\ \bibinfo {author} {\bibfnamefont {D.~A.}\ \bibnamefont {Weitz}},\ }\bibfield  {title} {\bibinfo {title} {Two-{{Point Microrheology}} of {{Inhomogeneous Soft Materials}}},\ }\href {https://doi.org/10.1103/PhysRevLett.85.888} {\bibfield  {journal} {\bibinfo  {journal} {Physical Review Letters}\ }\textbf {\bibinfo {volume} {85}},\ \bibinfo {pages} {888} (\bibinfo {year} {2000})}\BibitemShut {NoStop}%
\bibitem [{\citenamefont {Wilhelm}(2008)}]{wilhelmOutofEquilibriumMicrorheologyLiving2008}%
  \BibitemOpen
  \bibfield  {author} {\bibinfo {author} {\bibfnamefont {C.}~\bibnamefont {Wilhelm}},\ }\bibfield  {title} {\bibinfo {title} {Out-of-{{Equilibrium Microrheology}} inside {{Living Cells}}},\ }\href {https://doi.org/10.1103/PhysRevLett.101.028101} {\bibfield  {journal} {\bibinfo  {journal} {Physical Review Letters}\ }\textbf {\bibinfo {volume} {101}},\ \bibinfo {pages} {028101} (\bibinfo {year} {2008})}\BibitemShut {NoStop}%
\bibitem [{\citenamefont {Robert}\ \emph {et~al.}(2010)\citenamefont {Robert}, \citenamefont {Nguyen}, \citenamefont {Gallet},\ and\ \citenamefont {Wilhelm}}]{robertVivoDeterminationFluctuating2010}%
  \BibitemOpen
  \bibfield  {author} {\bibinfo {author} {\bibfnamefont {D.}~\bibnamefont {Robert}}, \bibinfo {author} {\bibfnamefont {T.-H.}\ \bibnamefont {Nguyen}}, \bibinfo {author} {\bibfnamefont {F.}~\bibnamefont {Gallet}},\ and\ \bibinfo {author} {\bibfnamefont {C.}~\bibnamefont {Wilhelm}},\ }\bibfield  {title} {\bibinfo {title} {In {{Vivo Determination}} of {{Fluctuating Forces}} during {{Endosome Trafficking Using}} a {{Combination}} of {{Active}} and {{Passive Microrheology}}},\ }\href {https://doi.org/10.1371/journal.pone.0010046} {\bibfield  {journal} {\bibinfo  {journal} {PLOS ONE}\ }\textbf {\bibinfo {volume} {5}},\ \bibinfo {pages} {e10046} (\bibinfo {year} {2010})}\BibitemShut {NoStop}%
\bibitem [{\citenamefont {Puertas}\ and\ \citenamefont {Voigtmann}(2014)}]{puertasMicrorheologyColloidalSystems2014}%
  \BibitemOpen
  \bibfield  {author} {\bibinfo {author} {\bibfnamefont {A.~M.}\ \bibnamefont {Puertas}}\ and\ \bibinfo {author} {\bibfnamefont {T.}~\bibnamefont {Voigtmann}},\ }\bibfield  {title} {\bibinfo {title} {Microrheology of colloidal systems},\ }\href {https://doi.org/10.1088/0953-8984/26/24/243101} {\bibfield  {journal} {\bibinfo  {journal} {Journal of Physics: Condensed Matter}\ }\textbf {\bibinfo {volume} {26}},\ \bibinfo {pages} {243101} (\bibinfo {year} {2014})}\BibitemShut {NoStop}%
\bibitem [{\citenamefont {Reichhardt}\ and\ \citenamefont {Reichhardt}(2015)}]{reichhardtActiveMicrorheologyActive2015}%
  \BibitemOpen
  \bibfield  {author} {\bibinfo {author} {\bibfnamefont {C.}~\bibnamefont {Reichhardt}}\ and\ \bibinfo {author} {\bibfnamefont {C.~J.~O.}\ \bibnamefont {Reichhardt}},\ }\bibfield  {title} {\bibinfo {title} {Active microrheology in active matter systems: {{Mobility}}, intermittency, and avalanches},\ }\href {https://doi.org/10.1103/PhysRevE.91.032313} {\bibfield  {journal} {\bibinfo  {journal} {Physical Review E}\ }\textbf {\bibinfo {volume} {91}},\ \bibinfo {pages} {032313} (\bibinfo {year} {2015})}\BibitemShut {NoStop}%
\bibitem [{\citenamefont {Wu}\ and\ \citenamefont {Libchaber}(2000)}]{wu}%
  \BibitemOpen
  \bibfield  {author} {\bibinfo {author} {\bibfnamefont {X.-L.}\ \bibnamefont {Wu}}\ and\ \bibinfo {author} {\bibfnamefont {A.}~\bibnamefont {Libchaber}},\ }\bibfield  {title} {\bibinfo {title} {Particle diffusion in a quasi-two-dimensional bacterial bath},\ }\href {https://doi.org/10.1103/PhysRevLett.84.3017} {\bibfield  {journal} {\bibinfo  {journal} {Phys. Rev. Lett.}\ }\textbf {\bibinfo {volume} {84}},\ \bibinfo {pages} {3017} (\bibinfo {year} {2000})}\BibitemShut {NoStop}%
\bibitem [{\citenamefont {Di~Leonardo}\ \emph {et~al.}(2010)\citenamefont {Di~Leonardo}, \citenamefont {Angelani}, \citenamefont {Dell’Arciprete}, \citenamefont {Ruocco}, \citenamefont {Iebba}, \citenamefont {Schippa}, \citenamefont {Conte}, \citenamefont {Mecarini}, \citenamefont {De~Angelis},\ and\ \citenamefont {Di~Fabrizio}}]{di2010bacterial}%
  \BibitemOpen
  \bibfield  {author} {\bibinfo {author} {\bibfnamefont {R.}~\bibnamefont {Di~Leonardo}}, \bibinfo {author} {\bibfnamefont {L.}~\bibnamefont {Angelani}}, \bibinfo {author} {\bibfnamefont {D.}~\bibnamefont {Dell’Arciprete}}, \bibinfo {author} {\bibfnamefont {G.}~\bibnamefont {Ruocco}}, \bibinfo {author} {\bibfnamefont {V.}~\bibnamefont {Iebba}}, \bibinfo {author} {\bibfnamefont {S.}~\bibnamefont {Schippa}}, \bibinfo {author} {\bibfnamefont {M.~P.}\ \bibnamefont {Conte}}, \bibinfo {author} {\bibfnamefont {F.}~\bibnamefont {Mecarini}}, \bibinfo {author} {\bibfnamefont {F.}~\bibnamefont {De~Angelis}},\ and\ \bibinfo {author} {\bibfnamefont {E.}~\bibnamefont {Di~Fabrizio}},\ }\bibfield  {title} {\bibinfo {title} {Bacterial ratchet motors},\ }\href@noop {} {\bibfield  {journal} {\bibinfo  {journal} {Proceedings of the National Academy of Sciences}\ }\textbf {\bibinfo {volume} {107}},\ \bibinfo {pages} {9541} (\bibinfo {year} {2010})}\BibitemShut {NoStop}%
\bibitem [{\citenamefont {Sokolov}\ \emph {et~al.}(2010)\citenamefont {Sokolov}, \citenamefont {Apodaca}, \citenamefont {Grzybowski},\ and\ \citenamefont {Aranson}}]{sokolov2010swimming}%
  \BibitemOpen
  \bibfield  {author} {\bibinfo {author} {\bibfnamefont {A.}~\bibnamefont {Sokolov}}, \bibinfo {author} {\bibfnamefont {M.~M.}\ \bibnamefont {Apodaca}}, \bibinfo {author} {\bibfnamefont {B.~A.}\ \bibnamefont {Grzybowski}},\ and\ \bibinfo {author} {\bibfnamefont {I.~S.}\ \bibnamefont {Aranson}},\ }\bibfield  {title} {\bibinfo {title} {Swimming bacteria power microscopic gears},\ }\href@noop {} {\bibfield  {journal} {\bibinfo  {journal} {Proceedings of the National Academy of Sciences}\ }\textbf {\bibinfo {volume} {107}},\ \bibinfo {pages} {969} (\bibinfo {year} {2010})}\BibitemShut {NoStop}%
\bibitem [{\citenamefont {Sokolov}\ and\ \citenamefont {Aranson}(2012)}]{Sokolov2012}%
  \BibitemOpen
  \bibfield  {author} {\bibinfo {author} {\bibfnamefont {A.}~\bibnamefont {Sokolov}}\ and\ \bibinfo {author} {\bibfnamefont {I.~S.}\ \bibnamefont {Aranson}},\ }\bibfield  {title} {\bibinfo {title} {Physical properties of collective motion in suspensions of bacteria},\ }\href {https://doi.org/10.1103/PhysRevLett.109.248109} {\bibfield  {journal} {\bibinfo  {journal} {Physical Review Letters}\ }\textbf {\bibinfo {volume} {109}},\ \bibinfo {pages} {1} (\bibinfo {year} {2012})}\BibitemShut {NoStop}%
\bibitem [{\citenamefont {Bechinger}\ \emph {et~al.}(2016)\citenamefont {Bechinger}, \citenamefont {Di~Leonardo}, \citenamefont {L{\"o}wen}, \citenamefont {Reichhardt}, \citenamefont {Volpe},\ and\ \citenamefont {Volpe}}]{bechingerActiveParticlesComplex2016}%
  \BibitemOpen
  \bibfield  {author} {\bibinfo {author} {\bibfnamefont {C.}~\bibnamefont {Bechinger}}, \bibinfo {author} {\bibfnamefont {R.}~\bibnamefont {Di~Leonardo}}, \bibinfo {author} {\bibfnamefont {H.}~\bibnamefont {L{\"o}wen}}, \bibinfo {author} {\bibfnamefont {C.}~\bibnamefont {Reichhardt}}, \bibinfo {author} {\bibfnamefont {G.}~\bibnamefont {Volpe}},\ and\ \bibinfo {author} {\bibfnamefont {G.}~\bibnamefont {Volpe}},\ }\bibfield  {title} {\bibinfo {title} {Active {{Particles}} in {{Complex}} and {{Crowded Environments}}},\ }\href {https://doi.org/10.1103/RevModPhys.88.045006} {\bibfield  {journal} {\bibinfo  {journal} {Reviews of Modern Physics}\ }\textbf {\bibinfo {volume} {88}},\ \bibinfo {pages} {045006} (\bibinfo {year} {2016})}\BibitemShut {NoStop}%
\bibitem [{\citenamefont {Anand}\ \emph {et~al.}(2024)\citenamefont {Anand}, \citenamefont {Ma}, \citenamefont {Guo}, \citenamefont {Martiniani},\ and\ \citenamefont {Cheng}}]{anandTransportEnergeticsBacterial2024a}%
  \BibitemOpen
  \bibfield  {author} {\bibinfo {author} {\bibfnamefont {S.}~\bibnamefont {Anand}}, \bibinfo {author} {\bibfnamefont {X.}~\bibnamefont {Ma}}, \bibinfo {author} {\bibfnamefont {S.}~\bibnamefont {Guo}}, \bibinfo {author} {\bibfnamefont {S.}~\bibnamefont {Martiniani}},\ and\ \bibinfo {author} {\bibfnamefont {X.}~\bibnamefont {Cheng}},\ }\href {https://doi.org/10.48550/arXiv.2308.08421} {\bibinfo {title} {Transport and {{Energetics}} of {{Bacterial Rectification}}}} (\bibinfo {year} {2024}),\ \Eprint {https://arxiv.org/abs/2308.08421} {arXiv:2308.08421 [cond-mat]} \BibitemShut {NoStop}%
\bibitem [{\citenamefont {Diluzio}\ \emph {et~al.}(2005)\citenamefont {Diluzio}, \citenamefont {Turner}, \citenamefont {Mayer}, \citenamefont {Garstecki}, \citenamefont {Weibel}, \citenamefont {Berg},\ and\ \citenamefont {Whitesides}}]{Diluzio2005}%
  \BibitemOpen
  \bibfield  {author} {\bibinfo {author} {\bibfnamefont {W.~R.}\ \bibnamefont {Diluzio}}, \bibinfo {author} {\bibfnamefont {L.}~\bibnamefont {Turner}}, \bibinfo {author} {\bibfnamefont {M.}~\bibnamefont {Mayer}}, \bibinfo {author} {\bibfnamefont {P.}~\bibnamefont {Garstecki}}, \bibinfo {author} {\bibfnamefont {D.~B.}\ \bibnamefont {Weibel}}, \bibinfo {author} {\bibfnamefont {H.~C.}\ \bibnamefont {Berg}},\ and\ \bibinfo {author} {\bibfnamefont {G.~M.}\ \bibnamefont {Whitesides}},\ }\bibfield  {title} {\bibinfo {title} {{Escherichia coli swim on the right-hand side}},\ }\href@noop {} {\bibfield  {journal} {\bibinfo  {journal} {Nature}\ }\textbf {\bibinfo {volume} {435}},\ \bibinfo {pages} {1271} (\bibinfo {year} {2005})}\BibitemShut {NoStop}%
\bibitem [{\citenamefont {Riedel}\ \emph {et~al.}(2005)\citenamefont {Riedel}, \citenamefont {Kruse},\ and\ \citenamefont {Howard}}]{Riedel2005}%
  \BibitemOpen
  \bibfield  {author} {\bibinfo {author} {\bibfnamefont {I.~H.}\ \bibnamefont {Riedel}}, \bibinfo {author} {\bibfnamefont {K.}~\bibnamefont {Kruse}},\ and\ \bibinfo {author} {\bibfnamefont {J.}~\bibnamefont {Howard}},\ }\bibfield  {title} {\bibinfo {title} {{Biophysics: A self-organized vortex array of hydrodynamically entrained sperm cells}},\ }\href@noop {} {\bibfield  {journal} {\bibinfo  {journal} {Science}\ }\textbf {\bibinfo {volume} {309}},\ \bibinfo {pages} {300} (\bibinfo {year} {2005})}\BibitemShut {NoStop}%
\bibitem [{\citenamefont {Drescher}\ \emph {et~al.}(2009)\citenamefont {Drescher}, \citenamefont {Leptos}, \citenamefont {Tuval}, \citenamefont {Ishikawa}, \citenamefont {Pedley},\ and\ \citenamefont {Goldstein}}]{Drescher2009}%
  \BibitemOpen
  \bibfield  {author} {\bibinfo {author} {\bibfnamefont {K.}~\bibnamefont {Drescher}}, \bibinfo {author} {\bibfnamefont {K.~C.}\ \bibnamefont {Leptos}}, \bibinfo {author} {\bibfnamefont {I.}~\bibnamefont {Tuval}}, \bibinfo {author} {\bibfnamefont {T.}~\bibnamefont {Ishikawa}}, \bibinfo {author} {\bibfnamefont {T.~J.}\ \bibnamefont {Pedley}},\ and\ \bibinfo {author} {\bibfnamefont {R.~E.}\ \bibnamefont {Goldstein}},\ }\bibfield  {title} {\bibinfo {title} {{Dancing Volvox : Hydrodynamic Bound States of Swimming Algae}},\ }\href@noop {} {\bibfield  {journal} {\bibinfo  {journal} {Physical review letters}\ }\textbf {\bibinfo {volume} {102}},\ \bibinfo {pages} {168101} (\bibinfo {year} {2009})}\BibitemShut {NoStop}%
\bibitem [{\citenamefont {Beppu}\ \emph {et~al.}(2021)\citenamefont {Beppu}, \citenamefont {Izri}, \citenamefont {Sato}, \citenamefont {Yamanishi}, \citenamefont {Sumino},\ and\ \citenamefont {Maeda}}]{beppu2021edge}%
  \BibitemOpen
  \bibfield  {author} {\bibinfo {author} {\bibfnamefont {K.}~\bibnamefont {Beppu}}, \bibinfo {author} {\bibfnamefont {Z.}~\bibnamefont {Izri}}, \bibinfo {author} {\bibfnamefont {T.}~\bibnamefont {Sato}}, \bibinfo {author} {\bibfnamefont {Y.}~\bibnamefont {Yamanishi}}, \bibinfo {author} {\bibfnamefont {Y.}~\bibnamefont {Sumino}},\ and\ \bibinfo {author} {\bibfnamefont {Y.~T.}\ \bibnamefont {Maeda}},\ }\bibfield  {title} {\bibinfo {title} {Edge current and pairing order transition in chiral bacterial vortices},\ }\href {https://doi.org/https://doi.org/10.1073/pnas.2107461118} {\bibfield  {journal} {\bibinfo  {journal} {Proceedings of the National Academy of Sciences}\ }\textbf {\bibinfo {volume} {118}},\ \bibinfo {pages} {e2107461118} (\bibinfo {year} {2021})}\BibitemShut {NoStop}%
\bibitem [{\citenamefont {K{\"u}mmel}\ \emph {et~al.}(2013)\citenamefont {K{\"u}mmel}, \citenamefont {Ten~Hagen}, \citenamefont {Wittkowski}, \citenamefont {Buttinoni}, \citenamefont {Eichhorn}, \citenamefont {Volpe}, \citenamefont {L{\"o}wen},\ and\ \citenamefont {Bechinger}}]{Kummel2013}%
  \BibitemOpen
  \bibfield  {author} {\bibinfo {author} {\bibfnamefont {F.}~\bibnamefont {K{\"u}mmel}}, \bibinfo {author} {\bibfnamefont {B.}~\bibnamefont {Ten~Hagen}}, \bibinfo {author} {\bibfnamefont {R.}~\bibnamefont {Wittkowski}}, \bibinfo {author} {\bibfnamefont {I.}~\bibnamefont {Buttinoni}}, \bibinfo {author} {\bibfnamefont {R.}~\bibnamefont {Eichhorn}}, \bibinfo {author} {\bibfnamefont {G.}~\bibnamefont {Volpe}}, \bibinfo {author} {\bibfnamefont {H.}~\bibnamefont {L{\"o}wen}},\ and\ \bibinfo {author} {\bibfnamefont {C.}~\bibnamefont {Bechinger}},\ }\bibfield  {title} {\bibinfo {title} {Circular motion of asymmetric self-propelling particles},\ }\href {https://doi.org/10.1103/PhysRevLett.110.198302} {\bibfield  {journal} {\bibinfo  {journal} {Physical Review Letters}\ }\textbf {\bibinfo {volume} {110}},\ \bibinfo {pages} {198302} (\bibinfo {year} {2013})},\ \Eprint {https://arxiv.org/abs/1302.5787} {arXiv:1302.5787} \BibitemShut {NoStop}%
\bibitem [{\citenamefont {Nourhani}\ \emph {et~al.}(2016)\citenamefont {Nourhani}, \citenamefont {Ebbens}, \citenamefont {Gibbs},\ and\ \citenamefont {Lammert}}]{Nourhani2016}%
  \BibitemOpen
  \bibfield  {author} {\bibinfo {author} {\bibfnamefont {A.}~\bibnamefont {Nourhani}}, \bibinfo {author} {\bibfnamefont {S.~J.}\ \bibnamefont {Ebbens}}, \bibinfo {author} {\bibfnamefont {J.~G.}\ \bibnamefont {Gibbs}},\ and\ \bibinfo {author} {\bibfnamefont {P.~E.}\ \bibnamefont {Lammert}},\ }\bibfield  {title} {\bibinfo {title} {{Spiral diffusion of rotating self-propellers with stochastic perturbation}},\ }\href@noop {} {\bibfield  {journal} {\bibinfo  {journal} {Physical review E}\ }\textbf {\bibinfo {volume} {94}},\ \bibinfo {pages} {030601(R)} (\bibinfo {year} {2016})}\BibitemShut {NoStop}%
\bibitem [{\citenamefont {Soni}\ \emph {et~al.}(2019)\citenamefont {Soni}, \citenamefont {Bililign}, \citenamefont {Magkiriadou}, \citenamefont {Sacanna}, \citenamefont {Bartolo}, \citenamefont {Shelley},\ and\ \citenamefont {Irvine}}]{Soni2019}%
  \BibitemOpen
  \bibfield  {author} {\bibinfo {author} {\bibfnamefont {V.}~\bibnamefont {Soni}}, \bibinfo {author} {\bibfnamefont {E.~S.}\ \bibnamefont {Bililign}}, \bibinfo {author} {\bibfnamefont {S.}~\bibnamefont {Magkiriadou}}, \bibinfo {author} {\bibfnamefont {S.}~\bibnamefont {Sacanna}}, \bibinfo {author} {\bibfnamefont {D.}~\bibnamefont {Bartolo}}, \bibinfo {author} {\bibfnamefont {M.~J.}\ \bibnamefont {Shelley}},\ and\ \bibinfo {author} {\bibfnamefont {W.~T.~M.}\ \bibnamefont {Irvine}},\ }\bibfield  {title} {\bibinfo {title} {The odd free surface flows of a colloidal chiral fluid},\ }\href {https://doi.org/10.1038/s41567-019-0603-8} {\bibfield  {journal} {\bibinfo  {journal} {Nature Physics}\ }\textbf {\bibinfo {volume} {15}},\ \bibinfo {pages} {1188} (\bibinfo {year} {2019})}\BibitemShut {NoStop}%
\bibitem [{\citenamefont {Witten}\ and\ \citenamefont {Diamant}(2020)}]{Witten2020}%
  \BibitemOpen
  \bibfield  {author} {\bibinfo {author} {\bibfnamefont {T.~A.}\ \bibnamefont {Witten}}\ and\ \bibinfo {author} {\bibfnamefont {H.}~\bibnamefont {Diamant}},\ }\bibfield  {title} {\bibinfo {title} {{A review of shaped colloidal particles in fluids: Anisotropy and chirality}},\ }\href@noop {} {\bibfield  {journal} {\bibinfo  {journal} {Reports on progress in physics}\ }\textbf {\bibinfo {volume} {83}},\ \bibinfo {pages} {116601} (\bibinfo {year} {2020})}\BibitemShut {NoStop}%
\bibitem [{\citenamefont {Avron}\ \emph {et~al.}(1995)\citenamefont {Avron}, \citenamefont {Seiler},\ and\ \citenamefont {Zograf}}]{avron1995viscosity}%
  \BibitemOpen
  \bibfield  {author} {\bibinfo {author} {\bibfnamefont {J.~E.}\ \bibnamefont {Avron}}, \bibinfo {author} {\bibfnamefont {R.}~\bibnamefont {Seiler}},\ and\ \bibinfo {author} {\bibfnamefont {P.~G.}\ \bibnamefont {Zograf}},\ }\bibfield  {title} {\bibinfo {title} {Viscosity of quantum hall fluids},\ }\href@noop {} {\bibfield  {journal} {\bibinfo  {journal} {Physical review letters}\ }\textbf {\bibinfo {volume} {75}},\ \bibinfo {pages} {697} (\bibinfo {year} {1995})}\BibitemShut {NoStop}%
\bibitem [{\citenamefont {Avron}(1998)}]{Avr98}%
  \BibitemOpen
  \bibfield  {author} {\bibinfo {author} {\bibfnamefont {J.~E.}\ \bibnamefont {Avron}},\ }\bibfield  {title} {\bibinfo {title} {Odd viscosity},\ }\href@noop {} {\bibfield  {journal} {\bibinfo  {journal} {Journal of statistical physics}\ }\textbf {\bibinfo {volume} {92}},\ \bibinfo {pages} {543} (\bibinfo {year} {1998})}\BibitemShut {NoStop}%
\bibitem [{\citenamefont {Banerjee}\ \emph {et~al.}(2017)\citenamefont {Banerjee}, \citenamefont {Souslov}, \citenamefont {Abanov},\ and\ \citenamefont {Vitelli}}]{banerjee2017odd}%
  \BibitemOpen
  \bibfield  {author} {\bibinfo {author} {\bibfnamefont {D.}~\bibnamefont {Banerjee}}, \bibinfo {author} {\bibfnamefont {A.}~\bibnamefont {Souslov}}, \bibinfo {author} {\bibfnamefont {A.~G.}\ \bibnamefont {Abanov}},\ and\ \bibinfo {author} {\bibfnamefont {V.}~\bibnamefont {Vitelli}},\ }\bibfield  {title} {\bibinfo {title} {Odd viscosity in chiral active fluids},\ }\href {https://doi.org/https://doi.org/10.1038/s41467-017-01378-7} {\bibfield  {journal} {\bibinfo  {journal} {Nature communications}\ }\textbf {\bibinfo {volume} {8}},\ \bibinfo {pages} {1573} (\bibinfo {year} {2017})}\BibitemShut {NoStop}%
\bibitem [{\citenamefont {Epstein}\ and\ \citenamefont {Mandadapu}(2020)}]{Epstein2020}%
  \BibitemOpen
  \bibfield  {author} {\bibinfo {author} {\bibfnamefont {J.~M.}\ \bibnamefont {Epstein}}\ and\ \bibinfo {author} {\bibfnamefont {K.~K.}\ \bibnamefont {Mandadapu}},\ }\bibfield  {title} {\bibinfo {title} {{Time-reversal symmetry breaking in two-dimensional nonequilibrium viscous fluids}},\ }\href {https://doi.org/10.1103/PhysRevE.101.052614} {\bibfield  {journal} {\bibinfo  {journal} {Physical review E}\ }\textbf {\bibinfo {volume} {101}},\ \bibinfo {pages} {52614} (\bibinfo {year} {2020})},\ \Eprint {https://arxiv.org/abs/1907.10041} {1907.10041} \BibitemShut {NoStop}%
\bibitem [{\citenamefont {Hargus}\ \emph {et~al.}(2020)\citenamefont {Hargus}, \citenamefont {Klymko}, \citenamefont {Epstein},\ and\ \citenamefont {Mandadapu}}]{Hargus2020}%
  \BibitemOpen
  \bibfield  {author} {\bibinfo {author} {\bibfnamefont {C.}~\bibnamefont {Hargus}}, \bibinfo {author} {\bibfnamefont {K.}~\bibnamefont {Klymko}}, \bibinfo {author} {\bibfnamefont {J.~M.}\ \bibnamefont {Epstein}},\ and\ \bibinfo {author} {\bibfnamefont {K.~K.}\ \bibnamefont {Mandadapu}},\ }\bibfield  {title} {\bibinfo {title} {{Time reversal symmetry breaking and odd viscosity in active fluids: Green-Kubo and NEMD results}},\ }\href@noop {} {\bibfield  {journal} {\bibinfo  {journal} {The journal of chemical physics}\ }\textbf {\bibinfo {volume} {152}} (\bibinfo {year} {2020})}\BibitemShut {NoStop}%
\bibitem [{\citenamefont {Han}\ \emph {et~al.}(2021)\citenamefont {Han}, \citenamefont {Fruchart}, \citenamefont {Scheibner}, \citenamefont {Vaikuntanathan}, \citenamefont {De~Pablo},\ and\ \citenamefont {Vitelli}}]{han2021fluctuating}%
  \BibitemOpen
  \bibfield  {author} {\bibinfo {author} {\bibfnamefont {M.}~\bibnamefont {Han}}, \bibinfo {author} {\bibfnamefont {M.}~\bibnamefont {Fruchart}}, \bibinfo {author} {\bibfnamefont {C.}~\bibnamefont {Scheibner}}, \bibinfo {author} {\bibfnamefont {S.}~\bibnamefont {Vaikuntanathan}}, \bibinfo {author} {\bibfnamefont {J.~J.}\ \bibnamefont {De~Pablo}},\ and\ \bibinfo {author} {\bibfnamefont {V.}~\bibnamefont {Vitelli}},\ }\bibfield  {title} {\bibinfo {title} {Fluctuating hydrodynamics of chiral active fluids},\ }\href@noop {} {\bibfield  {journal} {\bibinfo  {journal} {Nature Physics}\ }\textbf {\bibinfo {volume} {17}},\ \bibinfo {pages} {1260} (\bibinfo {year} {2021})}\BibitemShut {NoStop}%
\bibitem [{\citenamefont {Fruchart}\ \emph {et~al.}(2023)\citenamefont {Fruchart}, \citenamefont {Scheibner},\ and\ \citenamefont {Vitelli}}]{fruchartOddViscosityOdd2023}%
  \BibitemOpen
  \bibfield  {author} {\bibinfo {author} {\bibfnamefont {M.}~\bibnamefont {Fruchart}}, \bibinfo {author} {\bibfnamefont {C.}~\bibnamefont {Scheibner}},\ and\ \bibinfo {author} {\bibfnamefont {V.}~\bibnamefont {Vitelli}},\ }\bibfield  {title} {\bibinfo {title} {Odd {{Viscosity}} and {{Odd Elasticity}}},\ }\href@noop {} {\bibfield  {journal} {\bibinfo  {journal} {Annual Review of Condensed Matter Physics}\ }\textbf {\bibinfo {volume} {14}},\ \bibinfo {pages} {471} (\bibinfo {year} {2023})},\ \Eprint {https://arxiv.org/abs/2207.00071} {arXiv:2207.00071} \BibitemShut {NoStop}%
\bibitem [{\citenamefont {Hargus}\ \emph {et~al.}(2021)\citenamefont {Hargus}, \citenamefont {Epstein},\ and\ \citenamefont {Mandadapu}}]{Hargus2021}%
  \BibitemOpen
  \bibfield  {author} {\bibinfo {author} {\bibfnamefont {C.}~\bibnamefont {Hargus}}, \bibinfo {author} {\bibfnamefont {J.~M.}\ \bibnamefont {Epstein}},\ and\ \bibinfo {author} {\bibfnamefont {K.~K.}\ \bibnamefont {Mandadapu}},\ }\bibfield  {title} {\bibinfo {title} {Odd diffusivity of chiral random motion},\ }\href {https://doi.org/10.1103/PhysRevLett.127.178001} {\bibfield  {journal} {\bibinfo  {journal} {Physical Review Letters}\ }\textbf {\bibinfo {volume} {127}},\ \bibinfo {pages} {178001} (\bibinfo {year} {2021})}\BibitemShut {NoStop}%
\bibitem [{\citenamefont {Muzzeddu}\ \emph {et~al.}(2022)\citenamefont {Muzzeddu}, \citenamefont {Vuijk}, \citenamefont {L{\"o}wen}, \citenamefont {Sommer},\ and\ \citenamefont {Sharma}}]{Muzzeddu2022}%
  \BibitemOpen
  \bibfield  {author} {\bibinfo {author} {\bibfnamefont {P.~L.}\ \bibnamefont {Muzzeddu}}, \bibinfo {author} {\bibfnamefont {H.~D.}\ \bibnamefont {Vuijk}}, \bibinfo {author} {\bibfnamefont {H.}~\bibnamefont {L{\"o}wen}}, \bibinfo {author} {\bibfnamefont {J.~U.}\ \bibnamefont {Sommer}},\ and\ \bibinfo {author} {\bibfnamefont {A.}~\bibnamefont {Sharma}},\ }\bibfield  {title} {\bibinfo {title} {Active chiral molecules in activity gradients},\ }\bibfield  {journal} {\bibinfo  {journal} {Journal of Chemical Physics}\ }\textbf {\bibinfo {volume} {157}},\ \href {https://doi.org/10.1063/5.0109817} {10.1063/5.0109817} (\bibinfo {year} {2022}),\ \Eprint {https://arxiv.org/abs/2207.00315} {arXiv:2207.00315} \BibitemShut {NoStop}%
\bibitem [{\citenamefont {Kalz}\ \emph {et~al.}(2022)\citenamefont {Kalz}, \citenamefont {Vuijk}, \citenamefont {Abdoli}, \citenamefont {Sommer}, \citenamefont {L{\"o}wen},\ and\ \citenamefont {Sharma}}]{Kalz2022}%
  \BibitemOpen
  \bibfield  {author} {\bibinfo {author} {\bibfnamefont {E.}~\bibnamefont {Kalz}}, \bibinfo {author} {\bibfnamefont {H.~D.}\ \bibnamefont {Vuijk}}, \bibinfo {author} {\bibfnamefont {I.}~\bibnamefont {Abdoli}}, \bibinfo {author} {\bibfnamefont {J.~U.}\ \bibnamefont {Sommer}}, \bibinfo {author} {\bibfnamefont {H.}~\bibnamefont {L{\"o}wen}},\ and\ \bibinfo {author} {\bibfnamefont {A.}~\bibnamefont {Sharma}},\ }\bibfield  {title} {\bibinfo {title} {Collisions {{Enhance Self-Diffusion}} in {{Odd-Diffusive Systems}}},\ }\href {https://doi.org/10.1103/PhysRevLett.129.090601} {\bibfield  {journal} {\bibinfo  {journal} {Physical Review Letters}\ }\textbf {\bibinfo {volume} {129}},\ \bibinfo {pages} {90601} (\bibinfo {year} {2022})},\ \Eprint {https://arxiv.org/abs/2206.13566} {2206.13566} \BibitemShut {NoStop}%
\bibitem [{\citenamefont {Vega~Reyes}\ \emph {et~al.}(2022)\citenamefont {Vega~Reyes}, \citenamefont {{L{\'o}pez-Casta{\~n}o}},\ and\ \citenamefont {{Rodr{\'i}guez-Rivas}}}]{vegareyesDiffusiveRegimesTwodimensional2022}%
  \BibitemOpen
  \bibfield  {author} {\bibinfo {author} {\bibfnamefont {F.}~\bibnamefont {Vega~Reyes}}, \bibinfo {author} {\bibfnamefont {M.~A.}\ \bibnamefont {{L{\'o}pez-Casta{\~n}o}}},\ and\ \bibinfo {author} {\bibfnamefont {{\'A}.}~\bibnamefont {{Rodr{\'i}guez-Rivas}}},\ }\bibfield  {title} {\bibinfo {title} {Diffusive regimes in a two-dimensional chiral fluid},\ }\href {https://doi.org/10.1038/s42005-022-01032-9} {\bibfield  {journal} {\bibinfo  {journal} {Communications Physics}\ }\textbf {\bibinfo {volume} {5}},\ \bibinfo {pages} {1} (\bibinfo {year} {2022})}\BibitemShut {NoStop}%
\bibitem [{\citenamefont {Yasuda}\ \emph {et~al.}(2022)\citenamefont {Yasuda}, \citenamefont {Ishimoto}, \citenamefont {Kobayashi}, \citenamefont {Lin}, \citenamefont {Sou}, \citenamefont {Hosaka},\ and\ \citenamefont {Komura}}]{Yasuda2022}%
  \BibitemOpen
  \bibfield  {author} {\bibinfo {author} {\bibfnamefont {K.}~\bibnamefont {Yasuda}}, \bibinfo {author} {\bibfnamefont {K.}~\bibnamefont {Ishimoto}}, \bibinfo {author} {\bibfnamefont {A.}~\bibnamefont {Kobayashi}}, \bibinfo {author} {\bibfnamefont {L.~S.}\ \bibnamefont {Lin}}, \bibinfo {author} {\bibfnamefont {I.}~\bibnamefont {Sou}}, \bibinfo {author} {\bibfnamefont {Y.}~\bibnamefont {Hosaka}},\ and\ \bibinfo {author} {\bibfnamefont {S.}~\bibnamefont {Komura}},\ }\bibfield  {title} {\bibinfo {title} {Time-correlation functions for odd {{Langevin}} systems},\ }\bibfield  {journal} {\bibinfo  {journal} {Journal of Chemical Physics}\ }\textbf {\bibinfo {volume} {157}},\ \href {https://doi.org/10.1063/5.0095969} {10.1063/5.0095969} (\bibinfo {year} {2022}),\ \Eprint {https://arxiv.org/abs/2202.03225} {arXiv:2202.03225} \BibitemShut {NoStop}%
\bibitem [{\citenamefont {Ghimenti}\ \emph {et~al.}(2023)\citenamefont {Ghimenti}, \citenamefont {Berthier}, \citenamefont {Szamel},\ and\ \citenamefont {{van Wijland}}}]{ghimentiSamplingEfficiencyTransverse2023}%
  \BibitemOpen
  \bibfield  {author} {\bibinfo {author} {\bibfnamefont {F.}~\bibnamefont {Ghimenti}}, \bibinfo {author} {\bibfnamefont {L.}~\bibnamefont {Berthier}}, \bibinfo {author} {\bibfnamefont {G.}~\bibnamefont {Szamel}},\ and\ \bibinfo {author} {\bibfnamefont {F.}~\bibnamefont {{van Wijland}}},\ }\bibfield  {title} {\bibinfo {title} {Sampling {{Efficiency}} of {{Transverse Forces}} in {{Dense Liquids}}},\ }\href {https://doi.org/10.1103/PhysRevLett.131.257101} {\bibfield  {journal} {\bibinfo  {journal} {Physical Review Letters}\ }\textbf {\bibinfo {volume} {131}},\ \bibinfo {pages} {257101} (\bibinfo {year} {2023})}\BibitemShut {NoStop}%
\bibitem [{\citenamefont {Ghimenti}\ \emph {et~al.}(2024{\natexlab{a}})\citenamefont {Ghimenti}, \citenamefont {Berthier}, \citenamefont {Szamel},\ and\ \citenamefont {van Wijland}}]{ghimenti2024irreversible}%
  \BibitemOpen
  \bibfield  {author} {\bibinfo {author} {\bibfnamefont {F.}~\bibnamefont {Ghimenti}}, \bibinfo {author} {\bibfnamefont {L.}~\bibnamefont {Berthier}}, \bibinfo {author} {\bibfnamefont {G.}~\bibnamefont {Szamel}},\ and\ \bibinfo {author} {\bibfnamefont {F.}~\bibnamefont {van Wijland}},\ }\bibfield  {title} {\bibinfo {title} {Irreversible boltzmann samplers in dense liquids: Weak-coupling approximation and mode-coupling theory},\ }\href@noop {} {\bibfield  {journal} {\bibinfo  {journal} {Physical Review E}\ }\textbf {\bibinfo {volume} {110}},\ \bibinfo {pages} {034604} (\bibinfo {year} {2024}{\natexlab{a}})}\BibitemShut {NoStop}%
\bibitem [{\citenamefont {Ghimenti}\ \emph {et~al.}(2024{\natexlab{b}})\citenamefont {Ghimenti}, \citenamefont {Berthier}, \citenamefont {Szamel},\ and\ \citenamefont {van Wijland}}]{ghimenti2024transverse}%
  \BibitemOpen
  \bibfield  {author} {\bibinfo {author} {\bibfnamefont {F.}~\bibnamefont {Ghimenti}}, \bibinfo {author} {\bibfnamefont {L.}~\bibnamefont {Berthier}}, \bibinfo {author} {\bibfnamefont {G.}~\bibnamefont {Szamel}},\ and\ \bibinfo {author} {\bibfnamefont {F.}~\bibnamefont {van Wijland}},\ }\bibfield  {title} {\bibinfo {title} {Transverse forces and glassy liquids in infinite dimensions},\ }\href@noop {} {\bibfield  {journal} {\bibinfo  {journal} {Physical Review E}\ }\textbf {\bibinfo {volume} {109}},\ \bibinfo {pages} {064133} (\bibinfo {year} {2024}{\natexlab{b}})}\BibitemShut {NoStop}%
\bibitem [{\citenamefont {Langer}\ \emph {et~al.}(2024)\citenamefont {Langer}, \citenamefont {Sharma}, \citenamefont {Metzler},\ and\ \citenamefont {Kalz}}]{langerDanceOdddiffusiveParticles2024}%
  \BibitemOpen
  \bibfield  {author} {\bibinfo {author} {\bibfnamefont {A.}~\bibnamefont {Langer}}, \bibinfo {author} {\bibfnamefont {A.}~\bibnamefont {Sharma}}, \bibinfo {author} {\bibfnamefont {R.}~\bibnamefont {Metzler}},\ and\ \bibinfo {author} {\bibfnamefont {E.}~\bibnamefont {Kalz}},\ }\bibfield  {title} {\bibinfo {title} {Dance of odd-diffusive particles: {{A Fourier}} approach},\ }\href {https://doi.org/10.1103/PhysRevResearch.6.043036} {\bibfield  {journal} {\bibinfo  {journal} {Physical Review Research}\ }\textbf {\bibinfo {volume} {6}},\ \bibinfo {pages} {043036} (\bibinfo {year} {2024})}\BibitemShut {NoStop}%
\bibitem [{\citenamefont {Muzzeddu}\ \emph {et~al.}(2024)\citenamefont {Muzzeddu}, \citenamefont {Kalz}, \citenamefont {Gambassi}, \citenamefont {Sharma},\ and\ \citenamefont {Metzler}}]{muzzedduSelfdiffusionAnomaliesOdd2024}%
  \BibitemOpen
  \bibfield  {author} {\bibinfo {author} {\bibfnamefont {P.~L.}\ \bibnamefont {Muzzeddu}}, \bibinfo {author} {\bibfnamefont {E.}~\bibnamefont {Kalz}}, \bibinfo {author} {\bibfnamefont {A.}~\bibnamefont {Gambassi}}, \bibinfo {author} {\bibfnamefont {A.}~\bibnamefont {Sharma}},\ and\ \bibinfo {author} {\bibfnamefont {R.}~\bibnamefont {Metzler}},\ }\href {https://doi.org/10.48550/arXiv.2411.15552} {\bibinfo {title} {Self-diffusion anomalies of an odd tracer in soft-core media}} (\bibinfo {year} {2024}),\ \Eprint {https://arxiv.org/abs/2411.15552} {arXiv:2411.15552 [cond-mat]} \BibitemShut {NoStop}%
\bibitem [{\citenamefont {Kalz}\ \emph {et~al.}(2024)\citenamefont {Kalz}, \citenamefont {Vuijk}, \citenamefont {Sommer}, \citenamefont {Metzler},\ and\ \citenamefont {Sharma}}]{kalz2024oscillatory}%
  \BibitemOpen
  \bibfield  {author} {\bibinfo {author} {\bibfnamefont {E.}~\bibnamefont {Kalz}}, \bibinfo {author} {\bibfnamefont {H.~D.}\ \bibnamefont {Vuijk}}, \bibinfo {author} {\bibfnamefont {J.-U.}\ \bibnamefont {Sommer}}, \bibinfo {author} {\bibfnamefont {R.}~\bibnamefont {Metzler}},\ and\ \bibinfo {author} {\bibfnamefont {A.}~\bibnamefont {Sharma}},\ }\bibfield  {title} {\bibinfo {title} {Oscillatory force autocorrelations in equilibrium odd-diffusive systems},\ }\href@noop {} {\bibfield  {journal} {\bibinfo  {journal} {Physical Review Letters}\ }\textbf {\bibinfo {volume} {132}},\ \bibinfo {pages} {057102} (\bibinfo {year} {2024})}\BibitemShut {NoStop}%
\bibitem [{\citenamefont {Guo}\ \emph {et~al.}(2025)\citenamefont {Guo}, \citenamefont {Li},\ and\ \citenamefont {Ai}}]{guoDiffusionActiveParticles2025}%
  \BibitemOpen
  \bibfield  {author} {\bibinfo {author} {\bibfnamefont {R.-x.}\ \bibnamefont {Guo}}, \bibinfo {author} {\bibfnamefont {J.-j.}\ \bibnamefont {Li}},\ and\ \bibinfo {author} {\bibfnamefont {B.-q.}\ \bibnamefont {Ai}},\ }\bibfield  {title} {\bibinfo {title} {Diffusion of active particles driven by odd interactions},\ }\href {https://doi.org/10.1103/PhysRevE.111.014105} {\bibfield  {journal} {\bibinfo  {journal} {Physical Review E}\ }\textbf {\bibinfo {volume} {111}},\ \bibinfo {pages} {014105} (\bibinfo {year} {2025})}\BibitemShut {NoStop}%
\bibitem [{\citenamefont {Kogan}(2016)}]{Kogan2016}%
  \BibitemOpen
  \bibfield  {author} {\bibinfo {author} {\bibfnamefont {E.}~\bibnamefont {Kogan}},\ }\bibfield  {title} {\bibinfo {title} {Lift force due to odd {{Hall}} viscosity},\ }\href {https://doi.org/10.1103/PhysRevE.94.043111} {\bibfield  {journal} {\bibinfo  {journal} {Physical Review E}\ }\textbf {\bibinfo {volume} {94}},\ \bibinfo {pages} {043111} (\bibinfo {year} {2016})}\BibitemShut {NoStop}%
\bibitem [{\citenamefont {Reichhardt}\ and\ \citenamefont {Reichhardt}(2019)}]{Reichhardt2019}%
  \BibitemOpen
  \bibfield  {author} {\bibinfo {author} {\bibfnamefont {C.}~\bibnamefont {Reichhardt}}\ and\ \bibinfo {author} {\bibfnamefont {C.~J.~O.}\ \bibnamefont {Reichhardt}},\ }\bibfield  {title} {\bibinfo {title} {Active microrheology, {{Hall}} effect, and jamming in chiral fluids},\ }\href {https://doi.org/10.1103/physreve.100.012604} {\bibfield  {journal} {\bibinfo  {journal} {Physical Review E}\ }\textbf {\bibinfo {volume} {100}},\ \bibinfo {pages} {012604} (\bibinfo {year} {2019})},\ \Eprint {https://arxiv.org/abs/1901.11107} {arXiv:1901.11107} \BibitemShut {NoStop}%
\bibitem [{\citenamefont {Hosaka}\ \emph {et~al.}(2021)\citenamefont {Hosaka}, \citenamefont {Komura},\ and\ \citenamefont {Andelman}}]{hosakaNonreciprocalResponseTwodimensional2021}%
  \BibitemOpen
  \bibfield  {author} {\bibinfo {author} {\bibfnamefont {Y.}~\bibnamefont {Hosaka}}, \bibinfo {author} {\bibfnamefont {S.}~\bibnamefont {Komura}},\ and\ \bibinfo {author} {\bibfnamefont {D.}~\bibnamefont {Andelman}},\ }\bibfield  {title} {\bibinfo {title} {Nonreciprocal response of a two-dimensional fluid with odd viscosity},\ }\href {https://doi.org/10.1103/PhysRevE.103.042610} {\bibfield  {journal} {\bibinfo  {journal} {Physical Review E}\ }\textbf {\bibinfo {volume} {103}},\ \bibinfo {pages} {042610} (\bibinfo {year} {2021})}\BibitemShut {NoStop}%
\bibitem [{\citenamefont {Poggioli}\ and\ \citenamefont {Limmer}(2023)}]{Poggioli2023odd}%
  \BibitemOpen
  \bibfield  {author} {\bibinfo {author} {\bibfnamefont {A.~R.}\ \bibnamefont {Poggioli}}\ and\ \bibinfo {author} {\bibfnamefont {D.~T.}\ \bibnamefont {Limmer}},\ }\bibfield  {title} {\bibinfo {title} {Odd mobility of a passive tracer in a chiral active fluid},\ }\href@noop {} {\bibfield  {journal} {\bibinfo  {journal} {Physical Review Letters}\ }\textbf {\bibinfo {volume} {130}},\ \bibinfo {pages} {158201} (\bibinfo {year} {2023})}\BibitemShut {NoStop}%
\bibitem [{\citenamefont {Batton}\ and\ \citenamefont {Rotskoff}(2024)}]{battonMicroscopicOriginTunable2024}%
  \BibitemOpen
  \bibfield  {author} {\bibinfo {author} {\bibfnamefont {C.~H.}\ \bibnamefont {Batton}}\ and\ \bibinfo {author} {\bibfnamefont {G.~M.}\ \bibnamefont {Rotskoff}},\ }\bibfield  {title} {\bibinfo {title} {Microscopic origin of tunable assembly forces in chiral active environments},\ }\href {https://doi.org/10.1039/D4SM00247D} {\bibfield  {journal} {\bibinfo  {journal} {Soft Matter}\ }\textbf {\bibinfo {volume} {20}},\ \bibinfo {pages} {4111} (\bibinfo {year} {2024})}\BibitemShut {NoStop}%
\bibitem [{\citenamefont {Yang}\ \emph {et~al.}(2021)\citenamefont {Yang}, \citenamefont {Zhu}, \citenamefont {Liu}, \citenamefont {Liu}, \citenamefont {Shi}, \citenamefont {Chen}, \citenamefont {Zheng}, \citenamefont {Ye},\ and\ \citenamefont {Yang}}]{yangTopologicallyProtectedTransport2021}%
  \BibitemOpen
  \bibfield  {author} {\bibinfo {author} {\bibfnamefont {Q.}~\bibnamefont {Yang}}, \bibinfo {author} {\bibfnamefont {H.}~\bibnamefont {Zhu}}, \bibinfo {author} {\bibfnamefont {P.}~\bibnamefont {Liu}}, \bibinfo {author} {\bibfnamefont {R.}~\bibnamefont {Liu}}, \bibinfo {author} {\bibfnamefont {Q.}~\bibnamefont {Shi}}, \bibinfo {author} {\bibfnamefont {K.}~\bibnamefont {Chen}}, \bibinfo {author} {\bibfnamefont {N.}~\bibnamefont {Zheng}}, \bibinfo {author} {\bibfnamefont {F.}~\bibnamefont {Ye}},\ and\ \bibinfo {author} {\bibfnamefont {M.}~\bibnamefont {Yang}},\ }\bibfield  {title} {\bibinfo {title} {Topologically {{Protected Transport}} of {{Cargo}} in a {{Chiral Active Fluid Aided}} by {{Odd-Viscosity-Enhanced Depletion Interactions}}},\ }\href {https://doi.org/10.1103/PhysRevLett.126.198001} {\bibfield  {journal} {\bibinfo  {journal} {Physical Review Letters}\ }\textbf {\bibinfo {volume} {126}},\ \bibinfo {pages} {198001} (\bibinfo {year} {2021})}\BibitemShut {NoStop}%
\bibitem [{\citenamefont {Lier}\ \emph {et~al.}(2023)\citenamefont {Lier}, \citenamefont {Duclut}, \citenamefont {Bo}, \citenamefont {Armas}, \citenamefont {J{\"u}licher},\ and\ \citenamefont {Sur{\'o}wka}}]{lierLiftForceOdd2023}%
  \BibitemOpen
  \bibfield  {author} {\bibinfo {author} {\bibfnamefont {R.}~\bibnamefont {Lier}}, \bibinfo {author} {\bibfnamefont {C.}~\bibnamefont {Duclut}}, \bibinfo {author} {\bibfnamefont {S.}~\bibnamefont {Bo}}, \bibinfo {author} {\bibfnamefont {J.}~\bibnamefont {Armas}}, \bibinfo {author} {\bibfnamefont {F.}~\bibnamefont {J{\"u}licher}},\ and\ \bibinfo {author} {\bibfnamefont {P.}~\bibnamefont {Sur{\'o}wka}},\ }\bibfield  {title} {\bibinfo {title} {Lift force in odd compressible fluids},\ }\href {https://doi.org/10.1103/PhysRevE.108.L023101} {\bibfield  {journal} {\bibinfo  {journal} {Physical Review E}\ }\textbf {\bibinfo {volume} {108}},\ \bibinfo {pages} {L023101} (\bibinfo {year} {2023})}\BibitemShut {NoStop}%
\bibitem [{\citenamefont {Kalz}\ \emph {et~al.}(2025)\citenamefont {Kalz}, \citenamefont {Ravichandir}, \citenamefont {Birkenmeier}, \citenamefont {Metzler},\ and\ \citenamefont {Sharma}}]{kalzReversalTracerAdvection2025}%
  \BibitemOpen
  \bibfield  {author} {\bibinfo {author} {\bibfnamefont {E.}~\bibnamefont {Kalz}}, \bibinfo {author} {\bibfnamefont {S.}~\bibnamefont {Ravichandir}}, \bibinfo {author} {\bibfnamefont {J.}~\bibnamefont {Birkenmeier}}, \bibinfo {author} {\bibfnamefont {R.}~\bibnamefont {Metzler}},\ and\ \bibinfo {author} {\bibfnamefont {A.}~\bibnamefont {Sharma}},\ }\href {https://doi.org/10.48550/arXiv.2503.04544} {\bibinfo {title} {Reversal of tracer advection and {{Hall}} drift in an interacting chiral fluid}} (\bibinfo {year} {2025}),\ \Eprint {https://arxiv.org/abs/2503.04544} {arXiv:2503.04544 [cond-mat]} \BibitemShut {NoStop}%
\bibitem [{\citenamefont {Hargus}\ \emph {et~al.}(2025)\citenamefont {Hargus}, \citenamefont {Deshpande}, \citenamefont {Omar},\ and\ \citenamefont {Mandadapu}}]{hargusFluxHypothesisOdd2025}%
  \BibitemOpen
  \bibfield  {author} {\bibinfo {author} {\bibfnamefont {C.}~\bibnamefont {Hargus}}, \bibinfo {author} {\bibfnamefont {A.}~\bibnamefont {Deshpande}}, \bibinfo {author} {\bibfnamefont {A.~K.}\ \bibnamefont {Omar}},\ and\ \bibinfo {author} {\bibfnamefont {K.~K.}\ \bibnamefont {Mandadapu}},\ }\bibfield  {title} {\bibinfo {title} {Flux {{Hypothesis}} for {{Odd Transport Phenomena}}},\ }\href {https://doi.org/10.1103/PhysRevLett.134.097105} {\bibfield  {journal} {\bibinfo  {journal} {Physical Review Letters}\ }\textbf {\bibinfo {volume} {134}},\ \bibinfo {pages} {097105} (\bibinfo {year} {2025})}\BibitemShut {NoStop}%
\bibitem [{\citenamefont {Hargus}\ \emph {et~al.}()\citenamefont {Hargus}, \citenamefont {Ghimenti}, \citenamefont {Tailleur},\ and\ \citenamefont {Van~Wijland}}]{companionPRL}%
  \BibitemOpen
  \bibfield  {author} {\bibinfo {author} {\bibfnamefont {C.}~\bibnamefont {Hargus}}, \bibinfo {author} {\bibfnamefont {F.}~\bibnamefont {Ghimenti}}, \bibinfo {author} {\bibfnamefont {J.}~\bibnamefont {Tailleur}},\ and\ \bibinfo {author} {\bibfnamefont {F.}~\bibnamefont {Van~Wijland}},\ }\href@noop {} {\bibinfo {title} {Odd dynamics of passive objects in a chiral active bath (submitted to phys. rev. lett.)}}\BibitemShut {NoStop}%
\bibitem [{\citenamefont {Liu}\ \emph {et~al.}(2021)\citenamefont {Liu}, \citenamefont {Biroli}, \citenamefont {Reichman},\ and\ \citenamefont {Szamel}}]{liu2021dynamics}%
  \BibitemOpen
  \bibfield  {author} {\bibinfo {author} {\bibfnamefont {C.}~\bibnamefont {Liu}}, \bibinfo {author} {\bibfnamefont {G.}~\bibnamefont {Biroli}}, \bibinfo {author} {\bibfnamefont {D.~R.}\ \bibnamefont {Reichman}},\ and\ \bibinfo {author} {\bibfnamefont {G.}~\bibnamefont {Szamel}},\ }\bibfield  {title} {\bibinfo {title} {Dynamics of liquids in the large-dimensional limit},\ }\href@noop {} {\bibfield  {journal} {\bibinfo  {journal} {Physical Review E}\ }\textbf {\bibinfo {volume} {104}},\ \bibinfo {pages} {054606} (\bibinfo {year} {2021})}\BibitemShut {NoStop}%
\bibitem [{\citenamefont {J~Evans}\ and\ \citenamefont {P~Morriss}(2007)}]{j2007statistical}%
  \BibitemOpen
  \bibfield  {author} {\bibinfo {author} {\bibfnamefont {D.}~\bibnamefont {J~Evans}}\ and\ \bibinfo {author} {\bibfnamefont {G.}~\bibnamefont {P~Morriss}},\ }\href@noop {} {\emph {\bibinfo {title} {Statistical mechanics of nonequilbrium liquids}}}\ (\bibinfo  {publisher} {ANU Press},\ \bibinfo {year} {2007})\BibitemShut {NoStop}%
\bibitem [{\citenamefont {Mazur}\ and\ \citenamefont {Oppenheim}(1970)}]{mazur1970molecular}%
  \BibitemOpen
  \bibfield  {author} {\bibinfo {author} {\bibfnamefont {P.}~\bibnamefont {Mazur}}\ and\ \bibinfo {author} {\bibfnamefont {I.}~\bibnamefont {Oppenheim}},\ }\bibfield  {title} {\bibinfo {title} {Molecular theory of brownian motion},\ }\href@noop {} {\bibfield  {journal} {\bibinfo  {journal} {Physica}\ }\textbf {\bibinfo {volume} {50}},\ \bibinfo {pages} {241} (\bibinfo {year} {1970})}\BibitemShut {NoStop}%
\bibitem [{\citenamefont {Kim}\ and\ \citenamefont {Oppenheim}(1971)}]{kim1971molecular}%
  \BibitemOpen
  \bibfield  {author} {\bibinfo {author} {\bibfnamefont {S.}~\bibnamefont {Kim}}\ and\ \bibinfo {author} {\bibfnamefont {I.}~\bibnamefont {Oppenheim}},\ }\bibfield  {title} {\bibinfo {title} {Molecular theory of brownian motion of a rigid rotor},\ }\href@noop {} {\bibfield  {journal} {\bibinfo  {journal} {Physica}\ }\textbf {\bibinfo {volume} {54}},\ \bibinfo {pages} {593} (\bibinfo {year} {1971})}\BibitemShut {NoStop}%
\bibitem [{\citenamefont {Kim}\ and\ \citenamefont {Oppenheim}(1972)}]{kim1972molecular}%
  \BibitemOpen
  \bibfield  {author} {\bibinfo {author} {\bibfnamefont {S.}~\bibnamefont {Kim}}\ and\ \bibinfo {author} {\bibfnamefont {I.}~\bibnamefont {Oppenheim}},\ }\bibfield  {title} {\bibinfo {title} {Molecular theory of brownian motion in external fields},\ }\href@noop {} {\bibfield  {journal} {\bibinfo  {journal} {Physica}\ }\textbf {\bibinfo {volume} {57}},\ \bibinfo {pages} {469} (\bibinfo {year} {1972})}\BibitemShut {NoStop}%
\bibitem [{\citenamefont {Fily}\ \emph {et~al.}(2018)\citenamefont {Fily}, \citenamefont {Kafri}, \citenamefont {Solon}, \citenamefont {Tailleur},\ and\ \citenamefont {Turner}}]{Fily2018}%
  \BibitemOpen
  \bibfield  {author} {\bibinfo {author} {\bibfnamefont {Y.}~\bibnamefont {Fily}}, \bibinfo {author} {\bibfnamefont {Y.}~\bibnamefont {Kafri}}, \bibinfo {author} {\bibfnamefont {A.~P.}\ \bibnamefont {Solon}}, \bibinfo {author} {\bibfnamefont {J.}~\bibnamefont {Tailleur}},\ and\ \bibinfo {author} {\bibfnamefont {A.}~\bibnamefont {Turner}},\ }\bibfield  {title} {\bibinfo {title} {Mechanical pressure and momentum conservation in dry active matter},\ }\bibfield  {journal} {\bibinfo  {journal} {Journal of Physics A: Mathematical and Theoretical}\ }\textbf {\bibinfo {volume} {51}},\ \href {https://doi.org/10.1088/1751-8121/aa99b6} {10.1088/1751-8121/aa99b6} (\bibinfo {year} {2018})\BibitemShut {NoStop}%
\bibitem [{\citenamefont {van Beijeren}(1982)}]{VanBeijeren1982}%
  \BibitemOpen
  \bibfield  {author} {\bibinfo {author} {\bibfnamefont {H.}~\bibnamefont {van Beijeren}},\ }\bibfield  {title} {\bibinfo {title} {{Transport properties of stochastic Lorentz models}},\ }\href {https://doi.org/10.1103/RevModPhys.54.195} {\bibfield  {journal} {\bibinfo  {journal} {Rev. Mod. Phys.}\ }\textbf {\bibinfo {volume} {54}},\ \bibinfo {pages} {195} (\bibinfo {year} {1982})}\BibitemShut {NoStop}%
\bibitem [{\citenamefont {Granek}\ \emph {et~al.}(2022)\citenamefont {Granek}, \citenamefont {Kafri},\ and\ \citenamefont {Tailleur}}]{Granek2022}%
  \BibitemOpen
  \bibfield  {author} {\bibinfo {author} {\bibfnamefont {O.}~\bibnamefont {Granek}}, \bibinfo {author} {\bibfnamefont {Y.}~\bibnamefont {Kafri}},\ and\ \bibinfo {author} {\bibfnamefont {J.}~\bibnamefont {Tailleur}},\ }\bibfield  {title} {\bibinfo {title} {Anomalous transport of tracers in active baths},\ }\bibfield  {journal} {\bibinfo  {journal} {Physical Review Letters}\ }\textbf {\bibinfo {volume} {129}},\ \href {https://doi.org/10.1103/PhysRevLett.129.038001} {10.1103/PhysRevLett.129.038001} (\bibinfo {year} {2022})\BibitemShut {NoStop}%
\bibitem [{\citenamefont {Kubo}(1957)}]{kubo1957a}%
  \BibitemOpen
  \bibfield  {author} {\bibinfo {author} {\bibfnamefont {R.}~\bibnamefont {Kubo}},\ }\bibfield  {title} {\bibinfo {title} {Statistical-{{Mechanical Theory}} of {{Irreversible Processes}}. ({{I}}). {{General Theory}} and {{Simple Applications}} to {{Magnetic}} and {{Conduction Problems}}},\ }\href {https://doi.org/10.1143/JPSJ.12.570} {\bibfield  {journal} {\bibinfo  {journal} {Journal of the Physical Society of Japan}\ }\textbf {\bibinfo {volume} {12}},\ \bibinfo {pages} {570} (\bibinfo {year} {1957})}\BibitemShut {NoStop}%
\bibitem [{\citenamefont {Kubo}\ \emph {et~al.}(1957)\citenamefont {Kubo}, \citenamefont {Yokota},\ and\ \citenamefont {Nakajima}}]{Kubo1957b}%
  \BibitemOpen
  \bibfield  {author} {\bibinfo {author} {\bibfnamefont {R.}~\bibnamefont {Kubo}}, \bibinfo {author} {\bibfnamefont {M.}~\bibnamefont {Yokota}},\ and\ \bibinfo {author} {\bibfnamefont {S.}~\bibnamefont {Nakajima}},\ }\bibfield  {title} {\bibinfo {title} {Statistical-{{Mechanical Theory}} of {{Irreversible Processes}} ({{II}}). {{Response}} to {{Thermal Disturbance}}.},\ }\href {https://doi.org/10.1143/JPSJ.12.1203} {\bibfield  {journal} {\bibinfo  {journal} {Journal of the Physical Society of Japan}\ }\textbf {\bibinfo {volume} {12}},\ \bibinfo {pages} {1203} (\bibinfo {year} {1957})}\BibitemShut {NoStop}%
\bibitem [{\citenamefont {Chun}\ \emph {et~al.}(2018)\citenamefont {Chun}, \citenamefont {Durang},\ and\ \citenamefont {Noh}}]{chun2018emergence}%
  \BibitemOpen
  \bibfield  {author} {\bibinfo {author} {\bibfnamefont {H.-M.}\ \bibnamefont {Chun}}, \bibinfo {author} {\bibfnamefont {X.}~\bibnamefont {Durang}},\ and\ \bibinfo {author} {\bibfnamefont {J.~D.}\ \bibnamefont {Noh}},\ }\bibfield  {title} {\bibinfo {title} {Emergence of nonwhite noise in langevin dynamics with magnetic lorentz force},\ }\href@noop {} {\bibfield  {journal} {\bibinfo  {journal} {Physical Review E}\ }\textbf {\bibinfo {volume} {97}},\ \bibinfo {pages} {032117} (\bibinfo {year} {2018})}\BibitemShut {NoStop}%
\bibitem [{\citenamefont {Novikov}(1965)}]{novikov1965functionals}%
  \BibitemOpen
  \bibfield  {author} {\bibinfo {author} {\bibfnamefont {E.~A.}\ \bibnamefont {Novikov}},\ }\bibfield  {title} {\bibinfo {title} {Functionals and the random-force method in turbulence theory},\ }\href@noop {} {\bibfield  {journal} {\bibinfo  {journal} {Sov. Phys. JETP}\ }\textbf {\bibinfo {volume} {20}},\ \bibinfo {pages} {1290} (\bibinfo {year} {1965})}\BibitemShut {NoStop}%
\bibitem [{\citenamefont {Van~Kampen}(1992)}]{van1992stochastic}%
  \BibitemOpen
  \bibfield  {author} {\bibinfo {author} {\bibfnamefont {N.~G.}\ \bibnamefont {Van~Kampen}},\ }\href@noop {} {\emph {\bibinfo {title} {Stochastic processes in physics and chemistry}}},\ Vol.~\bibinfo {volume} {1}\ (\bibinfo  {publisher} {Elsevier},\ \bibinfo {year} {1992})\BibitemShut {NoStop}%
\bibitem [{\citenamefont {{Simulation and analysis code is publicly available}}()}]{github}%
  \BibitemOpen
  \bibfield  {author} {\bibinfo {author} {\bibnamefont {{Simulation and analysis code is publicly available}}},\ }\href@noop {} {}\bibinfo {note} {\href{https://github.com/chargus/chiral-active-bath}{www.github.com/chargus/chiral-active-bath.}}\BibitemShut {Stop}%
\bibitem [{\citenamefont {Plimpton}(1995)}]{lammps}%
  \BibitemOpen
  \bibfield  {author} {\bibinfo {author} {\bibfnamefont {S.~J.}\ \bibnamefont {Plimpton}},\ }\bibfield  {title} {\bibinfo {title} {Fast parallel algorithms for short-range molecular dynamics},\ }\href@noop {} {\bibfield  {journal} {\bibinfo  {journal} {Journal of computational physics}\ }\textbf {\bibinfo {volume} {117}},\ \bibinfo {pages} {1} (\bibinfo {year} {1995})},\ \bibinfo {note} {see also http://lammps.sandia.gov/}\BibitemShut {NoStop}%
\bibitem [{\citenamefont {O'Byrne}\ \emph {et~al.}(2022)\citenamefont {O'Byrne}, \citenamefont {Kafri}, \citenamefont {Tailleur},\ and\ \citenamefont {{van Wijland}}}]{obyrneTimeIrreversibilityActive2022}%
  \BibitemOpen
  \bibfield  {author} {\bibinfo {author} {\bibfnamefont {J.}~\bibnamefont {O'Byrne}}, \bibinfo {author} {\bibfnamefont {Y.}~\bibnamefont {Kafri}}, \bibinfo {author} {\bibfnamefont {J.}~\bibnamefont {Tailleur}},\ and\ \bibinfo {author} {\bibfnamefont {F.}~\bibnamefont {{van Wijland}}},\ }\bibfield  {title} {\bibinfo {title} {Time irreversibility in active matter, from micro to macro},\ }\href {https://doi.org/10.1038/s42254-021-00406-2} {\bibfield  {journal} {\bibinfo  {journal} {Nature Reviews Physics}\ }\textbf {\bibinfo {volume} {4}},\ \bibinfo {pages} {167} (\bibinfo {year} {2022})}\BibitemShut {NoStop}%
\bibitem [{\citenamefont {Granek}\ \emph {et~al.}(2024)\citenamefont {Granek}, \citenamefont {Kafri}, \citenamefont {Kardar}, \citenamefont {Ro}, \citenamefont {Tailleur},\ and\ \citenamefont {Solon}}]{granekInclusionsBoundariesDisorder2023}%
  \BibitemOpen
  \bibfield  {author} {\bibinfo {author} {\bibfnamefont {O.}~\bibnamefont {Granek}}, \bibinfo {author} {\bibfnamefont {Y.}~\bibnamefont {Kafri}}, \bibinfo {author} {\bibfnamefont {M.}~\bibnamefont {Kardar}}, \bibinfo {author} {\bibfnamefont {S.}~\bibnamefont {Ro}}, \bibinfo {author} {\bibfnamefont {J.}~\bibnamefont {Tailleur}},\ and\ \bibinfo {author} {\bibfnamefont {A.}~\bibnamefont {Solon}},\ }\bibfield  {title} {\bibinfo {title} {Colloquium: {{Inclusions}}, boundaries, and disorder in scalar active matter},\ }\href {https://doi.org/10.1103/RevModPhys.96.031003} {\bibfield  {journal} {\bibinfo  {journal} {Reviews of Modern Physics}\ }\textbf {\bibinfo {volume} {96}},\ \bibinfo {pages} {031003} (\bibinfo {year} {2024})}\BibitemShut {NoStop}%
\bibitem [{\citenamefont {Baek}\ \emph {et~al.}(2018)\citenamefont {Baek}, \citenamefont {Solon}, \citenamefont {Xu}, \citenamefont {Nikola},\ and\ \citenamefont {Kafri}}]{baekGenericLongrangeInteractions2018}%
  \BibitemOpen
  \bibfield  {author} {\bibinfo {author} {\bibfnamefont {Y.}~\bibnamefont {Baek}}, \bibinfo {author} {\bibfnamefont {A.~P.}\ \bibnamefont {Solon}}, \bibinfo {author} {\bibfnamefont {X.}~\bibnamefont {Xu}}, \bibinfo {author} {\bibfnamefont {N.}~\bibnamefont {Nikola}},\ and\ \bibinfo {author} {\bibfnamefont {Y.}~\bibnamefont {Kafri}},\ }\bibfield  {title} {\bibinfo {title} {Generic long-range interactions between passive bodies in an active fluid},\ }\href {https://doi.org/10.1103/PhysRevLett.120.058002} {\bibfield  {journal} {\bibinfo  {journal} {Physical Review Letters}\ }\textbf {\bibinfo {volume} {120}},\ \bibinfo {pages} {058002} (\bibinfo {year} {2018})},\ \Eprint {https://arxiv.org/abs/1709.02281} {arxiv:1709.02281 [cond-mat]} \BibitemShut {NoStop}%
\bibitem [{\citenamefont {Granek}\ \emph {et~al.}(2020)\citenamefont {Granek}, \citenamefont {Baek}, \citenamefont {Kafri},\ and\ \citenamefont {Solon}}]{Granek2020}%
  \BibitemOpen
  \bibfield  {author} {\bibinfo {author} {\bibfnamefont {O.}~\bibnamefont {Granek}}, \bibinfo {author} {\bibfnamefont {Y.}~\bibnamefont {Baek}}, \bibinfo {author} {\bibfnamefont {Y.}~\bibnamefont {Kafri}},\ and\ \bibinfo {author} {\bibfnamefont {A.~P.}\ \bibnamefont {Solon}},\ }\bibfield  {title} {\bibinfo {title} {Bodies in an interacting active fluid: Far-field influence of a single body and interaction between two bodies},\ }\href {https://doi.org/10.1088/1742-5468/ab7f34} {\bibfield  {journal} {\bibinfo  {journal} {Journal of Statistical Mechanics: Theory and Experiment}\ }\textbf {\bibinfo {volume} {2020}},\ \bibinfo {pages} {1} (\bibinfo {year} {2020})}\BibitemShut {NoStop}%
\end{thebibliography}%

\end{document}